\documentclass[aps,twocolumn,prc]{revtex4-1}
\usepackage[T1]{fontenc}
\usepackage[latin9]{inputenc}
\usepackage{amsthm}
\usepackage{amsmath}
\usepackage{graphicx}

\begin{document}

\title{Event-by-event fluctuations in perturbative QCD + saturation + hydro model: pinning down QCD matter shear viscosity in ultrarelativistic heavy-ion collisions}

\author{H.~Niemi${}^{a,b,c}$,  K.~J.~Eskola${}^{a,b}$, R.~Paatelainen${}^{a,b,d}$}
\affiliation{$^{a}$University of Jyv\"askyl\"a, Department of Physics, P.O. Box 35, FI-40014 University of Jyv\"askyl\"a, Finland}
\affiliation{$^b$Helsinki Institute of Physics, P.O.Box 64, FI-00014 University of
Helsinki, Finland}
\affiliation{$^c$Institut f\"ur Theoretische Physik, Johann Wolfgang Goethe-Universit\"at,
Max-von-Laue-Str. 1, D-60438 Frankfurt am Main, Germany}
\affiliation{$^d$Departamento de Fisica de Particulas, Universidade de Santiago de Compostela,
E-15782 Santiago de Compostela, Galicia, Spain} 

\begin{abstract}
We introduce an event-by-event perturbative-QCD + saturation + hydro ("EKRT") framework for ultrarelativistic heavy-ion collisions, where we compute the produced fluctuating QCD-matter energy densities from next-to-leading order perturbative QCD using a saturation conjecture to control soft particle production, and describe the space-time evolution of the QCD matter with dissipative fluid dynamics, event by event. We perform a simultaneous comparison of the centrality dependence of hadronic multiplicities, transverse momentum spectra, and flow coefficients of the azimuth-angle asymmetries, against the LHC and RHIC measurements. We compare also the computed event-by-event probability distributions of relative fluctuations of elliptic flow, and event-plane angle correlations, with the experimental data from Pb+Pb collisions at the LHC. We show how such a systematic multi-energy and multi-observable analysis tests the initial state calculation and the applicability region of hydrodynamics, and in particular how it constrains the temperature dependence of the shear viscosity-to-entropy ratio of QCD matter in its different phases in a remarkably consistent manner.
\end{abstract} 
 
 
\maketitle 

%%%%%%%%%%%%%%%%%%%%% SECTION %%%%%%%%%%%%%%%%%%%%%
\section{Introduction}

The main goal of ultrarelativistic heavy-ion collisions at the Large Hadron Collider (LHC) and the Relativistic Heavy-Ion Collider (RHIC) is to understand collectivity in the strong interaction sector of the Standard Model, and determine the properties such as temperature dependences of the shear and bulk viscosities in the different phases of QCD matter. Currently, with an increasing number of heavy-ion bulk observables from the LHC and RHIC to investigate, and with significant theoretical developments over the last decade both in computing the produced initial state from QCD and in describing the subsequent space-time evolution with dissipative fluid dynamics event by event, one is now more concretely approaching this ambitious goal.

Bulk (low-$p_T$) observables -- hadronic multiplicities, transverse momentum ($p_T$) spectra and especially the Fourier coefficients ($v_n$) of their azimuth-angle distributions -- measured in heavy-ion collisions at the LHC and RHIC, offer compelling evidence of a formation of a strongly collective locally nearly-thermalized low-viscosity hot QCD matter which undergoes both the quark-gluon plasma (QGP) and  hadron resonance gas phases. For recent reviews, see \cite{Heinz:2013th,Huovinen:2013wma,Niemi:2014lha}. The measurements are remarkably consistent with describing the space-time evolution of the formed system with dissipative relativistic fluid dynamics \cite{Romatschke:2007mq,Luzum:2008cw,Schenke:2010rr,Gale:2012rq,Song:2010mg,Song:2011qa,Shen:2010uy,Bozek:2009dw,Bozek:2012qs,Niemi:2011ix,Niemi:2012ry}. Consequently, relativistic fluid dynamics has established its role as a cornerstone in the analysis of heavy-ion bulk observables. 

One of the clearest signals of a collective behavior of the matter produced in nuclear collisions is the emergence of azimuthal asymmetries of the hadron transverse momentum spectra. In the fluid-dynamical limit the spatial inhomogeneities of the initial state are translated by the pressure gradients into the momentum space anisotropies of the spectra, and the effectiveness of this transition is essentially determined by the properties of the matter itself. It has turned out that the shear viscosity of the QCD matter strongly affects the final observed asymmetries, and therefore the measured azimuthal structure of the transverse momentum spectra (quantified by the $v_n$ coefficients)  gives the most direct constraints to  the shear viscosity.

As external input for solving the fluid-dynamical equations of motion, one needs to know the QCD equation of state (EoS) as well as event-by-event fluctuating initial conditions for the spatial distribution of energy (or entropy) density, the initial flow of the matter, and the starting time (space-time surface) for the evolution. The observable final-state momentum distributions of hadrons are obtained by computing the hadronic momentum distributions at the decoupling of the system and accounting for resonance decays after that. To model the dynamics of hadron gas, including its dissipation, decoupling and also the resonance decays, the fluid-dynamical evolution may also be coupled to a hadron cascade simulation at a suitably chosen space-time hypersurface. Such hybrid approaches have been developed e.g. in \cite{Bass:2000ib,Teaney:2000cw,Petersen:2008dd,Werner:2010aa,Song:2010aq,Song:2010mg,Werner:2012xh,Karpenko:2015xea}, see Ref.~\cite{Petersen:2014yqa} for a review. Common to the different dissipative fluid-dynamical settings, however, is that the initial conditions play a crucial role in determining the uncertainties to the QCD matter properties like the shear viscosity.

A traditionally used way to get a hold on the initial conditions (see e.g. \cite{Kolb:2001qz,Song:2010mg,Niemi:2011ix,Niemi:2012ry,Niemi:2012aj,Molnar:2014zha}) is to assume the initial energy (or entropy) densities to be a function of the Glauber model binary-collision and/or wounded-nucleon transverse densities and exploit the measured centrality dependence of various bulk observables (and more detailed observables such as relative EbyE fluctuations of $v_n$) for fixing the initial conditions in different centrality classes. A drawback in this is that there is essentially no predictability in the initial conditions when moving from one collision energy to another but the data fitting must be done for each cms-energy separately. Without considering the QCD dynamics responsible for the initial gluon and quark production one does not have enough dynamical control over the formation time of the hot system, either. In this case, the freedom in re-iterating the initial conditions complicates the determination of the matter properties such as the temperature dependence of the shear viscosity.

The uncertainties in the initial conditions, and thereby also in the QCD-matter viscosity determination, can be reduced if instead of fitting one can compute the initial conditions in a QCD-based framework. Steps into this direction include, e.g., the following approaches:  

\textbullet~ In the "IP-Glasma" initial conditions \cite{Schenke:2012wb,Gale:2012rq}, one combines the impact parameter dependent color-glass-condensate (CGC) saturation model (=IP-Sat model) with a pre-thermal classical evolution of the glasma gluon fields. Combined with the MUSIC fluid-dynamics code \cite{Schenke:2010rr,Schenke:2010nt}, such initial conditions have been particularly successful in explaining, e.g., the relative EbyE fluctuations of $v_n$ measured by ATLAS \cite{Aad:2013xma} and ALICE \cite{Timmins:2013hq}. This approach reproduces the measured $v_n$ and $v_n(p_T)$ systematics very well with an effective constant shear-viscosity-to-entropy ratio $\eta/s=0.12$ at RHIC and 0.2 at the LHC \cite{Gale:2012rq}. 

\textbullet~ The Monte Carlo version of the Kharzeev-Levin-Nardi ("MC-KLN") model \cite{Hirano:2005xf,Drescher:2006pi,Drescher:2007ax}, which is based on the CGC and $k_T$ factorization but where no pre-thermal evolution of the produced gluons is considered, has been used for obtaining the initial conditions in, e.g., \cite{Song:2010mg,Song:2013qma} for the VISHNU hybrid code \cite{Song:2010aq,Song:2012tv}. This setup gives a very good description of the measured multiplicities, $p_T$ spectra and elliptic flow of bulk hadrons at RHIC and LHC assuming a constant viscosity-to-entropy ratio in the QGP, $\eta/s=0.16$ \cite{Song:2013qma}. As discussed in \cite{Song:2010mg}, comparing the RHIC results obtained with the MC Glauber and MC-KLN initial conditions, one has arrived at an uncertainty interval $1<4\pi(\eta/s)_{\rm QGP}<2.5$. 

\textbullet~ The perturbative QCD + saturation model, often referred to as the Eskola-Kajantie-Ruuskanen-Tuominen ("EKRT") model \cite{Eskola:1999fc}, whose EbyE Next-to-leading order (NLO) extension we introduce here, combines the idea of the dominance of multiple few-GeV partonic jets, minijets, in high energy nuclear collisions \cite{Kajantie:1987pd,Eskola:1988yh} with a conjecture of saturation of gluon production to suppress the non-perturbative particle production \footnote{In this context, saturation was suggested in \cite{Eskola:1996ce}, originally the concept was introduced in \cite{Gribov:1984tu,Mueller:1985wy} and in the CGC context in \cite{McLerran:1993ni}.}. The original EKRT model \cite{Eskola:1999fc,Eskola:2001bf}, where the NLO effects in minijet transverse energy production \cite{Eskola:2000ji,Eskola:2000my} were only partially accounted for, and where only ideal 1 D and 1+1 D Bjorken hydrodynamics was applied, predicted the charged hadron multiplicities surprisingly correctly for central collisions both at the LHC \cite{Aamodt:2010pb} and RHIC \cite{Ruuskanen:2001gs}. Also the $p_T$  spectra of identified bulk hadrons at RHIC were reproduced very well \cite{Eskola:2002wx,Eskola:2005ue}. For predictions of elliptic flow in this framework, using 2+1 D ideal fluid dynamics, see \cite{Kolb:2001qz} for RHIC and \cite{Niemi:2008ta} for the LHC.

It is worth recalling here that the centrality dependence of multiplicities predicted by the EKRT model \cite{Eskola:2000xq} was first thought not to agree with the RHIC measurements, see e.g. \cite{Adcox:2000sp,Eskola:2001vs}. However, an excellent match with the data was eventually realized when the same (optical) Glauber model was used to calculate the number of participants also in the data analysis \cite{Miller:2007ri,Abelev:2008ab} --- compare Fig.~23(a) in \cite{Abelev:2008ab} and Fig.~22 (left) in \cite{Miller:2007ri} with Fig.~4 in \cite{Eskola:2000xq}. This observation also motivated us to develop the model further. In \cite{Renk:2011gj} we verified, albeit still using ideal hydrodynamics and leading order (LO) minijet cross sections, that the EKRT model was able to reproduce well the bulk (low-$p_T$) part of the LHC charged hadron $p_T$ spectrum in central Pb+Pb collisions. In \cite{Paatelainen:2012at} the model was then consistently brought to NLO, its model parameters were more precisely specified, the parameter correlations and propagation of nuclear parton distribution function (nPDF) uncertainties \cite{Eskola:2009uj} into the final multiplicities were studied, and the predictive power of the model was demonstrated.

Viscous fluid dynamics in the context of the NLO-improved EKRT model was introduced in \cite{Paatelainen:2013eea}, where we performed a simultaneous analysis of the centrality dependence of charged hadron multiplicities, $p_T$ spectra and elliptic flow, simultaneously for Pb+Pb collisions at the LHC and Au+Au at RHIC. The consistency of the EKRT results with the experimental data suggested, in terms of a linear parametrization assuming a minimum of $\eta/s$ at $T=180$~MeV, that $0.12<\eta/s< 0.12 + ({0.18}/320)(T/{\rm MeV}-180)$ in the QGP phase, and $\eta/s(T)=0.12 -({0.20}/{80})({T}/{\rm MeV}-180)$ in the hadron gas phase. Even though such a general behavior, a rising slope in $T$ in the QGP is expected on the basis of lattice QCD \cite{Nakamura:2004sy} and a decreasing one in the hadron gas on the basis of kinetic theory \cite{Csernai:2006zz}, we also had to conclude in \cite{Paatelainen:2013eea} that an equally good overall fit to the studied RHIC and LHC data can be obtained with a constant $\eta/s\approx 0.20$. In magnitude, this agrees with earlier studies \cite{Romatschke:2007mq,Luzum:2008cw,Schenke:2010rr,Schenke:2011bn,Schenke:2011tv,Gale:2012rq,Song:2010mg,Song:2011hk,Song:2011qa,Shen:2010uy,Shen:2011eg,Bozek:2009dw,Bozek:2011ua,Bozek:2012qs}. 
 
To pin down the possible temperature dependence of $\eta/s$ in the different phases of QCD matter, further constraints from analysing more detailed observables are needed. With this goal in mind, and especially for accessing higher Fourier flow-coefficients and their EbyE analysis, we introduce here for the first time an EbyE framework to the NLO-improved pQCD + saturation + viscous fluid dynamics model \cite{Paatelainen:2013eea}.  The following issues and observables are considered in what follows: 

In Sec.~II we define the 2+1 D equations of motion of longitudinally boost-invariant dissipative Israel-Stewart type transient fluid dynamics we use in this study, specify the parameters in our fluid dynamical setup, and discuss the applicability of fluid dynamics in general. We also specify the $\delta f$ corrections to the local equilibrium particle momentum distribution functions, which are applied in the computation of final state particle momentum distributions at decoupling. Unfortunately, we are not yet capable of performing a full statistical global analysis of the LHC and RHIC heavy-ion measurements to extract $\eta/s(T)$ and its uncertainty limits. However, as a step towards such an analysis, in order to demonstrate how sensitive (or, in some cases insensitive) the considered LHC and RHIC observables are to the shear viscosity, we study here the set of different parametrizations of $\eta/s(T)$ given in Sec.~\ref{sec:hydrosetup}.

In  Sec.~III we explain in detail how the NLO-improved pQCD + saturation initial conditions are obtained EbyE, first addressing the infrared (IR) and collinear (CL) safe NLO calculation of minijet transverse energy and the conjecture of saturation to obtain the saturation momentum $p_{\rm sat}$ locally in each transverse location. Accounting for the geometrical fluctuations of nucleon positions and exploiting the exclusive electroproduction measurement of $J/\psi$ mesons at HERA \cite{Chekanov:2004mw}, we build up the initial gluon clouds in the colliding nuclei. The key point enabling the EbyE framework in our case in practice, is the scaling of $p_{\rm sat}$ with the product of nuclear thickness functions of the colliding nuclei \cite{Eskola:2001rx,Paatelainen:2013eea}. From the local $p_{\rm sat}$ we then form the EbyE EKRT initial conditions, i.e., the energy densities and formation times locally in the transverse plane, addressing also the "pre-thermal" evolution to a constant longitudinal proper time $\tau_0=0.2$~fm at which we start the fluid dynamical simulation. Centrality selection and entropy production during the fluid-dynamical evolution in the EbyE case are demonstrated. Examples of the EKRT initial energy densities and eccentricities vs. centrality are given, and the effects of the key parameters in our framework on the centrality dependence of the initial state entropy, eccentricities, and $p_{\rm sat}$ are charted. 

Section IV summarizes the definitions of the flow-related observables, the $v_n$ coefficients from 2-, 3- and 4-particle cumulants, and event-plane angle correlations, which we compute in the EbyE EKRT framework and compare with experimental data.

Section V contains the results from the new EbyE EKRT framework. We perform a systematic multiobservable analysis, simultaneously for Pb+Pb collisions at the LHC and for the Au+Au collisions at the  top-energy of RHIC. We study the centrality dependence of charged hadron multiplicities, $p_T$ spectra, average $p_T$'s of the identified bulk hadrons, and in particular the charged hadron flow coefficients and event-plane angle correlations. Also the probability distributions of the relative fluctuations of elliptic flow ($\delta v_2$) are computed and compared with LHC data as well as with the relative initial eccentricity fluctuations ($\delta\epsilon_2, \delta\epsilon_{1,2}$) in our EbyE EKRT setup. The necessity of fluid dynamics in understanding the centrality systematics of these quantities is demonstrated.

In Sec.~VI we discuss the applicability limits of the pQCD + saturation + fluid dynamics framework in the light of the computed flow coefficients and event-plane angle correlations, demonstrating the effects of the $\delta f$ corrections and showing where these effects start to become too large to be trusted. 

The main conclusions from our new EbyE EKRT framework, discussed  in Sec.~VII, can be summarized as follows: The computed centrality dependence of charged hadron multiplicities, low-$p_T$ spectra, flow coefficients at the LHC and RHIC, and even the event-plane angle correlations at the LHC all agree very well with experimental data for  $\eta/s(T)=param1$, i.e. when $\eta/s(T)$ is modestly rising with $T$ in the QGP and where $\eta/s(T)$ remains small in the hadron gas phase, see Fig.~\ref{fig:etapers}. An equally good overall agreement is obtained with a constant $\eta/s=0.2$. In particular, we strongly emphasize the necessity for a simultaneous analysis of LHC and RHIC observables, from which one can obtain sufficiently independent probes simultaneously for the computed initial states, for the QCD matter $\eta/s(T)$ and also for the applicability of the fluid-dynamical framework: especially, the measured centrality systematics of the probability distributions of $\delta v_2$ test the computed initial states, while the LHC and RHIC flow-coefficient systematics together with the LHC event-plane angle correlations constrain the $\eta/s(T)$ remarkably consistently.

%%%%%%%%%%%%%%%%%%%%% SECTION %%%%%%%%%%%%%%%%%%%%%
\section{Fluid dynamics}

Fluid dynamics emerges as an approximation to the spacetime evolution of the system when the microscopic scales are small compared to the macroscopic scales like the size of the system. Basic equations for fluid dynamics are the conservation laws $\partial_\mu T^{\mu\nu} = 0$, and $\partial_\mu N_i^\mu = 0$, where $T^{\mu\nu}$ is the energy-momentum tensor and $N_i^\mu$ are the possible additional conserved currents (charge, baryon number, particle number, etc). In general, $T^{\mu\nu}$ and $N^\mu$ can be decomposed w.r.t.\ the fluid 4-velocity $u^{\mu}$, defined in the Landau frame $e u^{\mu} = T^{\mu\nu}u_\nu$, as 
\begin{eqnarray}
T^{\mu\nu} &=& e u^\mu u^\nu - P \Delta^{\mu\nu} + \pi^{\mu\nu}, \label{eq:energymomentum}\\
N_i^{\mu} &=& n_i u^\mu + n_i^\mu,
\end{eqnarray}
where $e = T^{\mu\nu} u_\mu u_\nu$ is the local energy density, $P = P_0 + \Pi$ is the isotropic pressure (sum of equilibrium pressure $P_0$ and bulk viscous pressure $\Pi$), $\pi^{\mu\nu} =T^{\langle \mu\nu \rangle}$ is the shear-stress tensor, $n_i=N_i^\mu u_\mu$ are the local particle densities, and $n_i^\mu=N_i^{\langle\mu\rangle}$ are the particle diffusion currents. The angular brackets indicate the projection operators that take the symmetric and traceless part of the tensor that is orthogonal to the fluid velocity, i.e., $A^{\langle\mu\rangle} = \Delta^{\mu\nu}A_\nu$ and 
\begin{equation}
A^{\langle\mu\nu\rangle} = \frac{1}{2}\left[\Delta^{\mu}_{\alpha}\Delta^{\nu}_{\beta} + \Delta^{\mu}_{\beta}\Delta^{\nu}_{\alpha} - \frac{2}{3} \Delta^{\mu\nu} \Delta_{\alpha\beta}\right]  A^{\alpha\beta}, 
\end{equation} 
where $\Delta^{\mu\nu} = g^{\mu\nu} - u^\mu u^\nu$, and $g^{\mu\nu}$ is the metric tensor for which we use the $g^{\mu\nu} = \rm diag(+, -, - ,-)$ convention.

The conservation laws are completely general. However, they are not enough to solve the evolution of the system, but additional constraints are needed. In the fluid dynamical approximation these additional constraints are provided by the  evolution equations for the dissipative quantities like $\pi^{\mu\nu}$. For example, in the Navier-Stokes (NS) approximation the dissipative quantities are directly proportional to the gradients of the equilibrium fields (like temperature $T$, and fluid velocity), e.g., $\pi^{\mu\nu}_{\rm NS} = 2\eta(T, \{\mu_i\})\nabla^{\langle \mu}u^{\nu\rangle}$ and $\Pi_{\rm NS} = -\zeta(T, \{\mu_i\})\nabla_{\mu}u^{\mu}$, where $\nabla^\mu = \Delta^{\mu\nu}\partial_\nu$. The microscopic properties of the matter are then integrated into the coefficients $\eta(T, \{\mu_i\})$ and $\zeta(T, \{\mu_i\})$, which in general depend on the temperature $T$ and the chemical potentials $\{\mu_i\}$ associated with the conserved charges. It is, however, known that the relativistic NS theory is not intrinsically stable, i.e., even the hydrostatic equilibrium is linearly unstable \cite{Hiscock:1983zz, Pu:2009fj}. Therefore, the relativistic NS theory is not suitable for the full dynamical description of the system.

\subsection{Transient fluid dynamics}

The reason for the instability of the NS theory can be traced to the fact that the resulting equations of motion are parabolic. Therefore, in this theory the signal propagation speed is not limited, and can exceed the speed of light, rendering the theory acausal, which in turn makes the theory unstable \cite{Pu:2009fj}. This problem is solved in the Israel-Stewart theory \cite{Israel:1979wp} by taking into account a part of the microscopic transient dynamics, e.g. the shear-stress tensor relaxes towards the NS values within the relaxation time $\tau_\pi$ and not instantaneously like in the NS theory. The relaxation times $\tau_i$ are fundamental properties of the matter similarly to the transport coefficients introduced above, and in general they can depend on temperature and chemical potentials. 

In this work, we use the equations of motion (e.o.m.) derived from kinetic theory~\cite{Israel:1979wp, Denicol:2010xn, Denicol:2012cn, Denicol:2012es, Molnar:2013lta, Betz:2008me, Betz:2010cx}. Transient fluid dynamics can be derived from a microscopic theory by expanding around an equilibrium state and neglecting all the microscopic time scales except the slowest one \cite{Denicol:2011fa}. This procedure leads to relaxation type equations of motion for the dissipative quantities, e.g.\ the evolution equations for the shear-stress tensor read \cite{Denicol:2012cn, Molnar:2013lta}, 
\begin{eqnarray}
 \tau_\pi \frac{d}{d\tau}\pi^{\langle \mu \nu \rangle} + \pi^{\mu\nu} & = & 2\eta \sigma^{\mu\nu}  + c_1 \pi^{\mu\nu} \nabla^{\alpha} u_{\alpha} +  c_2 \pi_{\alpha} ^{\langle \mu} \sigma^{\nu\rangle \alpha} \notag \\ 
 & + & c_3 \pi_\alpha^{\langle \mu} \omega^{\nu\rangle \alpha} + c_4 \pi_\alpha^{\langle \mu} \pi^{\nu\rangle \alpha},
\label{eq:IShydro}
\end{eqnarray}
where the terms up to the first order in gradients (or Knudsen number, a ratio of microscopic and macroscopic time/length scales, such as ${\rm Kn}\sim \tau_\pi \nabla_{\mu}u^{\mu}$, \cite{Niemi:2014wta}), second order in inverse Reynolds number $\sim \pi^{\mu\nu}/P_0$, and product of inverse Reynolds and Knudsen number are included. Here $\sigma^{\mu\nu} = \nabla^{\langle \mu}u^{\nu\rangle}$, and $\omega^{\mu\nu} = \frac{1}{2}\left(\nabla^{\mu}u^{\nu} - \nabla^{\nu}u^{\mu}\right)$ is the vorticity tensor. For the purposes of this work, we shall neglect the effects of bulk viscous pressure and diffusion currents, i.e., $\Pi=0=n_i^{\mu }$. Thus, all dissipative effects originate in this work only from the dynamics of the shear-stress tensor. If one includes also the bulk viscosity, several new terms that couple the shear-stress tensor and bulk pressure appear also in the e.o.m.\ of the shear-stress tensor~\cite{Denicol:2012cn, Denicol:2014vaa}. The bulk viscosity can still be important around the phase-transition, even if the bulk viscosity is negligible in the QGP and the low-temperature hadronic phase. However, the magnitude and importance of a possible large bulk viscosity near the QCD phase transition has not yet been fully established~\cite{Monnai:2009ad, Song:2009rh, Bozek:2011ua, Dusling:2011fd, Noronha-Hostler:2013gga, Noronha-Hostler:2014dqa, Rose:2014fba, Ryu:2015vwa}. 

Besides affecting the spacetime evolution of the densities and velocity, viscosity also modifies the local particle distributions. For example, in the original work by Israel and Stewart \cite{Israel:1979wp} transient fluid dynamics was derived from the Boltzmann equation by using the so-called 14-moment approximation, where the distribution function due to the non-zero shear-stress tensor is written as 
\begin{equation}
f_i(x, p) = f_{0i}(x,p) + \delta f_i = f_{0i}(x,p)\left[1 + \frac{p_{i\mu} p_{i\nu} \pi^{\mu\nu}}{2T^2(e + P_0)}\right].
\label{eq:delta_f}
\end{equation}
Here $p_i^\mu$ is the 4-momentum of the particle and $f_{0i}$ is the equilibrium distribution function,
\begin{equation}
 f_{0i}\left( x,p\right) = \frac{g_i}{\left( 2\pi \right) ^{3}} \left[ \exp\left( \frac{p_{i}^{\mu }u_{\mu }-\mu_i }{T}\right) \pm 1\right]^{-1},
\label{eq_distribution}
\end{equation}
where $g_i$ is the degeneracy factor of hadron $i$. This form of $\delta f$ does not follow uniquely from the Boltzmann equation, but is rather the first term of the full moment expansion~\cite{Denicol:2012cn}. Nevertheless, most studies of relativistic heavy-ion collisions use this form, and also we adopt this procedure here. Currently, the momentum dependence of the $\delta f$ corrections remains one of the major uncertainties in the fluid dynamical models, see e.g.\ Refs.~\cite{Dusling:2009df, Molnar:2011kx, Molnar:2014fva} for studies of the effects of different forms of $\delta f$. For an approach to derive $\delta f$ corrections from a simplified microscopic theory, i.e., relaxation time approximation to the Boltzmann equation, see \cite{Jaiswal:2013vta, Bhalerao:2013pza}.

\subsection{Applicability of fluid dynamics}

Fluid dynamics becomes a good approximation when gradients are sufficiently small and the evolution of the macroscopic variables is slow compared to the microscopic time scales. The systems formed in heavy-ion collisions are, however, very small and their lifetime is short, and these conditions are not trivially fulfilled. The estimates of the Knudsen numbers, i.e. ratio of microscopic and macroscopic scales, reached in the collisions indicate that even with small values of shear viscosity, there can still be large corrections to the fluid dynamical evolution~\cite{Niemi:2014wta}. Especially in the low density hadronic matter, where viscosity is expected to become large~\cite{Prakash:1993bt,Csernai:2006zz,Gorenstein:2007mw,NoronhaHostler:2008ju,Wiranata:2013oaa}, the fluid dynamical treatment becomes less reliable. In particular, this is true for the decoupling from a fluid to free particles, a process that cannot even in principle be fully described by fluid dynamics. Therefore, even if the fluid dynamical models have been very successful in describing the low-$p_T$ hadron spectra measured at RHIC and LHC energies, it is still not clear in how detail one should trust the fluid dynamical description, and what are its limitations.

It is then clear that reaching the final goal of determining the transport properties of the matter from the experimental data requires that also the uncertainties related to the fluid dynamical evolution are systematically charted. There are currently a few ways of extending the applicability of fluid dynamics. For example, the moment expansion of the Boltzmann equation provides a way to include in principle arbitrary orders of the gradients into the description, and it has been shown that including all the second order terms consistently into the description is essential in describing the detailed structure of shock waves~\cite{Denicol:2012vq}. One of the characteristics of heavy-ion collisions is that the early expansion is highly asymmetric, i.e.\ the system starts with a fast longitudinal expansion, and transverse expansion develops only later. This kind of anisotropic expansion results in also highly anisotropic local momentum distributions, which can lead to a breaking of the usual fluid dynamical description. This is the motivation for the so-called anisotropic hydrodynamics \cite{Martinez:2012tu, Bazow:2013ifa, Florkowski:2013lya}, where the functional form of the expansion around the equilibrium state is designed to allow large deviations from an isotropic momentum distributions. Neither of these methods are, however, applied to a full description of heavy-ion collisions, yet.

One of the important conditions for the applicability of fluid dynamics is that different systems should be described by the same transport coefficients that can depend on temperature and chemical potentials, but not e.g.\ on the collision energy or the nuclear mass number. 

\subsection{Our fluid dynamical setup}
\label{sec:hydrosetup}

In this work we employ the setup previously used in Refs.~\cite{Niemi:2011ix, Niemi:2012ry, Niemi:2012aj, Paatelainen:2013eea}, where the longitudinal expansion is approximated by a scaling flow consistent with longitudinal boost-invariance. In this approximation the longitudinal flow velocity is given by $v_z = z/t$, and the components of the energy-momentum tensor, Eq.~\eqref{eq:energymomentum}, become independent of the spacetime rapidity $\eta_{s} = (1/2) \ln\left[(t+z)/(t-z) \right]$, i.e., they depend on the transverse coordinates, $\mathbf{r} = (x,y) $, and the longitudinal proper time, $\tau = \sqrt{t^2-z^2}$, only. From a numerical point of view, this reduces the (3+1)--dimensional problem to a (2+1)--dimensional one. 

The coefficients of the non-linear terms in the equations of motion for the shear-stress tensor, Eq.~\eqref{eq:IShydro}, are taken from the 14-moment approximation to the ultrarelativistic gas~\cite{Denicol:2010xn, Denicol:2012cn, Molnar:2013lta}, i.e., $c_1 = -(4/3)\tau_\pi$, $c_2 = -(10/7)\tau_\pi$, $c_3 = 2\tau_\pi$, and $c_4 = 9/(70 P_0)$, and the relation between the relaxation time $\tau_\pi$ and the shear viscosity is
\begin{equation}
 \tau_\pi = \frac{5\eta}{e+P_0}.
 \label{eq:relaxation_time}
\end{equation}

In thermodynamical equilibrium, the properties of the matter are essentially given by the EoS that gives pressure as a function of temperature. Here we use the $s95p$-PCE-v1 parametrization of lattice QCD results at zero net-baryon density~\cite{Huovinen:2009yb}. The high-temperature part of this EoS is from the hotQCD collaboration~\cite{Cheng:2007jq, Bazavov:2009zn} and it is smoothly connected to a hadron resonance gas, where resonances up to mass of $2$ GeV are included. The hadronic part of the EoS includes a chemical freeze-out at $T_{\mathrm{chem}}=175$ MeV, where all stable hadron ratios are fixed~\cite{Bebie:1991ij, Hirano:2002ds, Huovinen:2007xh}. A hadron is considered stable, if its lifetime is more than 10 fm. In the perfect fluid limit the construction of the chemical freeze-out also conserves the number of stable particles. However, in the viscous fluid there is still small (approximately $1\%$) entropy production below $T_{\mathrm{chem}}=175$ MeV, and this leads to a small increase in the number of particles during the evolution of chemically frozen hadronic matter.

%%%%%%%%%%%%%%%%%%%%% FIGURE %%%%%%%%%%%%%%%%%%%%%
\begin{figure}
\includegraphics[width=8.5cm]{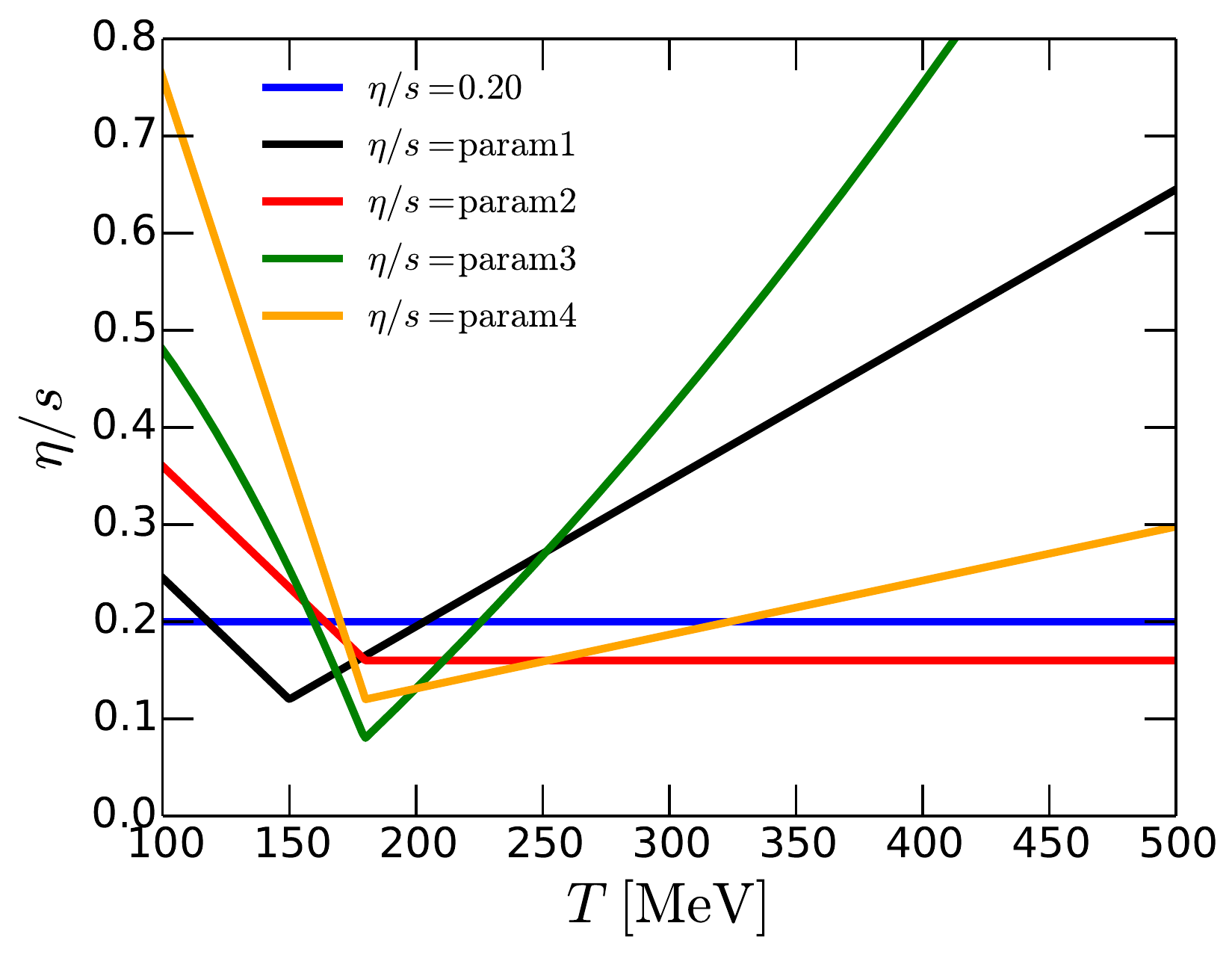}
\caption{(Color online) Parametrizations of the temperature dependence of the shear-viscosity to entropy ratio, labelled here in the order of increasing $\eta/s$  at $T=100$~MeV. For more details, see the text and Table~\ref{tab:etapers}.}
\label{fig:etapers}
\end{figure}
%%%%%%%%%%%%%%%%%%%%% FIGURE %%%%%%%%%%%%%%%%%%%%%

Once the transport coefficients and EoS above are given, the only degrees of freedom left are the shear viscosity to entropy density ratio $\eta/s(T)$ and the initial components $T^{\mu\nu}(\tau_0, \mathbf{r})$. In the boost-invariant approximation it is enough to specify $T^{\mu\nu}(\tau_0, \mathbf{r})$ in the transverse plane at some initial proper time $\tau_0$. The initial conditions calculated from the EbyE EKRT setup are discussed in detail in the next section.

As shown in Fig.~\ref{fig:etapers}, we parametrize the temperature dependence of the $\eta/s$ ratio in a similar manner as we did in \cite{Paatelainen:2013eea}, by assuming a minimum of $\eta/s$ at $T=T_{\rm min}$ to be somewhere in the cross-over temperature-region and a linearly rising (decreasing) behavior in the QGP (HRG) phase. Table~\ref{tab:etapers} shows the  corresponding parameters from which these linear slopes can be constructed. We have converged into these parametrizations iteratively, requiring them to reproduce the measured 2-particle cumulant elliptic flow $v_2\{2\}$ (see Sec.~\ref{sec:cumulants} for the definition) in mid-peripheral collisions at the LHC. In addition, we also exploit the \emph{HH-HQ} parametrization of Ref. \cite{Denicol:2010tr, Niemi:2011ix, Niemi:2012ry} (used later also in Ref.~\cite{Gale:2012rq}), which features a rapid growth of $\eta/s(T)$ in the QGP combined with a more modest decrease in the hadron gas phase. We label the above parametrizations here as $param1$, $param2$, $param3\equiv$ \emph{HH-HQ}, and $param4$, in the order of an increasing value of $\eta/s$ at $T_{\rm dec} = 100$ MeV. As we will show, a simultaneous comparison with the RHIC results is then necessary to see the sensitivity to $\eta/s(T)$. As indicated in Fig.~\ref{fig:etapers}, we perform the calculations also for a constant $\eta/s=0.2$, keeping also this value unchanged from the LHC to RHIC. The sensitivity of the computed $v_n$ to a constant $\eta/s=0.2\pm 0.1$ will be demonstrated.
\begin{table}[h]
\caption{The constant-slope parametrizations of $\eta/s(T)$, constructed so that they reproduce the LHC $v_n$ data.}
\begin{tabular}{ccccc}
\hline
\hline
			 &  $T_{\rm min}/{\rm MeV}$ & $(\eta/s)_{\rm min}$ & $\eta/s(100\,{\rm MeV})$ & $\eta/s(500\,{\rm MeV})$ \\
\hline
param1 & 150& 0.12& 0.24& 0.65\\
param2 & 180& 0.16& 0.36& 0.16\\
param4 & 180& 0.12& 0.76& 0.30\\
\hline
\hline
\end{tabular}
\label{tab:etapers}
\end{table}

Once the initial conditions, EoS, and the transport coefficients are given, the equations of motion for shear-stress tensor, Eq.~\eqref{eq:IShydro}, and the conservation laws form a closed system of equations that can be solved numerically to obtain the spacetime evolution of all the quantities appearing in the energy-momentum tensor, Eq.~\eqref{eq:energymomentum}. The numerical algorithm employed here to solve the equations of motion is introduced and discussed in Refs.~\cite{Molnar:2009tx, Niemi:2012ry}.

\subsection{The freeze-out stage}
\label{sec:freeze_out}

The fluid dynamical quantities are not directly comparable to the experimental data. Therefore, it is necessary to convert them into the experimentally observable hadron transverse momentum spectra. Here we employ the standard Cooper-Frye procedure~\cite{Cooper:1974mv}, where the spectrum is calculated as the number of particles crossing some surface $\Sigma$ whose normal vector is $d^{3}\Sigma_{\mu}$. This leads to a Lorentz-invariant spectrum for a hadron $i$,
\begin{equation}
 E\frac{d^3N_i}{d^3p} = \frac{1}{2}\frac{d^3N_i}{dy dp_T^2 d\phi} = \int_{\Sigma}d^3\Sigma_{\mu}(x)\,p^{\mu}f(x,p)\,,
   \label{Cooper-Frye}
\end{equation}
where $p^{\mu } = \left( E,\mathbf{p}\right)$ denotes the four-momentum of the hadron, and $f\left( x,p\right)$ is the single-particle distribution function, Eq.~\eqref{eq:delta_f}, of the hadron on the surface. In the boost-invariant approximation the spectrum is independent of the rapidity $y$. 

In this work we take the freeze-out surface to be a constant-temperature surface with $T_{\rm dec} = 100$ MeV, which gives a good agreement with the slopes of the measured charged hadron $p_T$ spectra. A more physical way would be to decouple the system dynamically on a surface where the expansion rate of the system becomes of the same magnitude as the average scattering or thermalization rate (here $\tau_\pi$), i.e. when ${\rm Kn} \sim 1$ \cite{Bondorf:1978kz, Dumitru:1999ud, Hung:1997du, Heinz:2007in, Eskola:2007zc, Holopainen:2012id, Molnar:2014zha}. However, in practice, the differences to the constant-temperature freeze-out are quite modest, especially near midrapidity.

In principle, the Cooper-Frye integral \eqref{Cooper-Frye} should be calculated for all the hadronic states included into the EoS, i.e., up to a mass $2$ GeV. However, in order to save computational time, we include here hadrons only up to a mass $1.5$ GeV. In practice, the effect on the final results shown here is negligible. All the strong and electromagnetic two-- and three--particle decays of the hadronic resonances (most of the hadrons in the EoS are unstable and decay before they can be observed) are calculated here according to Ref.~\cite{Sollfrank:1991xm}. In finding the constant temperature hypersurfaces, we employ the Cornelius algorithm~\cite{Huovinen:2012is}.

%%%%%%%%%%%%%%%%%%%%% SECTION %%%%%%%%%%%%%%%%%%%%%
\section{Initial conditions from the local EKRT saturation model}

Let us then discuss the details of the NLO-improved pQCD + local saturation framework \cite{Paatelainen:2012at,Paatelainen:2013eea}, which combines a NLO pQCD computation of the minijet transverse energy $E_T$ production with saturation of gluon production. First, we discuss the computation for averaged (smooth) initial conditions, after which we explain how the event-by-event setup utilizes these calculations.

\subsection{Minijet $E_T$ production in $A$+$A$ collisions}
\label{sec:pQCD}

For a given collision energy $\sqrt{s_{NN}}$ and nuclear mass number $A$ the initial minijet $E_T$ produced perturbatively into a rapidity window $\Delta y$ in $A$+$A$ collisions and above a transverse momentum scale $p_0 \gg \Lambda_{\rm QCD}$, can be computed as \cite{Paatelainen:2012at}
\begin{equation}
\frac{\mathrm{d}E_T}{\mathrm{d}^2\mathbf{r}}(p_0,\sqrt{s_{NN}},A,\mathbf{r},\mathbf{b};\beta) = T_A(\mathbf{r}_{1} )T_A(\mathbf{r}_{2}) \sigma\langle E_T,\rangle_{p_0,\Delta y, \beta} 
\end{equation}
where $\mathbf{r}_{1/2} = \mathbf{r} \pm \mathbf{b}/2$ with ${\bf r}=(x,y)$ denoting the transverse coordinate and $\mathbf{b}$ the impact parameter. The nuclear collision geometry is given by the nuclear thickness function 
\begin{equation}
T_A(\mathbf{r}) = \int_{-\infty}^{\infty} \mathrm{d}z \rho_A(\mathbf{r},z),
\end{equation} 
where the nuclear density $\rho_A(\mathbf{r},z)$ is parametrized with the standard Woods-Saxon (WS) profile 
\begin{equation}
\label{eq:WoodSaxon}
\rho_A(\mathbf{r},z) = \frac{n_0}{\exp\left (\frac{\sqrt{\vert \mathbf{r} \vert^2+ z^2} - R_A}{d} \right ) + 1},
\end{equation}
with the nuclear radius $R_A = (1.12A^{1/3} - 0.86A^{-1/3})$ fm, $d=0.54$ fm, and 
$n_0= 3A/(4\pi R_A^3)[(1+\pi^2d^2/R_A^2)]^{-1}\approx 0.17$ fm$^{-3}$.  According to collinear factorization and pQCD, the first $E_T$-moment of the minijet $E_T$ distribution, $\sigma\langle E_T\rangle_{p_0,\Delta y,\beta}$, in NLO is computed as \cite{Paatelainen:2012at,Eskola:1988yh,Eskola:2000my,Eskola:2000ji}
\begin{equation}
\label{eq:sigmaET}
\sigma\langle E_T\rangle_{p_0,\Delta y,\beta} \equiv \int_{0}^{\sqrt{s_{NN}}} \mathrm{d}E_T E_T\frac{\mathrm{d}\sigma}{\mathrm{d}E_T}\bigg\vert_{p_0,\Delta y,\beta},
\end{equation}
where the  semi-inclusive $E_T$ distribution of minijets in a rapidity interval $\Delta y$ in $N$+$N$ collisions is given by
\begin{equation}
\label{eq:incET}
\frac{\mathrm{d}\sigma}{\mathrm{d}E_T}\bigg\vert_{p_0,\Delta y,\beta} = \sum_{n=2}^{3}\frac{1}{n!}\int \mathrm{d[PS]}_n\frac{\mathrm{d}\sigma^{2\rightarrow n}}{\mathrm{d[PS]}_n}\mathcal{S}_n.
\end{equation}
Here, the $n$-particle momentum phase-space integration $\mathrm{d[PS]}_n$ takes place in $4-2\varepsilon$ dimensions, and we have introduced a compact notation for the differential NLO partonic cross sections $\mathrm{d}\sigma^{2\rightarrow 2}/\mathrm{d[PS]}_2$ and  $\mathrm{d}\sigma^{2\rightarrow 3}/\mathrm{d[PS]}_3$, corresponding to the $(2\rightarrow 2)$ and $(2 \rightarrow 3)$ scatterings, respectively. A detailed discussion of the $\mathrm{d}\sigma^{2\rightarrow n}/\mathrm{d[PS]}_n$ which consist of (NLO, $\overline{\mathrm{MS}}$ scheme) PDFs and squared spin- and color-summed/averaged scattering matrix elements, summed over all possible parton types, is given in \cite{Kunszt:1992tn,Eskola:2000my}. 

The IR and CL singularities present in the partonic cross sections at order $\alpha_{\rm s}^3$ are regulated by computing the $(2\rightarrow 2)$  and $(2\rightarrow 3)$ squared matrix elements in $4-2\epsilon$ dimensions. The ultraviolet divergences present in the $(2\rightarrow 2)$ parts are taken care of by renormalization using dimensional regularization and the $\overline{\mathrm{MS}}$ scheme. The full analytical calculation for these squared matrix elements was done first in \cite{Ellis:1985er}, and details of some of these rather complicated calculations are given in \cite{Paatelainen:2014fsa}. The phase space differentials ${\mathrm{d[PS]}_2}$ and ${\mathrm{d[PS]}_3}$ stand for
\begin{equation}
\begin{split}
\mathrm{d[PS]}_2 &= \mathrm{d}p_{T2}\mathrm{d}y_1\mathrm{d}y_2\mathrm{d}^{1-2\epsilon}\phi_2,\\
\mathrm{d[PS]}_3 &= \mathrm{d}p_{T2} \mathrm{d}p_{T3}\mathrm{d}y_1\mathrm{d}y_2\mathrm{d}y_3\mathrm{d}^{1-2\epsilon}\phi_2\mathrm{d}^{1-2\epsilon}\phi_3,
\end{split}
\end{equation}
where the appropriate kinematical variables for the two- and three-parton phase spaces are the transverse momenta $p_{Ti} = \vert \mathbf{p}_{Ti}\vert$, rapidities $y_i$ and azimuth angles $\phi_i$. For the two-parton final state, the transverse momentum conservation determines $p_{T1} = p_{T2}$ and $\phi_1 = \phi_2 +\pi$, and similarly for the three-parton final state $\mathbf{p}_{T1} = -(\mathbf{p}_{T2}  + \mathbf{p}_{T3})$.  The measurement functions ${\mathcal{S}}_2$ and ${\mathcal{S}}_3$ in Eq.~\eqref{eq:incET} specify the physical quantity to be computed.  As explained in \cite{Kunszt:1992tn},  the cancellation of the remaining IR and CL singularities between the UV renormalized squared  $(2\rightarrow 2)$ and $(2\rightarrow 3)$ matrix elements takes place only if the measurement function $S_3$ reduces to the $S_2$ in the soft (the energy of one of the final-state partons vanishes) and collinear (one of the final state particles becomes collinear with any other particle) limits. 

In our case, the measurement functions define the total minijet $E_T$ produced into a mid-rapidity window $\Delta y$ defined in the $(y,\phi)$-plane as 
\begin{equation}
\Delta y: \quad \vert y\vert \leq 0.5, \quad 0 \leq \phi \leq 2\pi.
\end{equation}
The minijet $E_T$ entering $\Delta y$ is defined here 
as a sum of the transverse momenta $p_{Ti}$ of those final-state partons whose rapidities are in $\Delta y$
\begin{equation}
\begin{split}
E_{T}   = \sum_{i=1}^{n=2,3} \Theta(y_i \in \Delta y)p_{Ti} ,
\end{split}
\end{equation}
where all partons are assumed massless and $\Theta$ is the standard step function. For computing the minijet $E_T$ distribution, our measurement functions must also specify which scatterings are to be considered hard and thus included in the perturbative calculation. We define the hard perturbative scatterings to be those with large enough  transverse momentum produced, regardless of where the partons go in rapidity,
\begin{equation}
\begin{split}
\sum_{i=1}^{n=2,3} p_{Ti} \ge 2p_0,
\end{split}
\end{equation}
where $p_0 \gg \Lambda_{\mathrm{QCD}}$.

Now, for the $(2\rightarrow 2)$ hard processes transverse momentum conservation ensures that if at least one parton falls into our rapidity acceptance, then $E_T\ge p_0$. However, in the $(2\rightarrow 3)$ case we may have processes which fulfil the requirement of being hard $(p_{T1} + p_{T2} + p_{T3} \geq 2p_0)$ but bring less than $p_0$ of $E_T$ in $\Delta y$. This happens, e.g., for configurations where two hard partons fall outside $\Delta y$ and only one softer parton with $p_T < p_0$ enters $\Delta y$. Therefore,  the remaining freedom in defining the measurement function $S_3$ is that in the $(2\rightarrow 3)$ case we may still restrict the amount of the minimum $E_T$ in $\Delta y$ in an IR/CL safe way. In \cite{Paatelainen:2012at} it was shown that in the  $S_3$ case in fact any minimum amount, $E_T \geq \beta p_0$, where  $0\le \beta \le 1$,  gives an equally good IR/CL safe restriction for the $E_T$ in $\Delta y$, which relaxes back to the $S_2$ case in the soft and collinear limits.

Thus, the IR- and CL-safe measurement functions $S_2$ and $S_3$ can now be written down by combining the definition of minijet $E_T$ in $\Delta y$, the definition of the hard perturbative scatterings and the restriction of minimum $E_T$ discussed above,
\begin{equation}
\begin{split}
\mathcal{S}_n =    \delta &\left (E_T - \biggl [\sum_{i=1}^{n}\Theta(y_i\in \Delta y)p_{Ti}\biggr]  \right )\\
& \times \Theta \left (\sum_{i=1}^{n}p_{Ti}  \geq 2p_0 \right )\times \Theta(E_T \geq \beta p_0), 
\end{split}
\end{equation}
where $\beta$ is a phenomenological parameter to be determined from the experimental data. Next, integrating the delta functions away in Eq.~\eqref{eq:sigmaET} we obtain
\begin{equation}
\label{eq:sigmaET2}
\sigma\langle E_T\rangle_{p_0,\Delta y,\beta} =  \sum_{n=2}^{3}\frac{1}{n!}\int \mathrm{d[PS]}_n\frac{\mathrm{d}\sigma^{2\rightarrow n}}{\mathrm{d[PS]}_n}\tilde{\mathcal{S}}_n,
\end{equation}
where the IR and CL safe measurement functions for the first $E_T$ moment are denoted by 
\begin{equation}
\begin{split}
\label{eq:mfunctions}
\tilde{\mathcal{S}}_n = 
\left(\sum_{i=1}^{n} \Theta(y_i \in \Delta y)p_{Ti}\right) 
\times 
\Theta \left (\sum_{i=1}^{n}p_{Ti}  \geq 2p_0 \right )\\
\times 
\Theta\left(\left[\sum_{i=1}^{n} \Theta(y_i \in \Delta y)p_{Ti}\right] \geq \beta p_0\right). 
\end{split}
\end{equation}

The numerical computation for the rather complicated six-dimensional integrals 
\footnote{The measurement functions in Eq.~\eqref{eq:mfunctions} are azimuthally symmetric and thus the $\phi_2$-integrals can be done trivially} 
in Eq.~\eqref{eq:sigmaET2} is performed with Monte Carlo integration, using an updated version of the code developed for \cite{Eskola:2000my,Eskola:2000ji,Paatelainen:2012at} where the $(2\rightarrow 3)$ parts and their partonic book-keeping are based on the Ellis-Kunszt-Soper jet code 
\footnote{S.~D.~Ellis, Z.~Kunszt, and D.~E.~Soper, \textit{JET version 3.4.}} 
\cite{Kunszt:1992tn}. For the DGLAP evolved nPDFs, we apply the NLO CTEQ6M free proton PDFs \cite{Pumplin:2002vw} together with the latest set of transverse-coordinate ($T_A({\bf r})$) dependent NLO EPS09s nuclear effects \cite{Helenius:2012wd}. The implementation of these spatial nuclear effects is done as instructed in \cite{Helenius:2012wd}, calculating the results directly for each {\bf r} and {\bf b}; for details, see \cite{Helenius:2012wd}. The renormalization scale $\mu_R$ and factorization scale $\mu_F$ are chosen equal, $\mu_R = \mu_F = \mu$.  We set the scale $\mu$ to be proportional to the total transverse momentum produced in the hard perturbative scattering, regardless of the partons being in $\Delta y$ or not: 
\begin{equation}
\begin{split}
\mu  = \frac{C}{2} \left( \sum_{i=1}^n p_{Ti} \right)
\end{split}
\end{equation}
where the constant $C$ is set to unity. 

\subsection{Local saturation of minijet $E_T$ production}
\label{sec:saturation}

As explained in \cite{Paatelainen:2012at}, the low-transverse-momentum parton (dominantly gluon) production can be conjectured to be controlled by saturation of minijet $E_T$ production. In this new EKRT approach the saturation takes place when $(3\rightarrow 2)$ and higher-order partonic processes start to dominate over the conventional $(2\rightarrow 2)$  processes (and $(2\rightarrow 3)$ at higher orders). Thus, at saturation, we require that the rapidity densities of the produced $E_T$ fulfil the condition
\begin{equation}
\label{eq:ETsat}
\frac{\mathrm{d}E_T}{\mathrm{d}^2\mathbf{r}\mathrm{d}y}(2\rightarrow 2)  \sim
\frac{\mathrm{d}E_T}{\mathrm{d}^2\mathbf{r}\mathrm{d}y}(3\rightarrow 2).
\end{equation}
To LO in $\alpha_{\rm s}$, the l.h.s. scales as
\begin{equation}
\label{eq:sat2to2}
\frac{\mathrm{d}E_T}{\mathrm{d}^2\mathbf{r}\mathrm{d}y}(2\rightarrow 2)   \sim (T_Ag)^2\left ( \frac{\alpha_{\rm s}^2}{p_0^2}\right )p_0,
\end{equation}
where we assign the factor $T_Ag$ for each of the incoming gluons, $\alpha_{\rm s}^2/p_0^2$ for the $\sigma(2\rightarrow  2)$ partonic cross section and the cut-off scale $p_0$ for the $E_T$. Here, $g$ denotes the gluon PDFs. Similarly, for the r.h.s $(3\rightarrow 2)$ term in Eq.~\eqref{eq:ETsat} we may write
\begin{equation}
\label{eq:sat2to3}
\frac{\mathrm{d}E_T}{\mathrm{d}^2\mathbf{r}\mathrm{d}y}(3\rightarrow 2)   \sim (T_Ag)^3\frac{1}{p_0^2}\left ( \frac{\alpha_{\rm s}^3}{p_0^2}\right )p_0,
\end{equation}
where the scale $p_0^{-2}$ is to compensate the fm$^{-2}$ dimension of the extra $T_A$ in Eq.~\eqref{eq:sat2to3}. Substituting the Eqs.~\eqref{eq:sat2to2} and \eqref{eq:sat2to3} into the saturation condition \eqref{eq:ETsat}, we get
\begin{equation}
(T_Ag)^2\left ( \frac{\alpha_{\rm s}^2}{p_0^2}\right )p_0 \sim (T_Ag)^3\frac{1}{p_0^2}\left ( \frac{\alpha_{\rm s}^3}{p_0^2}\right )p_0,
\label{eq:sat_scaling}
\end{equation}
which leads to a scaling $T_Ag \sim p_0^2/\alpha_{\rm s}$ for the gluon density probed at saturation. Feeding this scaling law back to the saturation condition in Eq.~\eqref{eq:ETsat}, we obtain a transversally local saturation criterion for the minijet $E_T$ production in $A$+$A$ collisions at non-zero impact parameters \cite{Paatelainen:2013eea},
\begin{equation}
\label{eq:satcri}
\frac{\mathrm{d}E_T}{\mathrm{d}^2\mathbf{r}}(p_0,\sqrt{s_{NN}},A,\Delta y,\mathbf{r},\mathbf{b};\beta) = \frac{K_{\textrm{sat}}}{\pi}p_0^3\Delta y,
\end{equation}
with an unknown (but to a first approximation $\alpha_{\rm s}$-independent) proportionality constant $K_{\textrm{sat}} \sim 1$, whose value needs to be determined from the data.  Once the saturation scale is obtained as the solution $p_0 = p_{\textrm{sat}}(\sqrt{s_{NN}},A,\Delta y,\mathbf{r},\mathbf{b};\beta,K_{\textrm{sat}})$ of Eq.~\eqref{eq:satcri}, we get the total amount of minijet transverse energy  $\mathrm{d}E_T(p_0=p_{\rm sat})/\mathrm{d}^2\mathbf{r}$ produced into a mid-rapidity window $\Delta y$.

\subsection{Numerical implementation}

The procedure to obtain the locally saturated NLO minijet $E_T$ is straightforward, but the challenges in the numerical implementation are worth mentioning. First, with the spatially dependent nPDFs the computation of the locally saturated NLO $\mathrm{d}E_T/\mathrm{d}^2\mathbf{r}(p_{\rm sat})$ becomes slow, mainly due to the multidimensional MC integrations in the $(2\rightarrow3)$ parts. Second, since $p_{\rm sat}$ can be determined from Eq.~\eqref{eq:satcri} through iteration only, we need $\mathrm{d}E_T/\mathrm{d}^2\mathbf{r}(p_0)$ for {\cal O}(10) different $p_0$'s at each {\bf r} for each {\bf b}. Third, the spatial $(x,y)$ grid for constructing initial conditions for fluid dynamics has to be dense enough, say $\Delta x=\Delta y= 0.4$~fm, and extend far enough, at least to $r\sim R_A$ where the approach can still be imagined to work. In one quarter-plane we then have to compute the saturated minijet $E_T$ in {\cal O}(250) different $(x,y)$ points for each {\bf b}. Fourth, and worst, we have to determine the free parameters $K_{\rm sat}$ and $\beta$ iteratively on the basis of the centrality dependence of the bulk data, i.e., after performing the hydrodynamic evolution for all centrality classes with initial conditions computed for each $K_{\rm sat},\beta$ pair with a given $\eta/s$. Thus, a blindly repeated NLO computation of locally saturated averaged initial conditions for such an iterative procedure becomes numerically too slow, and the EbyE framework would then seem just impossible. 

The first key-observation in circumventing the above critical slowness problems, made in \cite{Paatelainen:2013eea}, is that to a good approximation the "K-factor" 
\begin{equation}
K\equiv \sigma\langle E_T\rangle_{p_0,\Delta y,\beta}({\rm \scriptstyle NLO})/
\sigma\langle E_T\rangle_{p_0,\Delta y,\beta}({\rm \scriptstyle LO})
\label{eq:Kfactor}
\end{equation}
does not depend on the PDFs (free proton, nuclear or spatial). Then, the full NLO result can be approximated by implementing the spatial nPDFs into the fast LO part only, and using the K-factors to account for the NLO effects, i.e. 
$$
\sigma\langle E_T\rangle_{p_0,\Delta y,\beta}({\rm \scriptstyle NLO, EPS09s}) \approx 
\sigma\langle E_T\rangle_{p_0,\Delta y,\beta}({\rm \scriptstyle LO, EPS09s}) \times K,  
$$
where the K-factor has been computed only once, with the free-proton PDFs. According to the checks we have made over the $(x,y)$ plane, this approximates the full NLO result very well, within a few percents both at RHIC and LHC.

%%%%%%%%%%%%%%%%%%%%% FIGURE %%%%%%%%%%%%%%%%%%%%%
\begin{figure}
\includegraphics[width=8.0cm]{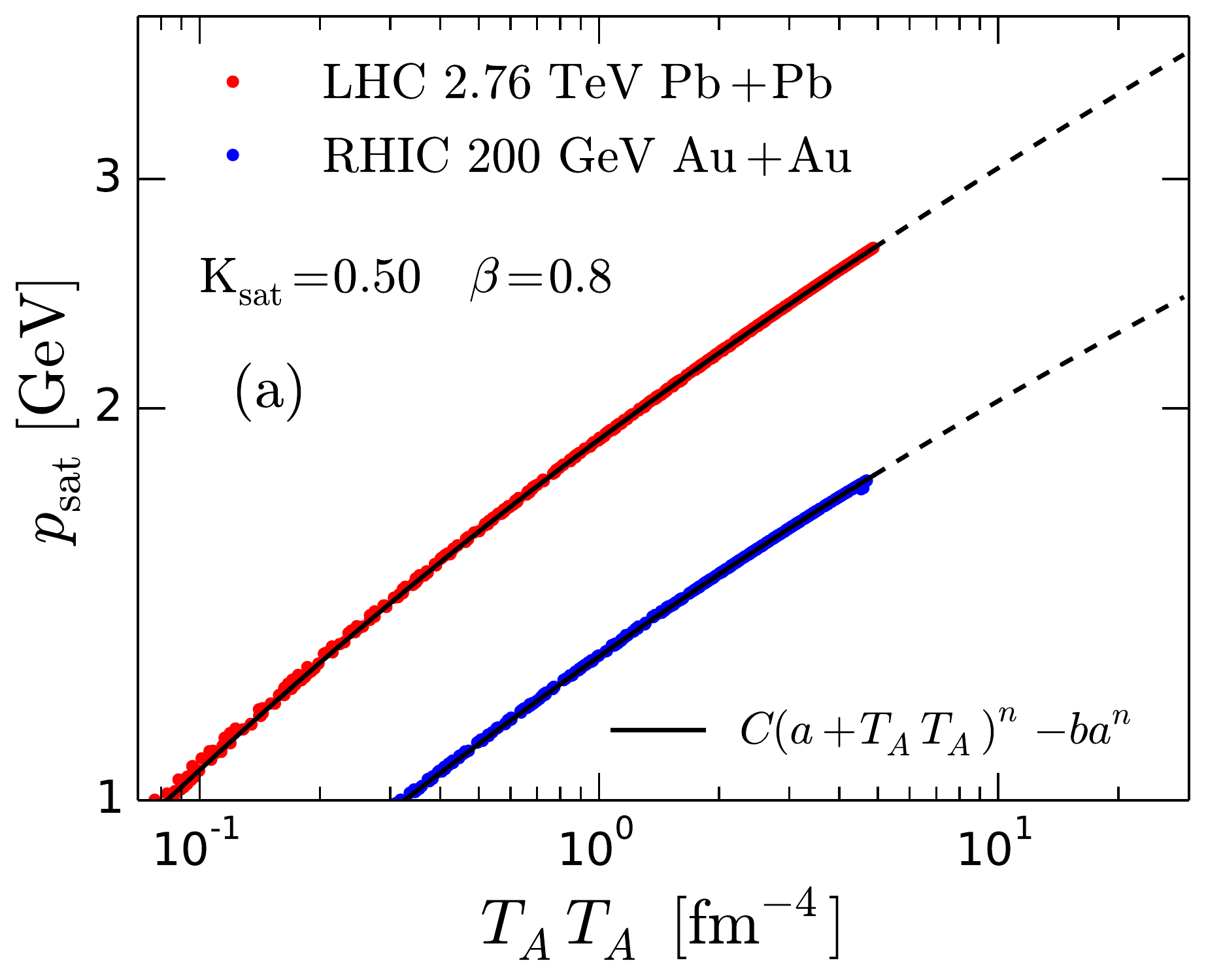}
\includegraphics[width=8.0cm]{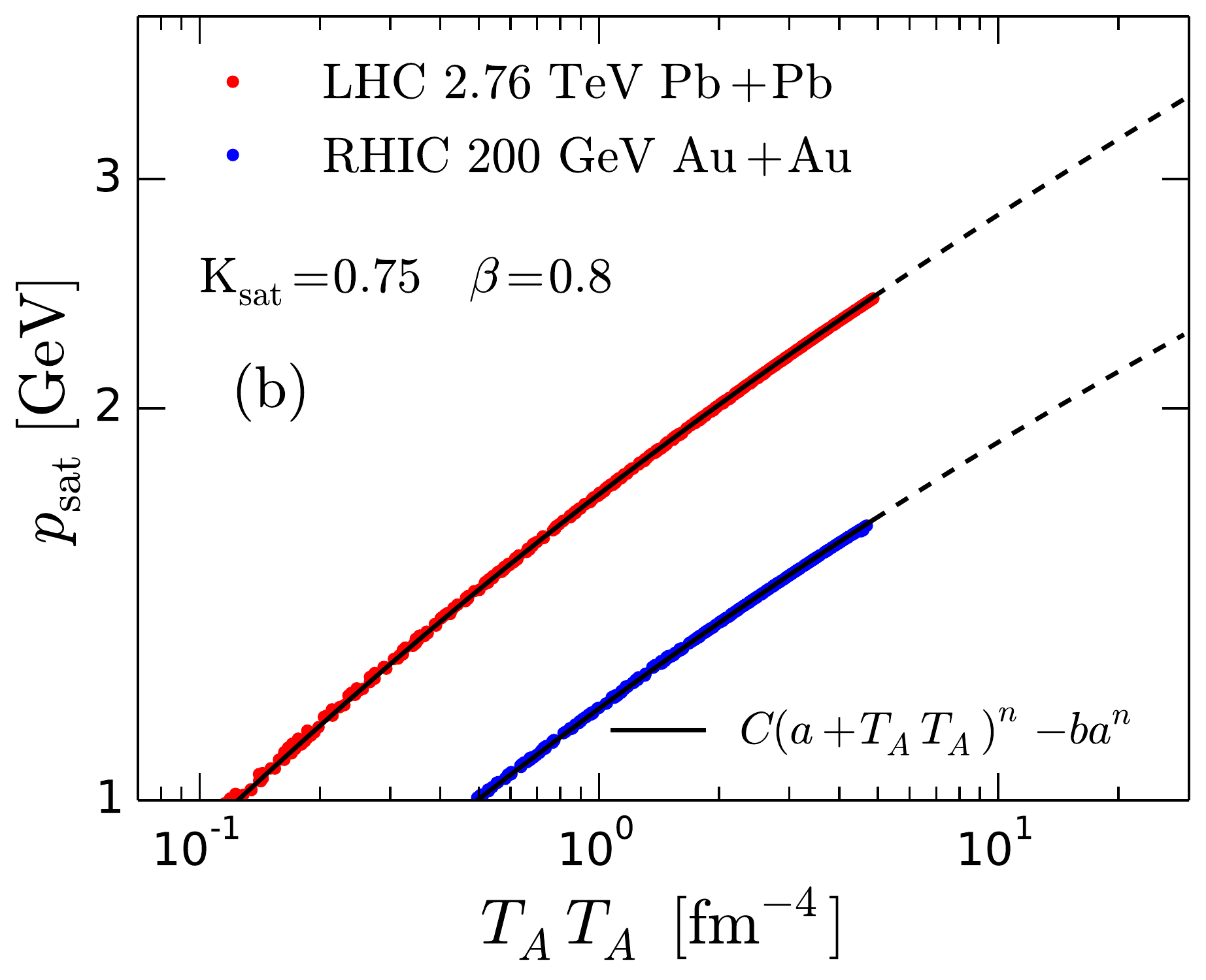}
\caption{(Color online) Saturation momentum $p_{\rm sat}$ as a function of nuclear overlap density $T_AT_A$ with $K_{\rm sat}=0.5$ (a) and $K_{\rm sat}=0.75$ (b) in Pb+Pb collisions at the LHC (red points), and in Au+Au collisions at RHIC (blue points), calculated with several different impact parameters. The dashed lines show the corresponding parametrization \eqref{eq:psat_param}, and its extrapolation to the typical highest $T_AT_A$'s we encounter in the EbyE analysis.}
\label{fig:psat_TATA}
\end{figure}
%%%%%%%%%%%%%%%%%%%%% FIGURE %%%%%%%%%%%%%%%%%%%%%

The second key-observation enabling the locally saturated EKRT framework is demonstrated in Fig.~\ref{fig:psat_TATA} which shows the calculated values of $p_{\rm sat}$ as a function of the nuclear overlap density,
\begin{equation}
 \rho_{AA}\left( \mathbf{r} \right) = T_A\left( \mathbf{r}-\frac{\mathbf{b}}{2} \right) T_A\left( \mathbf{r} + \frac{\mathbf{b}}{2} \right),
 \label{eq:rhoAB}
\end{equation}
in $\sqrt{s_{NN}} = 200$ GeV Au+Au collisions at RHIC and in $\sqrt{s_{NN}} = 2.76$  TeV Pb+Pb collisions at the LHC with $\beta = 0.8$ and $K_{\rm sat} = 0.5$ (a) and $0.75$ (b). The blue and red points in the figure are from the pQCD+saturation calculation at different transverse positions and with several different impact parameters. Thus, for a fixed cms-energy and collision system,  the computed local saturation scale $p_{\rm sat}(x,y)$ is to a very good approximation only a function of the $\rho_{AA}$, and furthermore, the function is the \textit{same} for all centrality classes. 

The emergence of such a scaling can be understood as follows. In the naive scaling limit, where the minijet $\sigma \langle E_T\rangle\propto p_0^{-1}$, the saturation criterion \eqref{eq:satcri} leads to the scaling $p_{\rm sat}^2\propto (\rho_{AA})^\delta$ with $\delta = 1/2$. As discussed in Ref.~\cite{Eskola:2001rx} (in LO, without nPDFs), corrections to the power $\delta$ can be traced back to the $x-$ and $Q^2$-slopes of the small-$x$ gluon distribution, phase-space integration and running of $\alpha_{\rm s}$. 

Figure \ref{fig:psat_TATA} now shows that also the NLO calculation with nPDFs preserves the power-law scaling property of $p_{\rm sat}$ extremely well for a fixed cms-energy and for a fixed nucleus $A$. The spatial effects in the nPDFs could still modify this scaling from one impact parameter to another. Figure \ref{fig:psat_TATA} shows, however, that the these effects  are so small that the $\rho_{AA}$ dependence of $p_{\rm sat}$ is to a good approximation universal over all centralities. Thus, we can very accurately parametrize the saturation scale as   
\begin{equation}
 p_{\rm sat}(\rho_{AA}) = C\left[a + \rho_{AA}\right]^n - b C a^n,
 \label{eq:psat_param}
\end{equation}
where $a$, $b$, $C$ and $n$ are parameters that depend on $A$, $\sqrt{s_{NN}}$, $K_{\rm sat}$ and $\beta$. For a given $A$ and $\sqrt{s_{NN}}$ the ($K_{\rm sat}, \beta$)--dependence can be parametrized by a polynomial, 
\begin{equation}
\begin{split}
 P_i(K_{\rm sat}, \beta) &= a_{i0} + a_{i1} K_{\rm sat} + a_{i2} \beta \\ 
&+ a_{i3}K_{\rm sat}\beta + a_{i4}\beta^2 + a_{i5} K_{\rm sat}^2.
\end{split} 
\end{equation}
The coefficients $a_{ij}$ for the parameters $a$, $b$, $C$ and $n$ are listed in Tables~\ref{tab:psat_parametrization_LHC_betalow}, \ref{tab:psat_parametrization_LHC_betahigh}, \ref{tab:psat_parametrization_RHIC_betalow}, and \ref{tab:psat_parametrization_RHIC_betahigh} for $\sqrt{s_{NN}} = 2.76$ TeV Pb+Pb collisions and $\sqrt{s_{NN}} = 200$ GeV Au+Au collisions. Note that the parametrizations are found separately for $\beta<0.9$ and $\beta>0.9$. Armed with the above parametrization of $p_{\rm sat}(\rho_{AA})$, we have been able to chart the $K_{\rm sat},\beta$ plane for finding the initial conditions discussed next, and develop the EbyE framework. 
\begin{table}[h]
\caption{The parametrization of $p_{\rm sat}(K_{\rm sat}, \beta)$ for $\sqrt{s_{NN}} = 2.76$ TeV Pb+Pb collisions for $K_{\rm sat} \in [0.4, 2.0]$ and $\beta<0.9$}
\begin{tabular}{c|cccc}
\hline
\hline
$P_i\rightarrow$ & $C$          & $n$                  & $a$             & $b$ \\  
\hline
$a_{i0}$	& 3.9027590    & 0.1312476            &  -0.0044020     &  0.8537670 \\
$a_{i1}$	& -0.6277216   & -0.0157637           &  0.0220154      &  -0.0580163 \\
$a_{i2}$	& 1.0703962    & -0.0362980           &  -0.0005974     &  0.0957157 \\
$a_{i3}$	& 0.0692793    & -0.0022506           &  0.0125320      &  -0.0016413 \\
$a_{i4}$	& -1.9808449   &  0.0615129           &  -0.0032844     &  -0.1788390 \\
$a_{i5}$	& 0.1106879    &  0.0052116           &  -0.0033841     &  0.0220187 \\
\hline
\hline
\end{tabular}
\label{tab:psat_parametrization_LHC_betalow}
\end{table}

\begin{table}[h]
\caption{The parametrization of $p_{\rm sat}(K_{\rm sat}, \beta)$ for $\sqrt{s_{NN}} = 2.76$ TeV Pb+Pb collisions for $K_{\rm sat} \in [0.4, 2.0]$ and $\beta>0.9$}
\begin{tabular}{c|cccc}
\hline
\hline
$P_i\rightarrow$ & $C$          & $n$                  & $a$             & $b$ \\  
\hline
$a_{i0}$	 &  27.3259359  	& -1.9924684       &  0.1038047       &  0.5211725 \\
$a_{i1}$	 &  -0.3371381          & 0.0835716        &  0.0539039       &  -0.6286044 \\
$a_{i2}$	 &  -42.6176287         & 4.1698751        &  -0.2099840      &  2.5059182 \\
$a_{i3}$	 &  -0.1844621          & -0.1206132       &  -0.0144174      &  0.7131778 \\
$a_{i4}$	 &  17.6786774          &  -1.9891770      &  0.0950212       &  -2.5125962 \\
$a_{i5}$	 &  0.3092463           &  0.0003279       &  0.0014117       &  0.0150475 \\
\hline
\hline
\end{tabular}
\label{tab:psat_parametrization_LHC_betahigh}
\end{table}

\begin{table}[h]
\caption{The parametrization of $p_{\rm sat}(K_{\rm sat}, \beta)$ for $\sqrt{s_{NN}} = 200$ GeV Au+Au collisions for $K_{\rm sat} \in [0.4, 2.0]$ and $\beta<0.9$}
\begin{tabular}{c|cccc}
\hline
\hline
$P_i\rightarrow$ & $C$          & $n$                  & $a$             & $b$ \\  
\hline
$a_{i0}$	 &  10.3313939          & 0.0303079       &  -0.0070317       &  0.9381026  \\
$a_{i1}$	 &  -0.3165983          & -0.0024562      &  0.1561924        &  -0.0005718  \\
$a_{i2}$	 &  -12.8128174         & 0.0139955       &  -0.0026174       &  0.0376918  \\
$a_{i3}$	 &  -0.0273664          & -0.0017971      &  -0.0369552       &  0.0072667  \\
$a_{i4}$	 &  4.6810067           &  0.0923750      &  -0.0174187       &  -0.3018326  \\
$a_{i5}$	 &  0.0527041           &  0.0005875      &  -0.0226980       &  0.0013976  \\
\hline
\hline
\end{tabular}
\label{tab:psat_parametrization_RHIC_betalow}
\end{table}

\begin{table}[h]
\caption{The parametrization of $p_{\rm sat}(K_{\rm sat}, \beta)$ for $\sqrt{s_{NN}} = 200$ GeV Au+Au collisions for $K_{\rm sat} \in [0.4, 2.0]$ and $\beta>0.9$}
\begin{tabular}{c|cccc}
\hline
\hline
$P_i\rightarrow$ & $C$          & $n$                  & $a$             & $b$ \\  
\hline
$a_{i0}$	&  91.4314177             & -0.4406026      &  0.7332375     &  3.0875818  \\
$a_{i1}$	&  2.5123667              & 0.0782859       &  0.2132747     &  -0.2205018  \\
$a_{i2}$	&  -165.8206094           & 0.6486681       &  -1.5009886    &  -3.8563125  \\
$a_{i3}$	&  -2.6487281             & -0.1005554      &  -0.0219393    &  0.2777689  \\
$a_{i4}$	&  77.0170469             &  -0.0909378     &  0.7419402     &  1.5327054  \\
$a_{i5}$	&  0.2192064              &  0.0004503      &  -0.0336409    &  -0.0006138  \\
\hline
\hline
\end{tabular}
\label{tab:psat_parametrization_RHIC_betahigh}
\end{table}

\subsection{Initial state for fluid dynamical evolution}
\label{sec:iniforhydro}

As initial conditions, our boost-invariant dissipative fluid-dynamical modeling requires the transverse energy density $e(\mathbf{r},\tau_0)$, transverse velocity $\mathbf{v}_T(\mathbf{r},\tau_0)$ and initial shear stress tensor $\pi^{\mu\nu}(\mathbf{r},\tau_0)$ at a constant initialization proper time $\tau_0$ of fluid-dynamics. 

In this work the initial transverse velocity and shear stress tensor are chosen to be zero. The transverse profile for the local initial energy density at the formation (production) of the system is computed similarly as in Refs.~\cite{Eskola:2001bf,Eskola:2005ue,Paatelainen:2013eea},
\begin{equation}
e(\mathbf{r},\tau_{\mathrm{s}}(\mathbf{r})) = \frac{\mathrm{d}E_T}{\mathrm{d}^2\mathbf{r}}\frac{1}{\tau_{\mathrm{s}} (\mathbf{r}) \Delta y } = \frac{K_{\textrm{sat}}}{\pi}[p_{\textrm{sat}}(\mathbf{r})]^4,
\end{equation}
where the local formation time of the minijet plasma at each transverse point $\mathbf{r}$ is given by $\tau_{\rm s}(\mathbf{r}) = 1/p_{\textrm{sat}}(\mathbf{r})$. Since for the fluid-dynamical evolution we need the initial state at a fixed time, the computed energy densities have to be evolved to the same $\tau_0$ at each $\mathbf{r}$. To do this, we first set a minimum scale $p^{\textrm{min}}_{\textrm{sat}} = 1$ GeV for which we assume that we can still trust the pQCD calculation. This corresponds to a maximum formation time $\tau_0 = 1/p^{\textrm{min}}_{\textrm{sat}} \approx 0.2$ fm in our pQCD+saturation setup. Next, the uncertainties in the "pre-thermal" evolution from $\tau_{\mathrm{s}}(\mathbf{r})$ to $\tau_0$ can be studied by considering the two limits:  1) the Bjorken free streaming (FS) scaling 
\begin{equation}
e(\mathbf{r},\tau_0) = e(\mathbf{r},\tau_{\rm s}(\mathbf{r})) \left (\frac{\tau_{\rm s}(\mathbf{r})}{\tau_0} \right )
\end{equation}
which preserves the transverse energy, and 2) the Bjorken hydrodynamic scaling solution (BJ)
\begin{equation}
e(\mathbf{r},\tau_0) = e(\mathbf{r},\tau_{\rm s}(\mathbf{r})) \left (\frac{\tau_{\rm s}(\mathbf{r})}{\tau_0} \right )^{4/3},
\end{equation}
where a maximum amount of energy is transfered into the longitudinal direction by the $P_0 dV$ work. As discussed in \cite{Paatelainen:2013eea}, due to the freedom we still have in fixing $(K_{\rm sat},\beta)$, our final results will be relatively insensitive to the pre-thermal evolution. For this reason, in the present study we stick to the latter (BJ) case.

Finally, we need the initial energy densities at the edges of the system which are outside the applicability region of our pQCD+saturation model, i.e. the energy densities below $e_{\rm min} = K_{\rm sat}[p_{\rm sat}^{\rm min}]^4$ at $\tau_0$. To obtain these, we smoothly connect the BJ-evolved energy density to the binary profile, i.e. the energy density profile is parametrized below $e_{\rm min}$ as $e = C(T_AT_A)^n$, where the power $n$ is given by   
\begin{equation}
n = \frac{1}{2}\biggl [(k+1) + (k-1)\tanh\left ( \frac{\sigma_{\rm NN}T_AT_A - g}{\delta}\right ) \biggr ],
\end{equation}
with the total inelastic nucleon-nucleon cross-section $\sigma_{\rm NN}$ and $g=\delta=0.5$ fm$^{-2}$. The parameters $C$ and $k$ are constants that ensure a smooth connection at $e=e_{\rm min}$.

\subsection{Averaged initial conditions}

\begin{figure*}
\includegraphics[width=8.0cm]{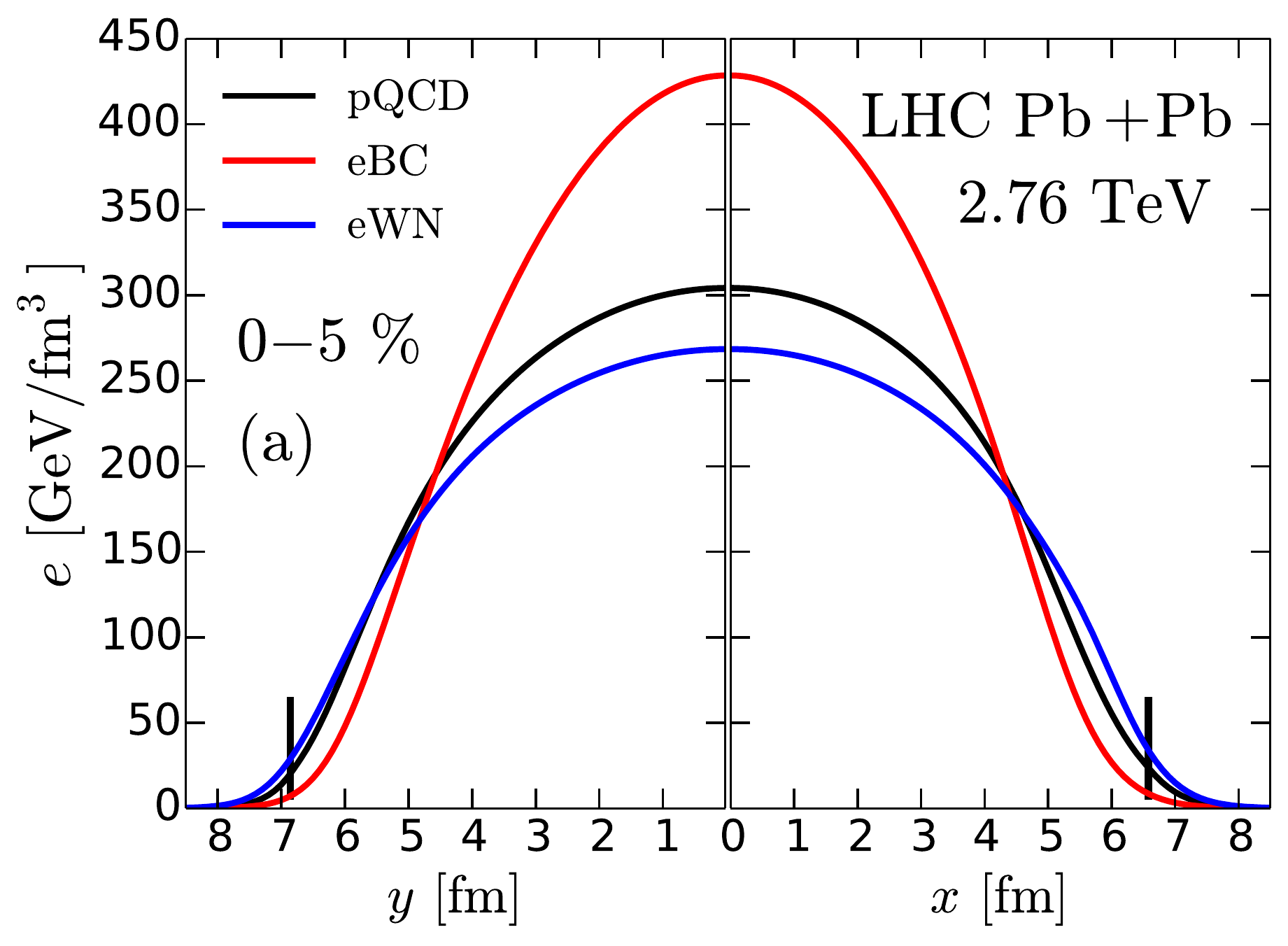}
\includegraphics[width=8.0cm]{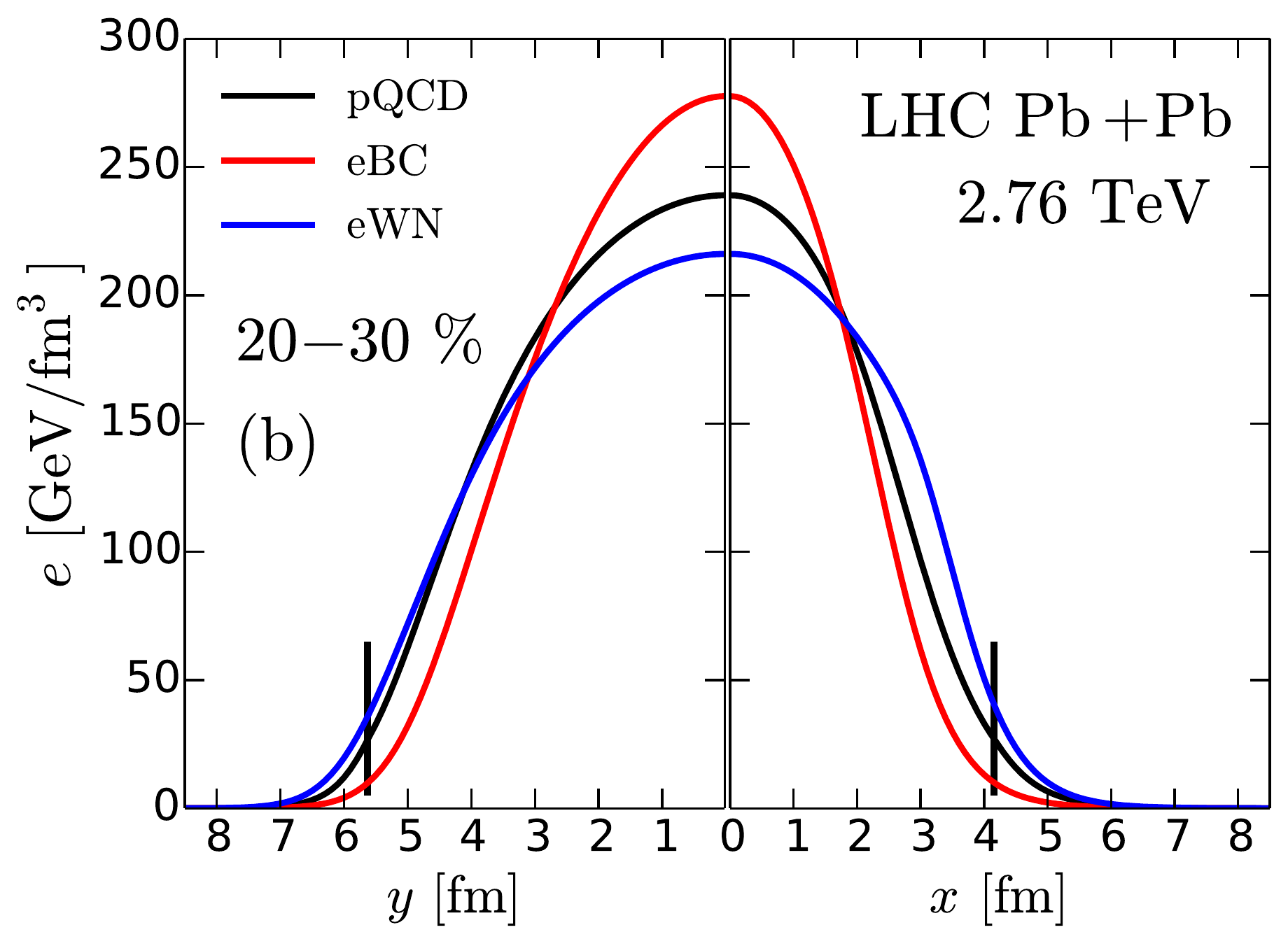}
\includegraphics[width=8.0cm]{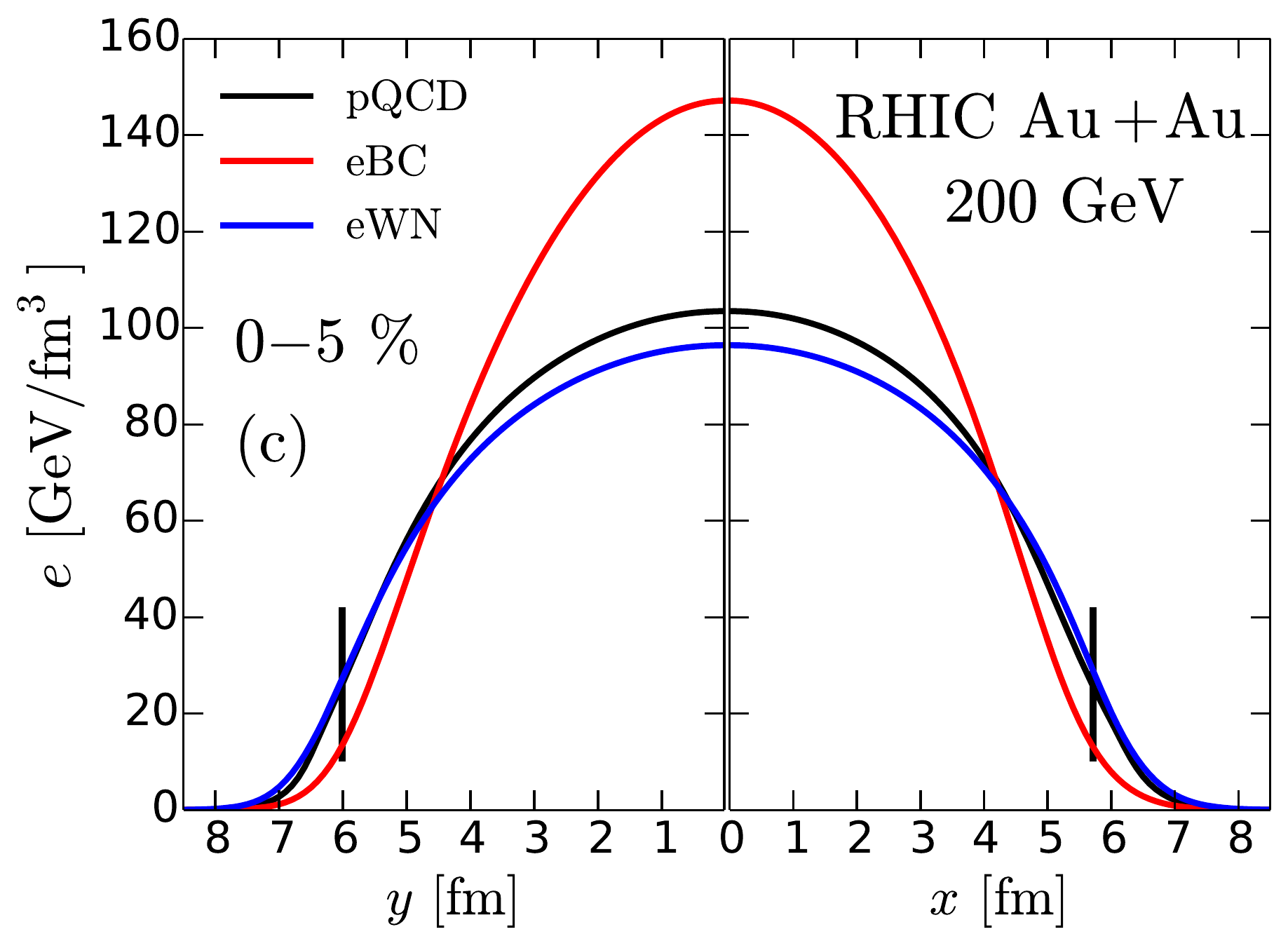}
\includegraphics[width=8.0cm]{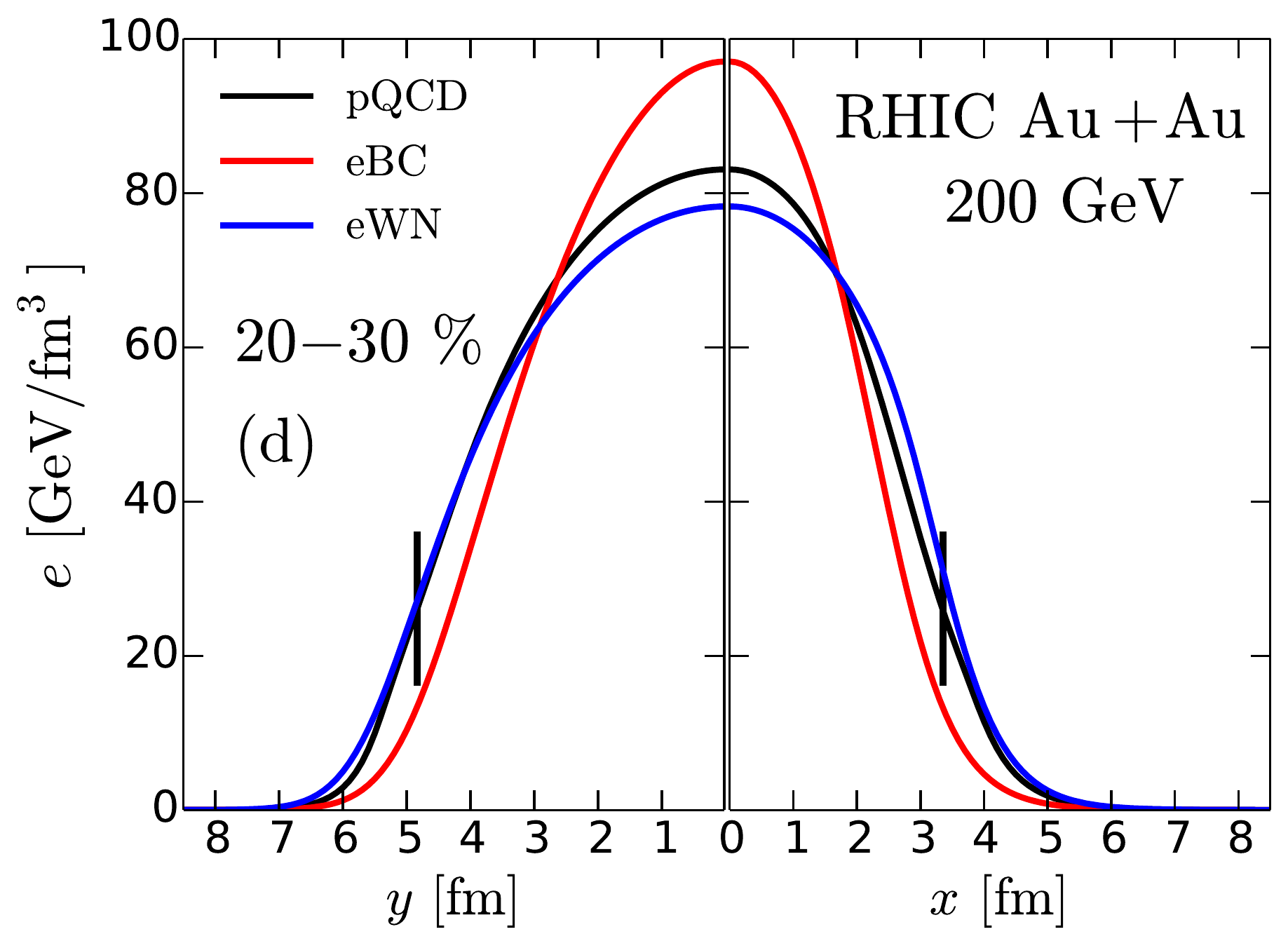}
\caption{(Color online) Energy density profiles in $\sqrt{s_{NN}}=2.76$ TeV Pb+Pb collisions at $\tau_0 = 0.20$~fm, in the $0-5$ \% (a) and in the $20-30$ \% centrality class (b) computed with $K_{\rm sat}=0.63$ and $\beta=0.8$ in the BJ prethermal evolution case. The small vertical lines show approximately where the matching to the $T_A T_A$ profile is done, i.e. at $p_{\rm sat} = 1$ GeV. The figures (c) and (d) show the same for $\sqrt{s_{NN}}=200$ GeV Au+Au collisions.}
\label{fig:edprofiles}
\end{figure*}

As an example we show the calculated initial energy density profiles in $\sqrt{s_{NN}}=2.76$ TeV Pb+Pb collisions at $\tau_0 = 0.20$ fm in $0-5$ \% and $20-30$ \% centrality classes in Figs.~\ref{fig:edprofiles}a and \ref{fig:edprofiles}b, respectively. The calculation of the nuclear overlap geometry and of the impact parameters corresponding to the centrality classes are in this case based on the optical Glauber model. For comparison, we also show the usual simple Glauber model based eBC and eWN profiles \cite{Kolb:2001qz}. The eBC and eWN profiles are normalized such that the initial entropy per unit spacetime rapidity, $dS_{i}/d\eta_{\rm s}$, which in the ideal fluid is directly proportional to the final hadron multiplicity, is the same as in the calculated initial state in $0-5$ \% centrality class. Overall, the energy density gradients from the EKRT model are slightly steeper than in the eWN profile, but not as steep as in the eBC profile. 

%%%%%%%%%%%%%%%%%%%%% FIGURE %%%%%%%%%%%%%%%%%%%%%
%%%%%%%%%%%%%%%%%%%%% FIGURE %%%%%%%%%%%%%%%%%%%%%

\begin{figure*}
\includegraphics[width=8.0cm]{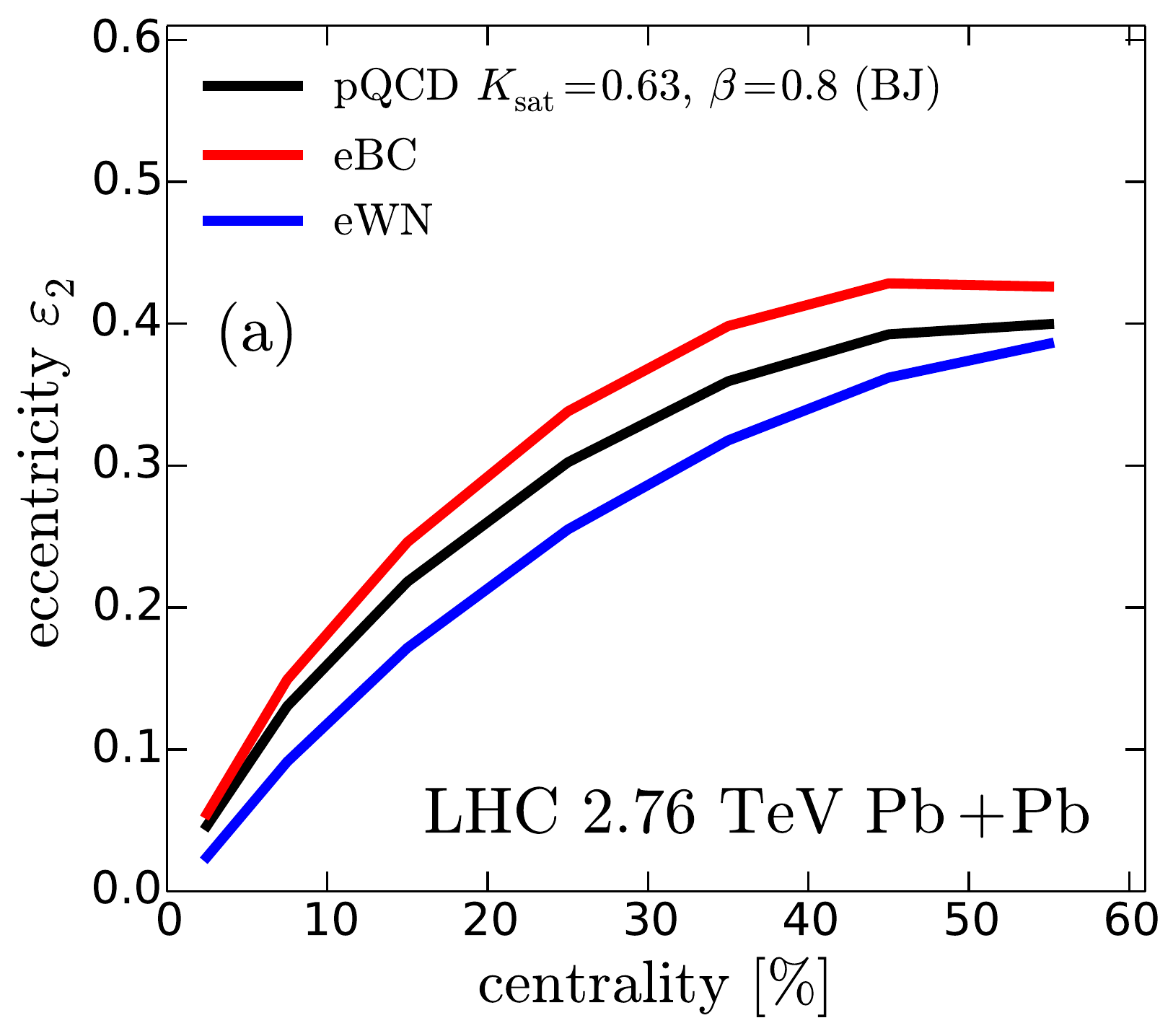}
\includegraphics[width=8.0cm]{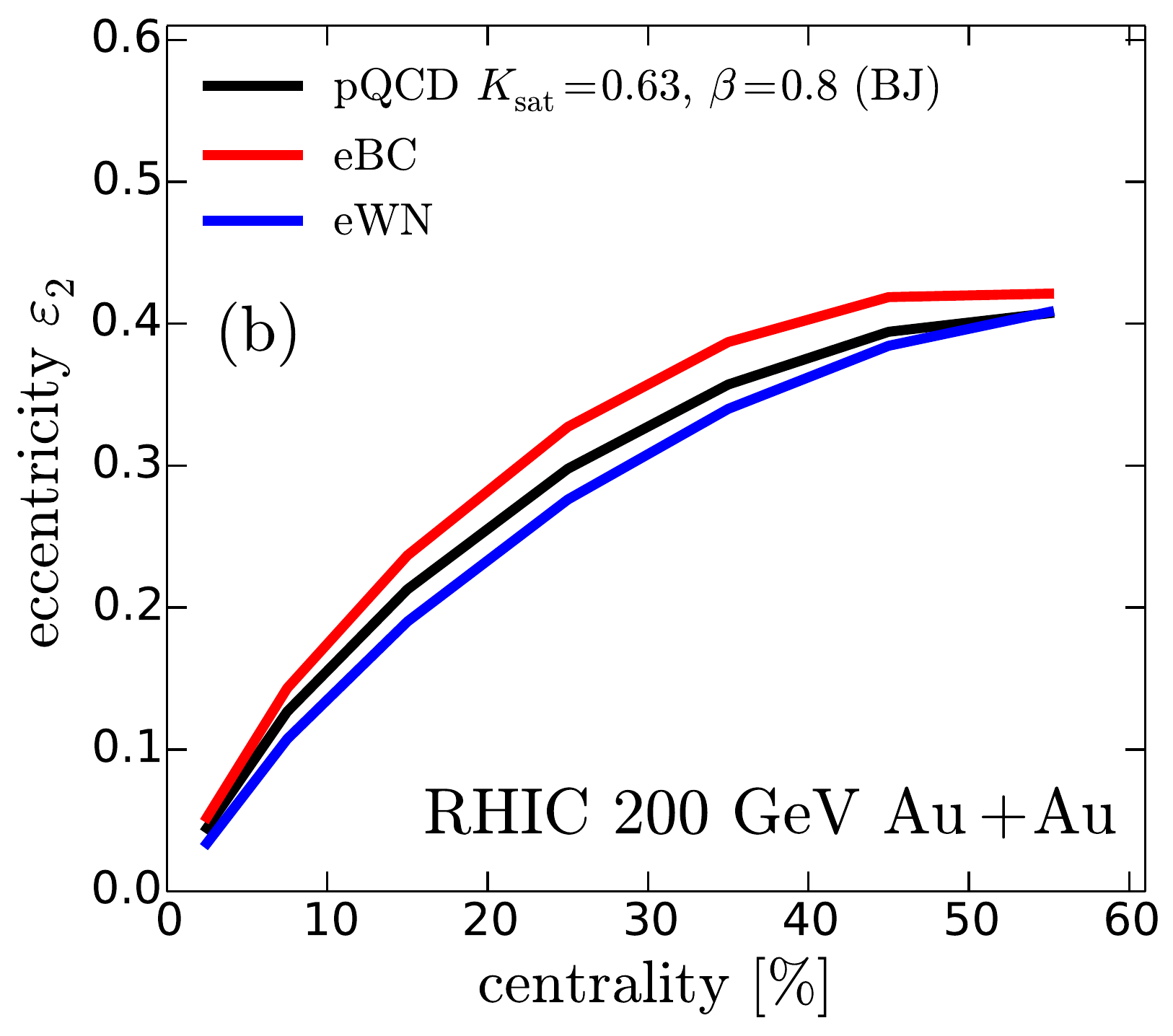}
\caption{(Color online) Initial eccentricity as a function of centrality in $\sqrt{s_{NN}}=2.76$ TeV Pb+Pb collisions (a) and in $\sqrt{s_{NN}}=200$ GeV Au+Au collisions (b) at $\tau_0 = 0.20$~fm, computed with $K_{\rm sat}=0.63$, $\beta=0.8$ and BJ prethermal evolution.}
\label{fig:eccentricity_ave}
\end{figure*}

The initial profiles can be further quantified by calculating the eccentricity,
\begin{equation}
  \varepsilon_{m, n} e^{in\Psi_{m,n}} = -\{r^m e^{in\phi}\}/ \{r^m\},
  \label{eq:eccentricities}
\end{equation}
where the curly brackets denote the average over the transverse plane, i.e., $\{\cdots\} = \int dxdy\, e(x,y,\tau_0) (\cdots)$, $r$ is the distance to the system's center of mass, and $e(x,y,\tau_0)$ is the energy density at the initial time $\tau_0$. The ``participant plane''-angle $\Psi_{m,n}$ can be calculated as
\begin{equation}
 \Psi_{m,n} = \frac{1}{n} {\rm atan2}\left(\{r^m \cos(n\phi)\}, \{r^m \sin(n\phi)\}  \right)+\frac{\pi}{n},
\end{equation}
where the ${\rm atan2}(x, y)$ function gives the angle in the correct quadrant of the transverse plane. In the absence of event-by-event fluctuations the event-plane angle $\Psi_{m, n} = 0$, if the $x$-axis is chosen in the direction of the impact parameter, and $\varepsilon_{m, n}=0$ for all odd $n$. Note, however, that later when we consider the event-by-event density fluctuations the phase and $\varepsilon_{m, n}$ for odd $n$ are not generally zero, but fluctuate from event to event. We also use a short-hand notation $\varepsilon_{n} \equiv \varepsilon_{n, n}$. The eccentricities $\varepsilon_2$ of the calculated initial profiles as a function of centrality are shown in Fig.~\ref{fig:eccentricity_ave}a. As before, we show the comparison to the eBC and eWN profiles, and we can immediately see that $\varepsilon_2$ of the pQCD-based initial conditions are between the eBC and eWN Glauber model limits.

The corresponding energy density profiles in $\sqrt{s_{NN}}=200$ GeV Au+Au collisions are shown in Figs.~\ref{fig:edprofiles}c and \ref{fig:edprofiles}d, and the initial eccentricities in Fig.~\ref{fig:eccentricity_ave}b. Overall the pQCD initial states are quite similar at RHIC and the LHC. The most notable change is that the eWN initial state is closer to the pQCD initial state at RHIC energy, as can be seen both in the energy density profiles and eccentricities.

The computed energy density profiles discussed above were used as initial conditions to fluid dynamical evolution in Ref.~\cite{Paatelainen:2013eea}. It was shown that this model can reproduce the centrality dependence of the multiplicity, $p_T$-spectra and elliptic flow coefficients simultaneously at the RHIC and LHC energies. However, in order to compare to the available experimental data in more detail, it is necessary to take into account the event-by-event nature of the collisions. Inclusion of the effects of the density fluctuations to the pQCD initial state is described next.

%%%%%%%%%%%%%%%%%%%%% FIGURE %%%%%%%%%%%%%%%%%%%%%
%%%%%%%%%%%%%%%%%%%%% FIGURE %%%%%%%%%%%%%%%%%%%%%

\subsection{Event-by-event density fluctuation}

The main source that drives the initial state density fluctuations are the random fluctuations in the positions of the nucleons inside the colliding nuclei. Therefore, the basic ingredient in modeling such fluctuations is the spatial distribution of nucleons inside the nuclei. These distributions are mainly constrained by the measured nuclear charge distributions. The nuclear charge density is frequently parametrized by the Woods-Saxon function \eqref{eq:WoodSaxon}, with $R$ and $d$ as the free parameters. However, the measured charge distribution is not the same as the nucleon position distribution, because the nucleons are not point-like particles, but have a finite size and charge distributions themselves. Thus, in principle, the Woods-Saxon parametrization for nucleon position should be constrained in such way that when folding with the nucleon charge profile it gives the measured nuclear charge distribution. The situation is complicated even more by the fact that protons and neutrons are not distributed in the same way, but especially in heavy nuclei the charge-neutral neutrons tend to form the outer layer of nuclei. The formation of this ``neutron skin'' should be taken into account when constraining the distributions. In our case, however, since we are mainly interested in gluons, whose distribution in protons and neutrons are similar, only the average nucleon distribution matters.

Here, in building the EbyE setup, we take the nucleon distribution in a Pb nucleus from Ref.~\cite{Gao:2013is}, which is already constrained by the charge distribution and available measurements of the neutron skin thickness. In practice, this nucleon density profile can be parametrized by the usual Woods-Saxon function, with $R=6.7$ fm and $d=0.55$ fm. It is noteworthy that the neutron skin and the finite size of the nucleons tend to affect the parametrizations in opposite direction: the final Woods-Saxon parameters are actually within the errors of the parameters given for the measured charge distribution~\cite{DeJager:1987qc}. Therefore, effectively we can take the Woods-Saxon parameters for the charge density and interpret the resulting profile as a nucleon position distribution. The theoretical models indicate that the neutron skin thickness varies only slowly with the nuclear mass number~\cite{Kortelainen:2013wpa}, and therefore we can expect that a similar cancellation happens also for other heavy nuclei. For Au nuclei we, therefore, take the Woods-Saxon parameters from the charge distribution and interpret the resulting distribution as a nuclear position distribution.

The nucleon positions inside the nuclei are then sampled according to the Woods-Saxon distribution by assuming them uncorrelated, i.e. sampling each nucleon position independently. In doing this, we keep in mind, however, that in principle the nucleon positions are correlated, e.g.\ two nucleons cannot overlap, but in practice the effect of the correlations is rather weak~\cite{Alvioli:2011sk}, except perhaps in ultracentral collisions~\cite{Denicol:2014ywa}. As a result we obtain an ensemble of nuclear configurations characterized by the nucleon positions $(x_{i},y_{i},z_{i})$. By randomly sampling the impact parameters from a distribution $dN/db^2\propto const$, we then get an ensemble of nuclear collisions.

\subsection{Nuclear and nucleon overlap densities}

In the EKRT minijet framework a nuclear collision is regarded as a collision of two gluon clouds rather than a collection of individual nucleon-nucleon collisions. The leading idea in our EbyE setup is that we first form the nuclear overlap density $\rho_{AA}$ locally in $\mathbf{r}=(x,y)$ for each nuclear collision event, accounting for the nucleon configurations in each collision. Then, the local saturation scales $p_{\rm sat}(\rho_{AA}(\mathbf{r}))$ in each event are obtained from  Eq.~\eqref{eq:psat_param}. The initial energy densities at fixed $\tau_0$ can then be computed, EbyE, as described in Sec.~\ref{sec:iniforhydro}.

We define the nuclear thickness function $T_{A}$ in each event as a sum of the corresponding nucleon thickness functions $T_{n}$, 
\begin{equation}
T_{A}(\mathbf{r})=\sum_{i=1}^{A}T_{n}(|\mathbf{r}-\mathbf{r}_{i}|),
\end{equation}
where the sum is over the nucleon positions in the nucleus $A$ and where the $T_n$ have been normalized to one. The nuclear overlap density $\rho_{AB}\left( \mathbf{r} \right)$ in each $A$+$B$ collision is then obtained from Eq.~\eqref{eq:rhoAB}. 

Since the minijet production considered here is dominated by gluonic channels, the $T_n$ above is to be understood as the gluonic thickness function rather than the one obtained from the (better known) charge densities of nucleons. To obtain the gluonic $T_n$ needed here, we exploit exclusive electroproduction of $J/\psi$  at HERA, $\gamma^*+p\rightarrow J/\psi+p$, for which ZEUS has measured the differential cross section near $t=0$ to be $d\sigma/dt \propto \exp(-b|t|)\propto |G|^2$ with a slope $b=4.72$~GeV$^{-2}$ \cite{Chekanov:2004mw}. Taking a 2-dimensional Fourier transformation of the corresponding 2-gluon form factor $G$ leads to a Gaussian distribution for $T_n$, 
\begin{equation}
T_{n}(r)=\frac{1}{2\pi\sigma^{2}}e^{-\frac{r^{2}}{2\sigma^{2}}},\label{eq:gaussian}
\end{equation}
where the width parameter $\sigma=\sqrt b \approx 0.43$ fm.

\subsection{Centrality selection and sampling the nuclear collisions}

After sampling the nucleon configurations and the impact parameter we determine whether a nuclear collision occurs by using  the following geometric collision criterion: the $A$+$B$ collision takes place if the transverse distance between at least one of the nucleons from $A$ and one from $B$ is shorter than $\sqrt{\sigma_{\rm NN}/\pi}$, where $\sigma_{\rm NN}$ is the total inelastic NN cross-section. At the LHC $\sigma_{\rm NN} = 64$ mb and at RHIC $\sigma_{\rm NN} = 42$ mb. We emphasize that $\sigma_{\rm NN}$ is here only used in the above collision trigger criterion, and that the calculation of the initial state is otherwise essentially independent of $\sigma_{\rm NN}$. Following this procedure, we create a large number of nuclear collision events, for which we then calculate the initial energy density profiles as described in the previous sections.

Next, the fluid-dynamical evolution is calculated separately for each event, after which we calculate the $p_T$-spectrum and multiplicities as described in Sec.~\ref{sec:freeze_out}. The events are then divided into centrality classes according to their final multiplicity (or equivalently the final total entropy). For example, the $0-5$ \% centrality class consists of the events with the highest multiplicity, the top 5 \% of the total number of events. 

%%%%%%%%%%%%%%%%%%%%% FIGURE %%%%%%%%%%%%%%%%%%%%%
\begin{figure}
\includegraphics[width=8.5cm]{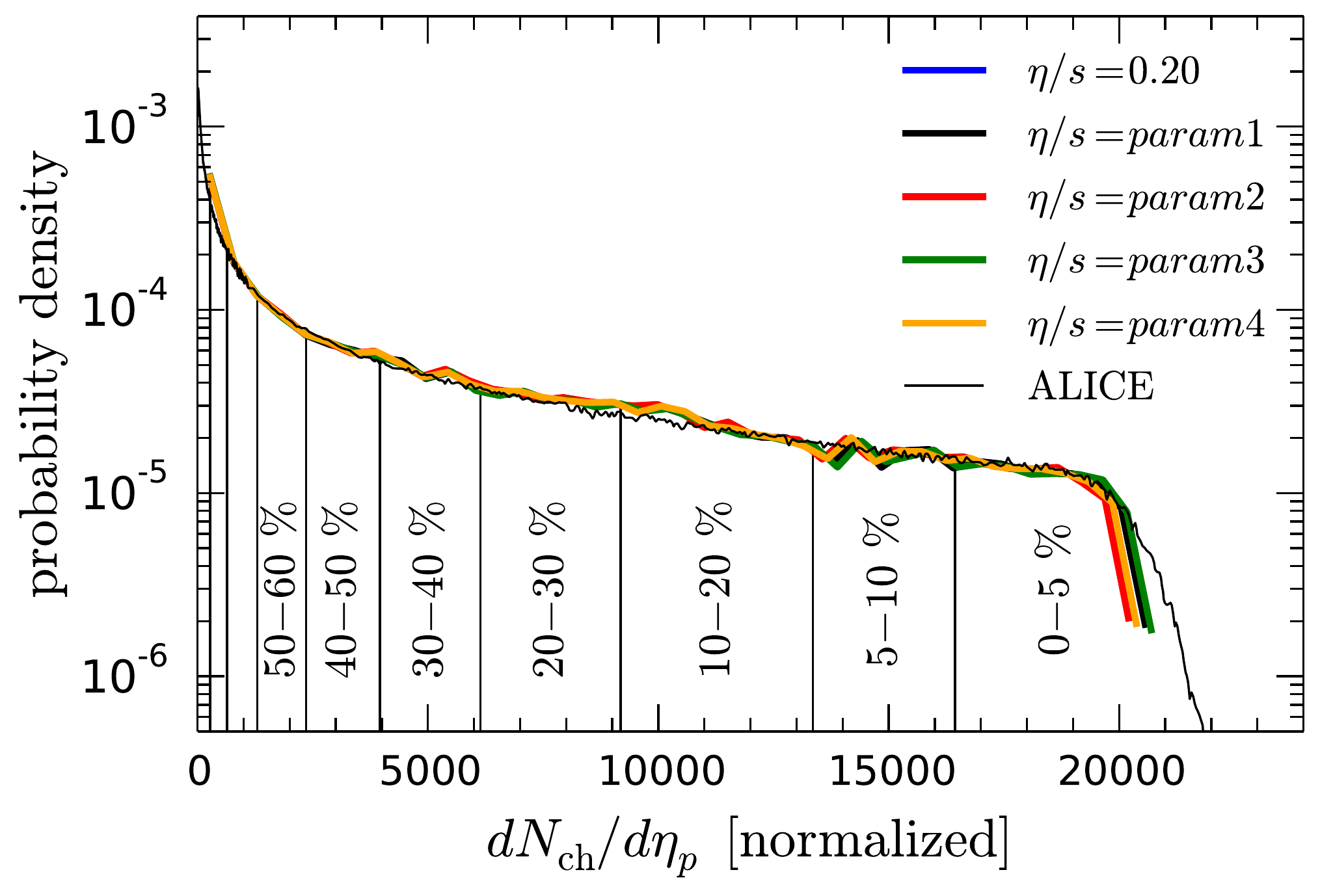}
\caption{(Color online) Probability distribution of the charged hadron multiplicity $dN_{\rm ch}/d\eta_p$ for the five different $\eta/s(T)$ cases of Fig.~\ref{fig:etapers}, compared with the parametrization of the ALICE VZERO amplitude read off from Ref.~\cite{Abelev:2013qoq}, in $\sqrt{s_{NN}}=2.76$ TeV Pb+Pb collisions.}
\label{fig:entropy_distribution}
\end{figure}
%%%%%%%%%%%%%%%%%%%%% FIGURE %%%%%%%%%%%%%%%%%%%%%
In Fig.~\ref{fig:entropy_distribution} we show the calculated probability distribution of the charged hadron multiplicity $dN_{\rm ch}/d\eta_{\rm p}$ compared to the parametrization of the ALICE measurement of VZERO amplitude, read off from Fig.~10 of Ref.~\cite{Abelev:2013qoq}, that is approximately proportional to the final state multiplicity. The distributions are scaled to have approximately the same average. As one can see from the figure, the agreement between our calculation and the ALICE measurement is very good, except in the very central collisions. This is, indeed, expected as this tail of the distribution is dominated by the dynamical multiplicity fluctuations which we do not yet include in the current EKRT framework. In our case, such dynamical fluctuations would mean that for the same value of the overlap density $\rho_{AA}=T_A T_A$ the saturation scale (i.e. gluon multiplicity), and hence entropy production, would be fluctuating from one event to another. 

%%%%%%%%%%%%%%%%%%%%% FIGURE %%%%%%%%%%%%%%%%%%%%%
\begin{figure}
\includegraphics[width=8.0cm]{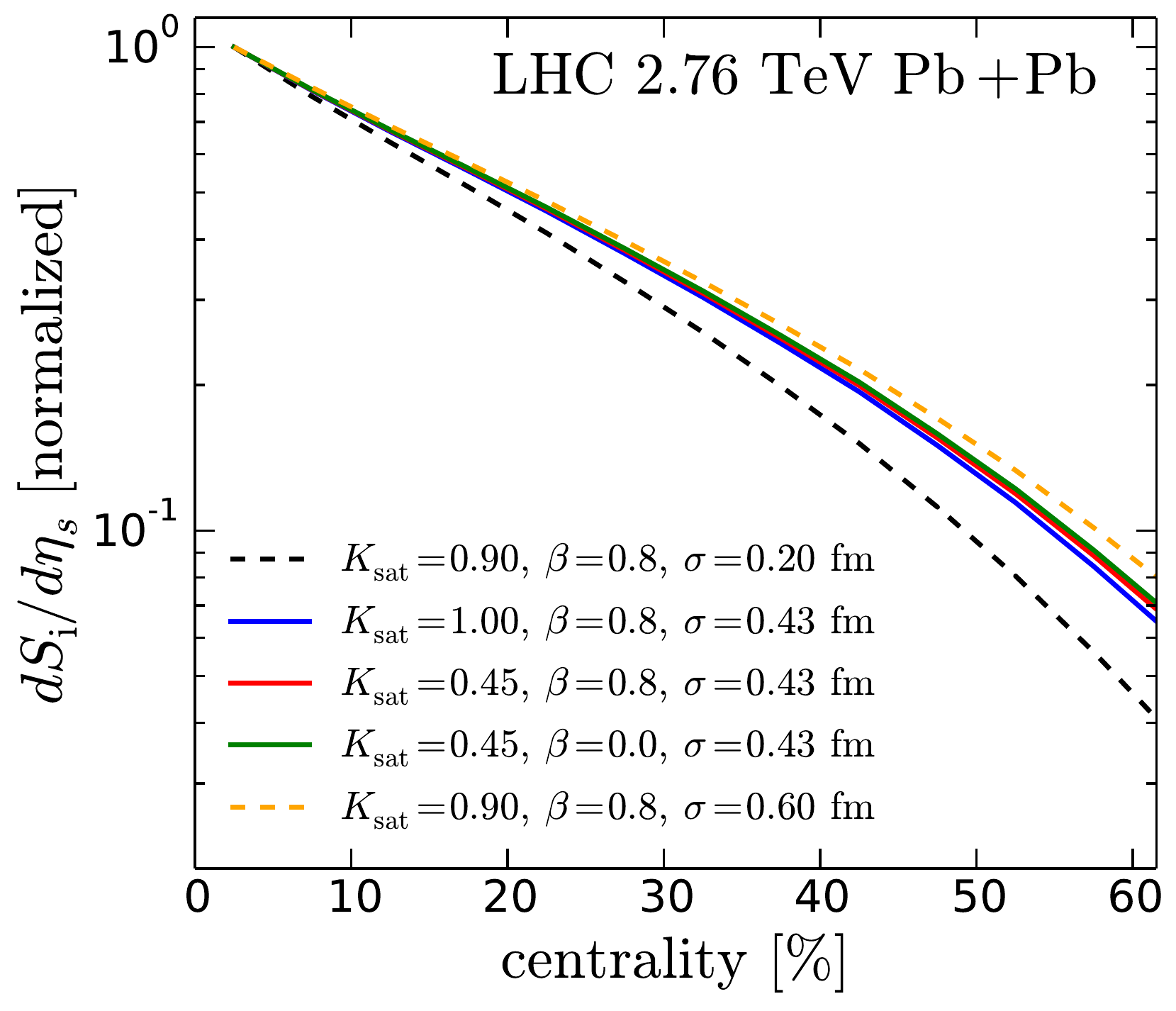}
\caption{(Color online) Centrality dependence of the initial entropy $dS_i/d\eta_{\rm s}$ in $\sqrt{s_{NN}}=2.76$ TeV Pb+Pb collisions at the LHC with different values of $K_{\rm sat}$, $\beta$ and $\sigma$. The curves are normalized such that such that the total entropy in $0-5$ \% centrality class is one.}
\label{fig:initial_entropy}
\end{figure}
%%%%%%%%%%%%%%%%%%%%% FIGURE %%%%%%%%%%%%%%%%%%%%%
Even without the fluid-dynamical evolution, it is possible to estimate the centrality dependence of multiplicity from the initial entropy. Because the hadron multiplicity is proportional to the final entropy (by a factor that depends on the decoupling temperature), the entropy production during the fluid dynamical evolution can significantly affect the multiplicity, but the its effect on the relative centrality dependence of the multiplicity is much weaker, see below. Once we have the energy density profiles, we can convert them to the entropy density profiles through the EoS, and calculate the spacetime-rapidity density of the total entropy, $dS_{i}/d\eta_{\rm s}$, as
\begin{equation}
 \frac{dS_{i}}{d\eta_{\rm s}} = \int dxdy\, \tau_0 s(x,y, \tau_0) \gamma,
 \label{eq:initial_entropy}
\end{equation}
where $s$ is the local entropy density. In our case the initial velocity is zero and $\gamma=\left( 1-\mathbf{v}_T^2 \right)^{1/2}=1$. Figure \ref{fig:initial_entropy} shows the normalized initial entropy as a function of centrality. The lines show calculations with different values of $K_{\rm sat}$, $\beta$ and $\sigma$. The actual entropy varies as the parameters are changed but to better compare the centrality dependence in different cases, we have normalized the results in this figure such that $dS_{i}/d\eta_{\rm s}=1$ in the $0-5$ \% centrality class in each case. As one can read from the figure, the centrality dependence changes only slightly with different values of $K_{\rm sat}$ and $\beta$, but the width of the nucleon gluon distribution affects it much more. These extremes, i.e. $\sigma = 0.60$ fm or $0.20$ fm are, however, not supported by the HERA/ZEUS data. Note that, when coupled with viscous fluid dynamics, the $K_{\rm sat}=0.45$ and $\beta=0.8$ case corresponds to the case $\eta/s=param3$ in the data comparison in Sec.~V, see Table \ref{tab:ksat}.

%%%%%%%%%%%%%%%%%%%%% FIGURE %%%%%%%%%%%%%%%%%%%%%
\begin{figure}
\includegraphics[width=8.0cm]{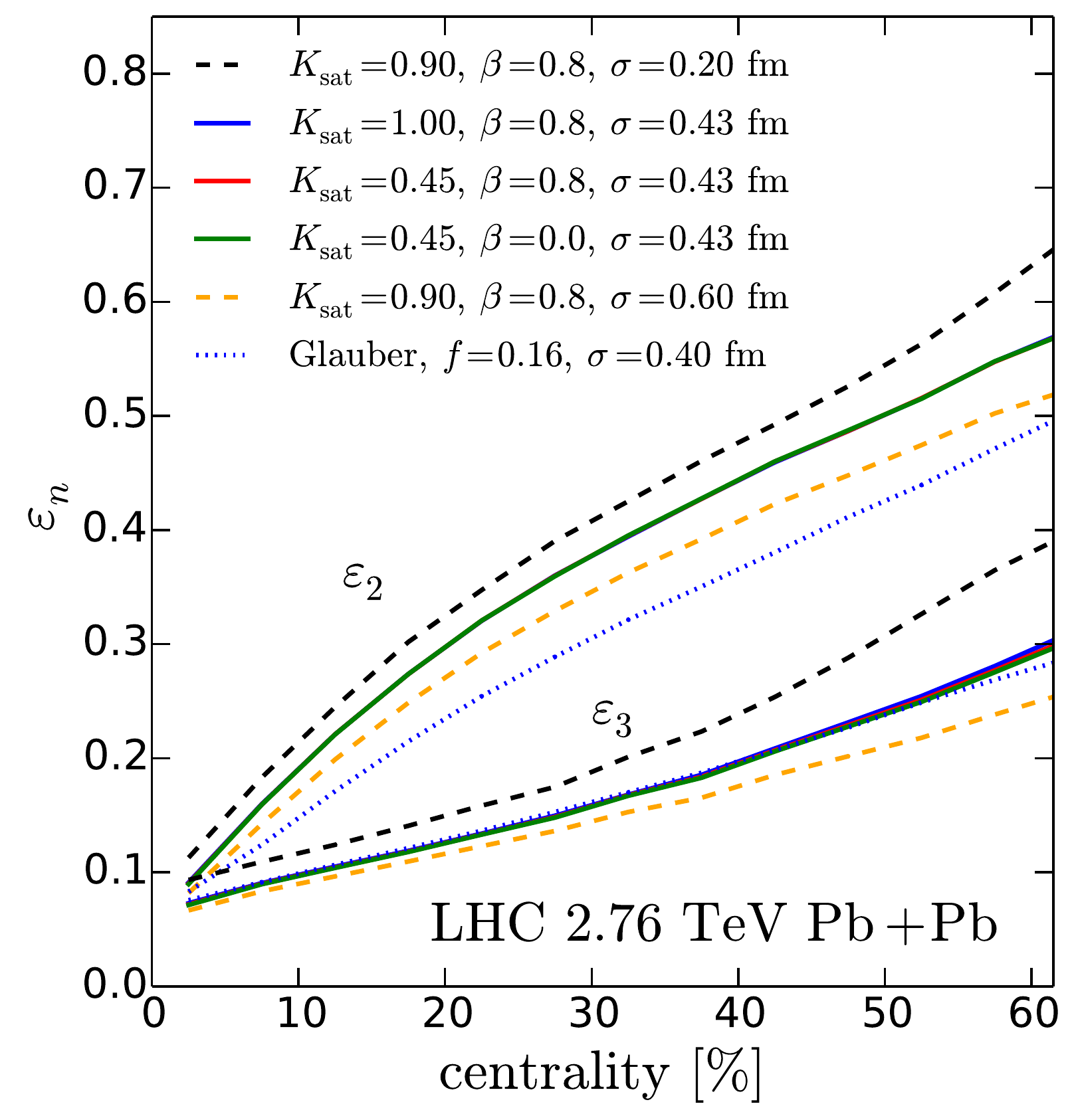}
\caption{(Color online) Centrality dependence of the initial eccentricity  in $\sqrt{s_{NN}}=2.76$ TeV Pb+Pb collisions at the LHC with different values of $K_{\rm sat}$, $\beta$ and $\sigma$ (solid and dashed lines). The Glauber-model case is shown for comparison (dotted lines).}
\label{fig:initial_eccentricity}
\end{figure}
%%%%%%%%%%%%%%%%%%%%% FIGURE %%%%%%%%%%%%%%%%%%%%%
Figure \ref{fig:initial_eccentricity} shows the initial eccentricities $\varepsilon_2$ and $\varepsilon_3$ for the same cases as above. In addition we show the eccentricity from the usual Glauber model initial state, i.e. a mixture of the eWN and eBC initial densities, $\epsilon \propto f\rho_{\rm bin} + (1-f)\rho_{\rm wn}$, with $f=0.16$. Similarly to the initial entropy case, there is practically no sensitivity on $K_{\rm sat}$ and $\beta$, but a strong sensitivity on the value of $\sigma$. The pQCD + saturation initial conditions give values of $\varepsilon_2$ that are significantly larger than those of the Glauber model, but the $\varepsilon_3$ values are very similar in Glauber model and pQCD+saturation initial conditions with $\sigma=0.43$ fm, i.e with the $\sigma$ value obtained from the HERA/ZEUS fit.

%%%%%%%%%%%%%%%%%%%%% FIGURE %%%%%%%%%%%%%%%%%%%%%
\begin{figure*}
\includegraphics[width=8.5cm]{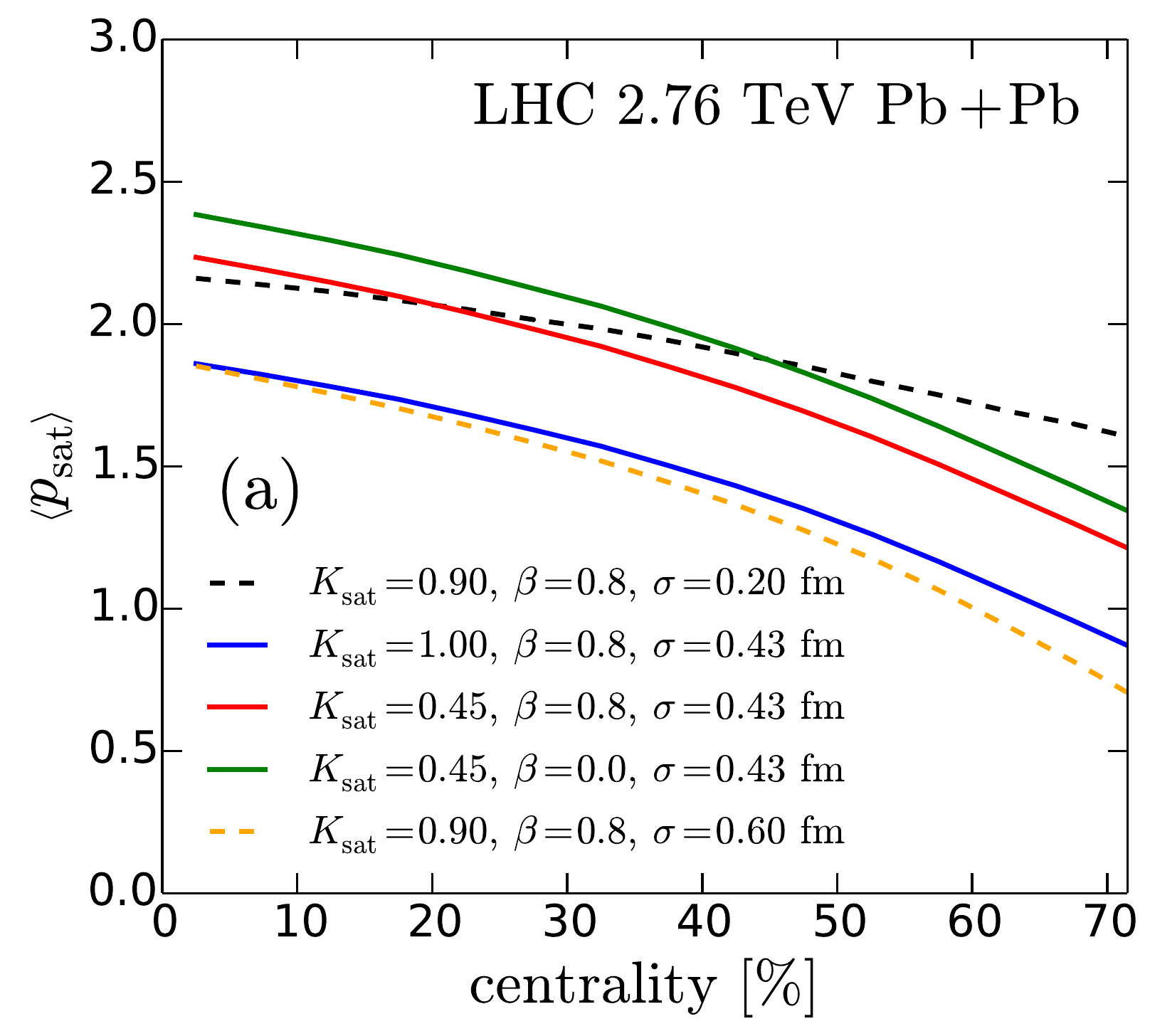}
\includegraphics[width=8.5cm]{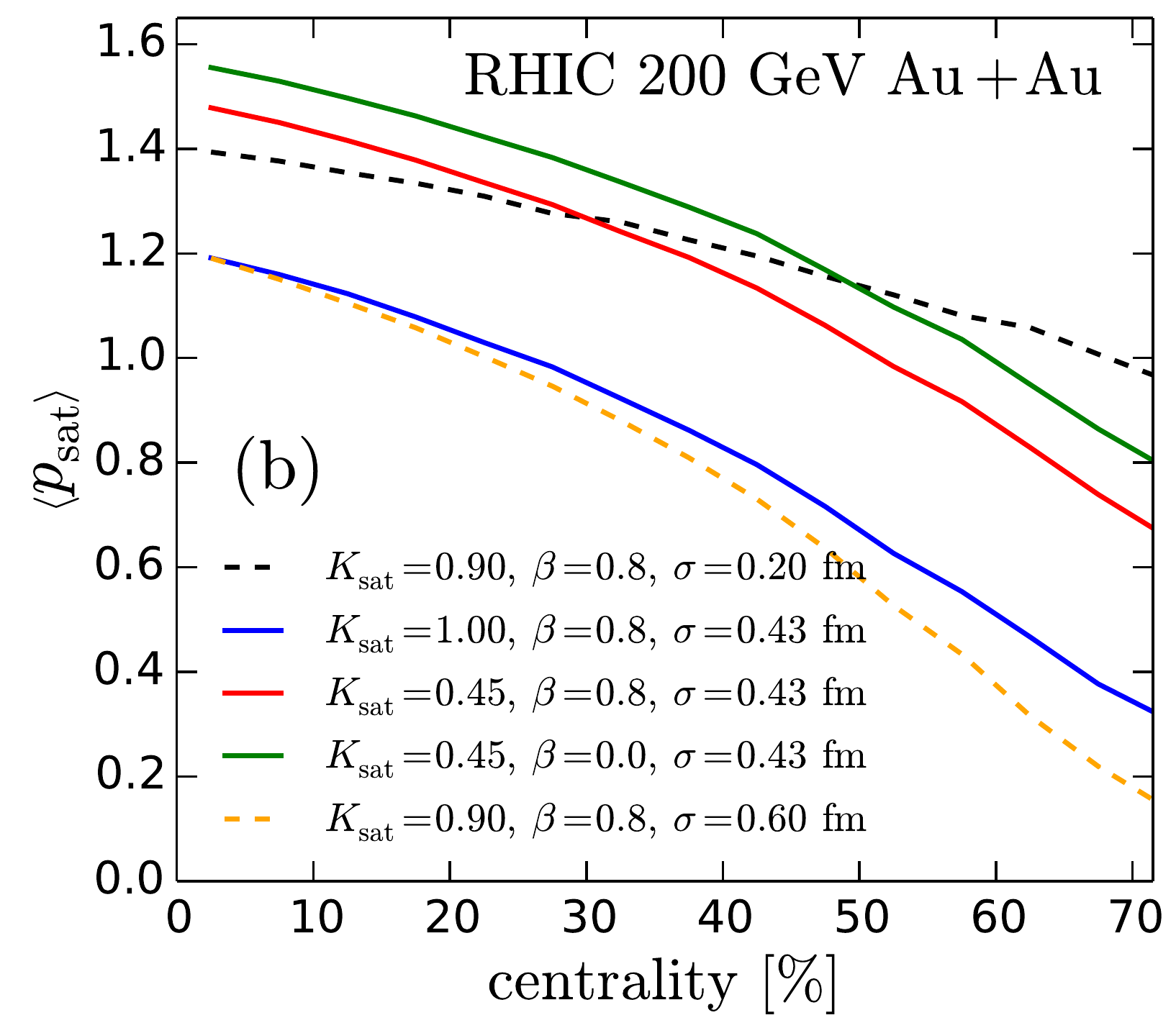}
\caption{(Color online) Average $p_{\rm sat}$ as a function of centrality  in $\sqrt{s_{NN}}=2.76$ TeV Pb+Pb collisions at the LHC (a), and in $\sqrt{s_{NN}}=200$ GeV Au+Au collisions at RHIC (b) with different values of $K_{\rm sat}$, $\beta$ and $\sigma$.}
\label{fig:psat_ave}
\end{figure*}
%%%%%%%%%%%%%%%%%%%%% FIGURE %%%%%%%%%%%%%%%%%%%%%
Figure \ref{fig:psat_ave}a shows the entropy weighted average saturation scale $p_{\rm sat}$ as a function of centrality in Pb+Pb collisions at the LHC, computed for the same values of $K_{\rm sat}$, $\beta$ and $\sigma$ as in the previous figures. Fig.~\ref{fig:psat_ave}b shows the same for Au+Au collisions at RHIC. Again, we see that the gluonic width $\sigma$ has the  largest effect on the centrality dependence (compare the dashed lines) while $\beta$ and $K_{\rm sat}$ affect more the normalization of $p_{\rm sat}$. The opposite systematics in $\beta$ and $K_{\rm sat}$ can be understood from Eq.~\eqref{eq:sat_scaling} at the naive scaling limit: $p_{\rm sat}\sim (K/K_{\rm sat})^{1/4}$, where the NLO/LO K-factor $K$ of Eq.~\eqref{eq:Kfactor} increases with decreasing  $\beta$. We also see that the average saturation scales remain above 1 GeV for a very wide range of centralities both at the LHC and RHIC. 

%%%%%%%%%%%%%%%%%%%%% FIGURE %%%%%%%%%%%%%%%%%%%%%
\begin{figure*}
\includegraphics[width=8.5cm]{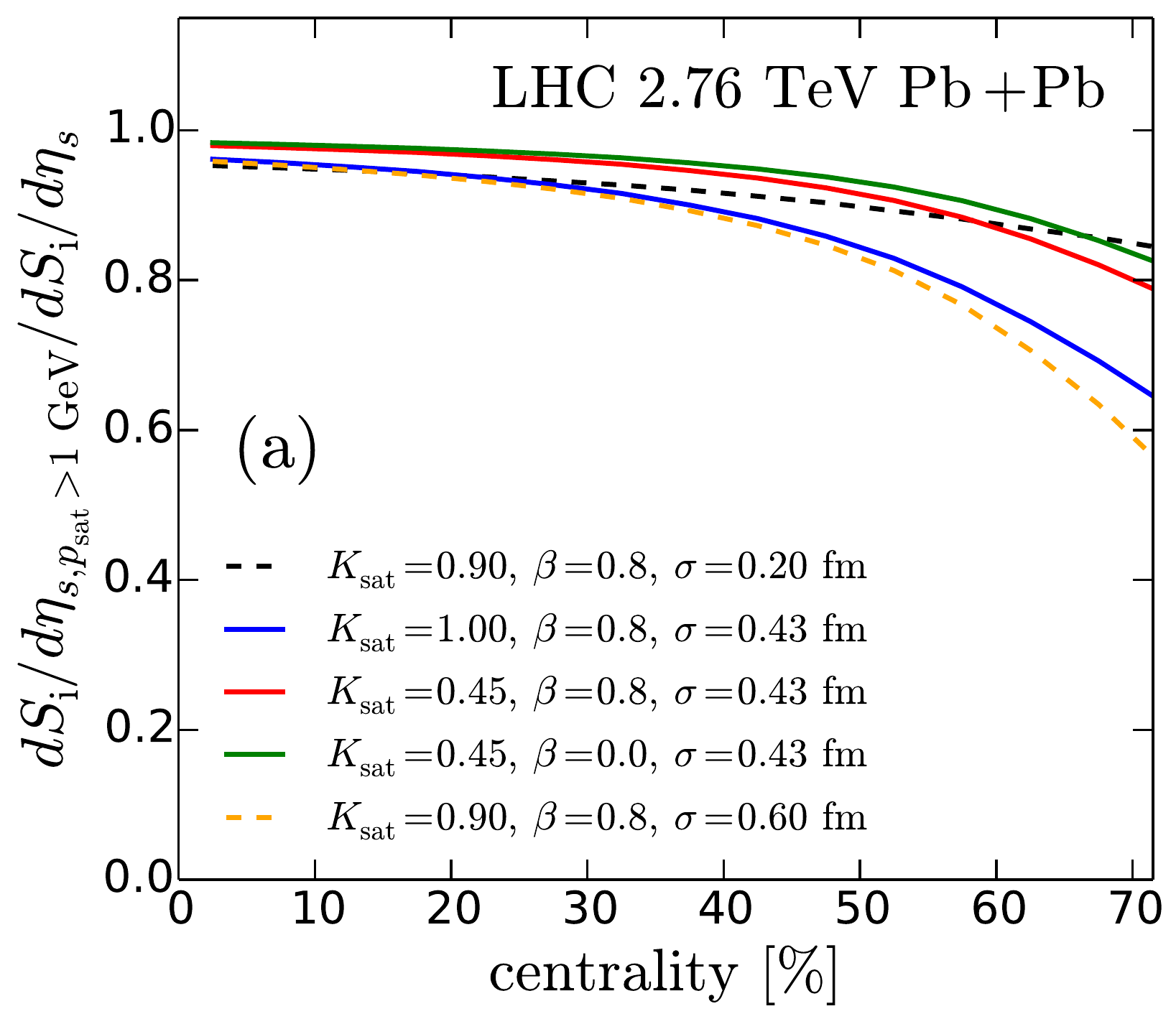}
\includegraphics[width=8.5cm]{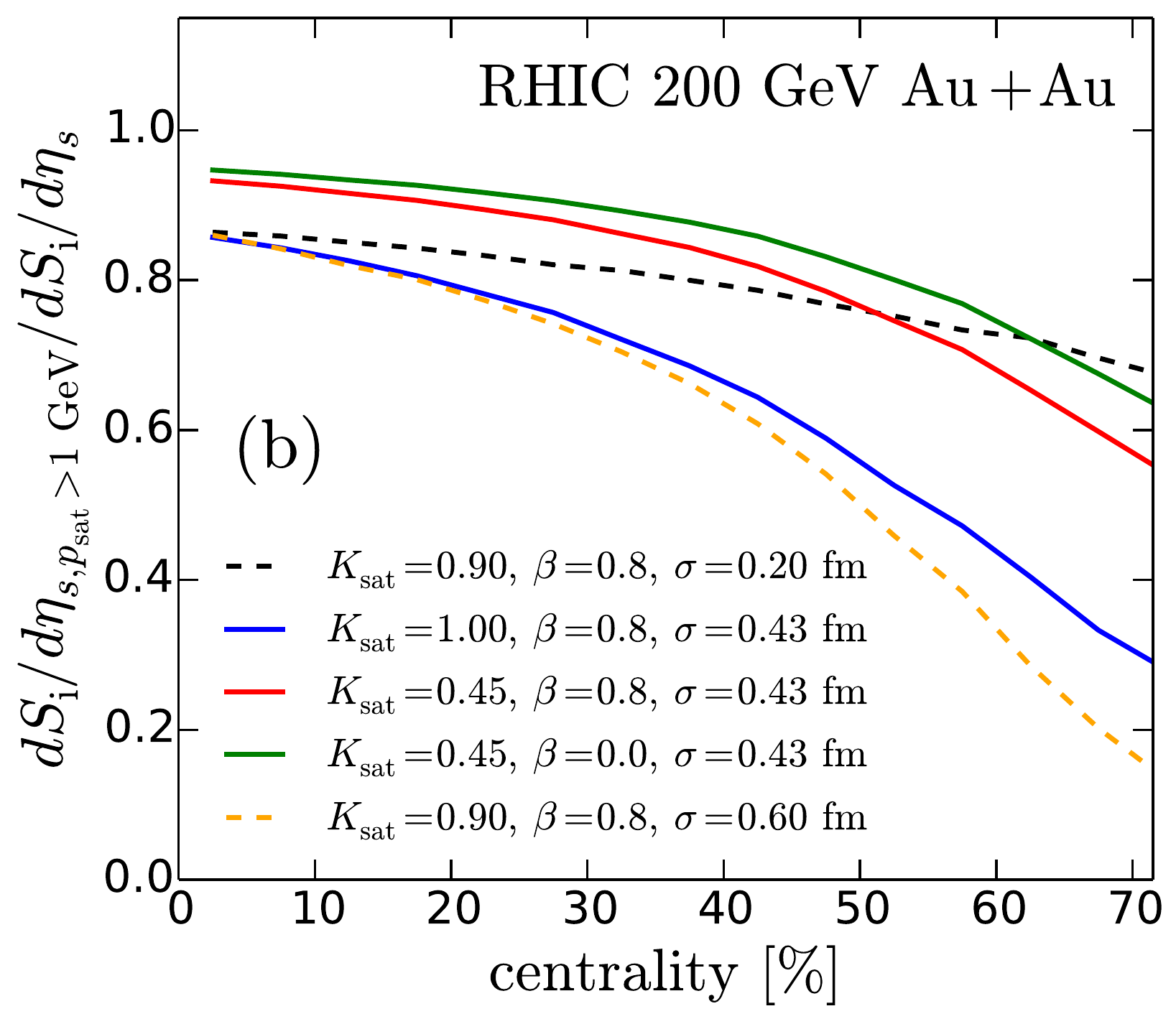}
\caption{(Color online) Fraction of $dS_i/d\eta_{\rm s}$ from the region $p_{\rm sat} \geq 1$ GeV as a function of centrality in $\sqrt{s_{NN}}=2.76$ TeV Pb+Pb collisions at the LHC (a), and in $\sqrt{s_{NN}}=200$ GeV Au+Au collisions at RHIC (b).}
\label{fig:pent_fraction}
\end{figure*}
%%%%%%%%%%%%%%%%%%%%% FIGURE %%%%%%%%%%%%%%%%%%%%%%%%
Figure \ref{fig:pent_fraction}a shows the fraction of the initial $dS_i/d\eta_{\rm s}$ from the regions of the transverse plane where $p_{\rm sat} \geq 1$ GeV, both in Pb+Pb collisions at the LHC, and Fig.~\ref{fig:pent_fraction}b the same in Au+Au at RHIC, computed for the same values of $K_{\rm sat}$, $\beta$ and $\sigma$ as above. Note again that the case with $K_{\rm sat}=0.45$, $\beta=0.8$ will correspond to the $param3$ case in the data comparison ahead in Sec.~V. This figure, together with Fig.~\ref{fig:edprofiles}, indicates that pQCD + saturation indeed gives the dominant part of the initial conditions over a sufficiently wide range of centralities both at the LHC and RHIC, and that the additional phenomenology at the low-density edges of the system does not play a major role.

%%%%%%%%%%%%%%%%%%%%% FIGURE %%%%%%%%%%%%%%%%%%%%%
\begin{figure}
\includegraphics[width=8.5cm]{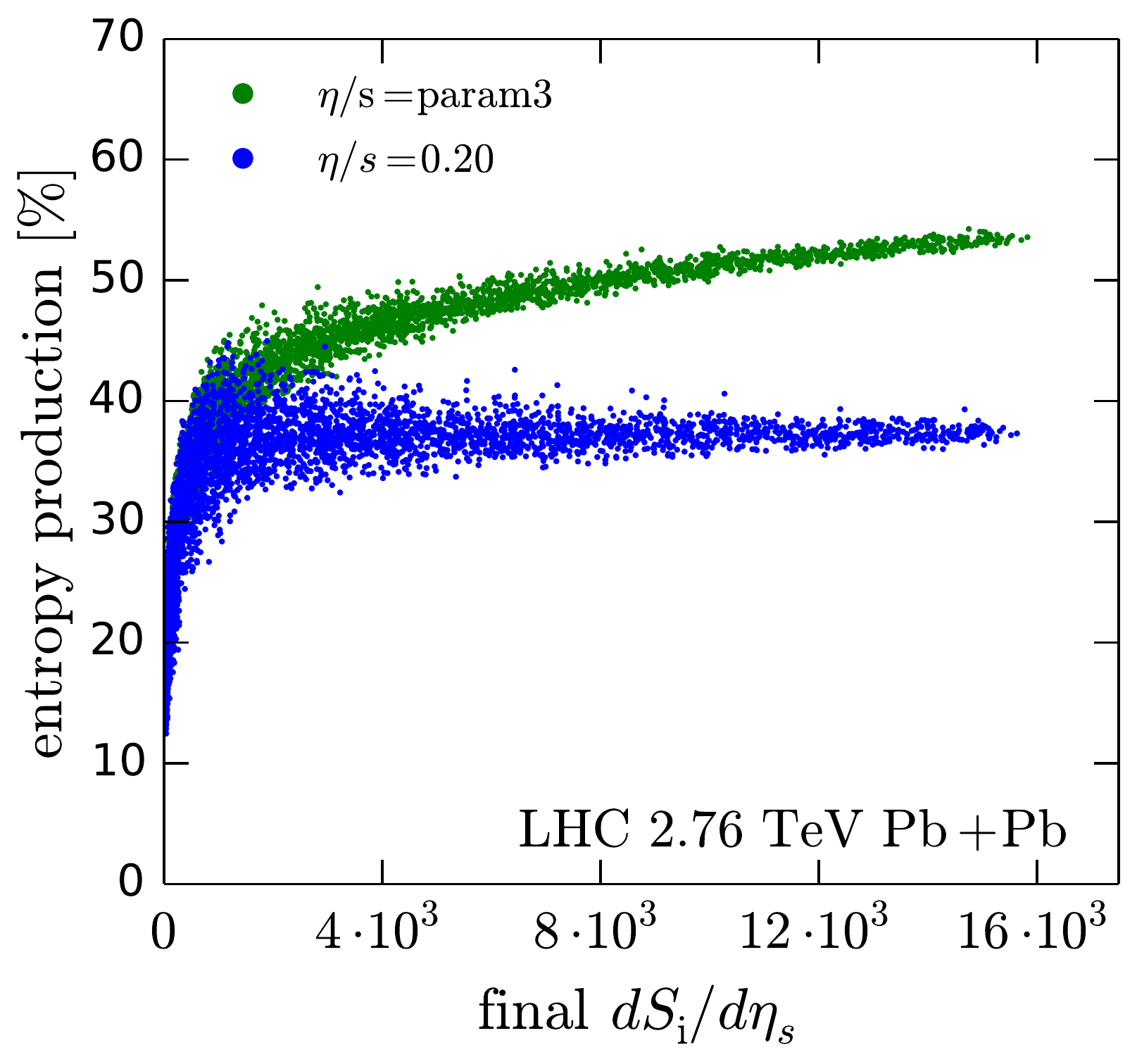}
\caption{(Color online) Entropy production as a function of the initial $dS_i/d\eta_{\rm s}s$ in $\sqrt{s_{NN}}=2.76$ TeV Pb+Pb collisions at the LHC, computed for $\eta/s(T)$ from $param3$ in Fig.~\ref{fig:etapers}
and for $\eta/s=0.20$.}
\label{fig:entropy_production}
\end{figure}
%%%%%%%%%%%%%%%%%%%%% FIGURE %%%%%%%%%%%%%%%%%%%%%
In Fig.~\ref{fig:entropy_production} we show the entropy production due to the viscous effects in fluid dynamics for Pb+Pb collisions at the LHC, computed for the $\eta/s=0.2$ and $param3$ cases (cf. Fig.~\ref{fig:etapers}). Since our starting time for the fluid dynamics is relatively small, $\tau_0=0.2$~fm, the entropy production becomes sensitive to the QGP viscosity: hence there is significantly more entropy produced for $param3$ in central collisions where the initial temperatures are highest. As we can see in the figure, the entropy production is rather significant but especially for the parametrizations where the QGP viscosity remains below that in $param3$ (and which will also reproduce the experimental data best) it can still be regarded as a correction. In practice, in order to get the same multiplicity, e.g. in the most central collisions, with all the different $\eta/s$ parametrizations, $K_{\rm sat}$ is adjusted for each $\eta/s(T)$ separately.

%%%%%%%%%%%%%%%%%%%%% SECTION %%%%%%%%%%%%%%%%%%%%%
\section{Flow coefficients and correlations}
Before comparing our results with the LHC and RHIC measurements, let us recapitulate the definitions of the various flow coefficients and correlations discussed in the next section. The azimuthal parts of the transverse momentum spectra are, traditionally, decomposed into the Fourier components $v_n$ and their phases or event-plane angles $\Psi_n$. For a single event these can be defined as
\begin{equation}
 v_n(p_T, y) e^{in\Psi_n(p_T, y)} = \langle e^{in\phi} \rangle_\phi,
 \label{eq:vnpt}
\end{equation}
where the angular brackets $\langle \cdots \rangle_\phi$ denote the average 
\begin{equation}
\langle \cdots \rangle_\phi = \left( \frac{dN}{dydp_T^2}\right)^{-1}\int d\phi \frac{dN}{dydp_T^2 d\phi}\left(\cdots\right).
\end{equation}
Similarly, the $p_T$-integrated flow coefficients are defined as
\begin{equation}
 v_n(y) e^{in\Psi_n(y)} = \langle e^{in\phi} \rangle_{\phi, p_T},
 \label{eq:vn}
\end{equation}
where the average is defined as  
\begin{equation}
\langle \cdots \rangle_{\phi,p_T} = \left( \frac{dN}{dy}\right)^{-1}\int d\phi dp_T^2\frac{dN}{dydp_T^2 d\phi}\left(\cdots\right).
\end{equation}
From here on we drop the $y$ from the arguments, as we are using the boost-invariant approximation, where the flow coefficients do not depend on the rapidity. In practice, the $p_T$-integration is never over the full $p_T$-range, but different experiments have different $p_T$-cuts in their analyses  --- a fact to be taken into account in the calculations as well. Also, in the case of unidentified charged hadrons the rapidity $y$ cannot be measured, but the spectra are averaged over some pseudorapidity range $\Delta \eta_{\rm ps}$ symmetric around $\eta=0$. In this case the spectra above are replaced by
\begin{multline}
  \frac{dN_{ch}}{d\eta_{\rm ps} dp_T^2 d\phi}\biggr|_{\Delta \eta_{\rm ps}} = \\
 \sum_i \frac{2}{\Delta \eta_{\rm ps}} \sinh^{-1}\left[\frac{p_T}{m_{T,i}} \sinh\left(\frac{\Delta \eta_{\rm ps}}{2}\right)\right] \frac{dN_i}{dy dp_T^2 d\phi},
\end{multline}
where the sum is over all the charged hadrons, $m_{T,i} = \sqrt{m_i^2 + p_T^2}$, and $m_i$ is the mass of the hadron $i$.

\subsection{Event-plane method}

In addition, it is also possible to define the so called event-plane flow coefficients as
\begin{equation}
 v_n\{{\rm EP}\}(p_T) = \langle \cos\left[ n\left(\phi - \Psi_n\{{\rm EP}\})\right)\right] \rangle_{\phi},
\end{equation}
where 
\begin{equation}
\begin{split}
&\Psi_n\{{\rm EP}\} = \\ 
&\frac{1}{n}{\rm atan2}\left(\langle w\cos(n\phi)\rangle_{\phi,p_T}, \langle w\sin(n\phi)\rangle_{\phi,p_T}\right),
\end{split}
\label{eq:EPangle}
\end{equation}
with $w$ being a weight factor, e.g. $w=p_T$. The problem with the event-plane method is that, although here it coincides with the previous definitions if $\Psi_n\{\rm EP\}$ is defined appropriately ($\Psi_n\{\rm EP\}=\Psi_n$ if $w=1$), in the experiments there is a finite number of particles in single event, resulting in a finite resolution in determining the event-plane angle. The finite event-plane resolution in turn introduces the ambiguity to the relation between the underlying flow coefficients $v_n$ and the measured event-averaged event-plane coefficients $\langle v_n\{\rm EP\} \rangle_{ev}$. In the high-resolution limit $\langle v_n\{\rm EP\} \rangle_{ev} \rightarrow \langle v_n \rangle_{ev}$, and in the low-resolution limit $\langle v_n\{\rm EP\} \rangle_{ev} \rightarrow \langle v_n^2 \rangle_{ev}^{1/2}$. In the presence of the flow fluctuations, these two averages are in general different. Typically, the real events are somewhere between these limits, and a consistent comparison to the data requires that the calculated events are analyzed similarly to the experiments~\cite{Holopainen:2010gz}, and even then the exact experimental configuration, e.g. non-uniform acceptance, that deviates from a theoretical perfect detector, can introduce ambiguity to the results~\cite{Luzum:2012da}. Thus, in this work, we do not consider the event-plane flow coefficients but rely on those obtained from the cumulants discussed next.

\subsection{Cumulants}
\label{sec:cumulants}

The ambiguity problem associated with the event-plane method can be resolved by using the $n$-particle cumulants. For example, the two-particle cumulant is defined as the correlation
\begin{equation}
 v_n\{2\}^2 = \langle e^{in(\phi_1 - \phi_2)}\rangle_\phi \equiv \frac{1}{N_{2}}\int d\phi_1 d\phi_2 \frac{dN_{2}}{d\phi_1 d\phi_2}e^{in(\phi_1 - \phi_2)},
 \label{eq:2plecumulant}
\end{equation}
where $dN_{2}/d\phi_1 d\phi_2$ is the two-particle spectrum (suppressing the possible rapidity and $p_T$ dependence), which can in general be decomposed as a sum of a product of single-particle spectra and a ``direct'' two-particle correlation $\delta_2(\phi_1, \phi_2)$,
\begin{equation}
 \frac{dN_{2}}{d\phi_1 d\phi_2} = \frac{dN}{d\phi_1}\frac{dN}{d\phi_2} + \delta_2(\phi_1, \phi_2).
\end{equation}
The direct correlations can result e.g. from a $\rho$-meson decaying into two pions, and these correlations are usually referred to as non-flow contributions. Using Eq.~\eqref{eq:vn}, the event-averaged two-particle cumulant can be written as
\begin{equation}
v_n\{2\} = \langle v_n^2 + \delta_2\rangle_{ev}^{1/2} \stackrel{{\rm flow}}{=} \langle v_n^2\rangle_{ev}^{1/2},
\end{equation}
where the last equality follows in the absence of the non-flow contributions, i.e. assuming that all the azimuthal correlations are due to the collective flow only. It turns out that the 2-particle cumulant always results in $\langle v_n^2\rangle_{ev}^{1/2}$ regardless of the event-plane resolution~\cite{Luzum:2012da}, therefore resolving the ambiguity in the event-plane method.

In our calculations, we use the single-particle spectra directly, i.e. we are not considering individual particles. Therefore, in our calculations the event-plane resolution is in principle (up to the numerical accuracy) infinite, and we do not need the corrections due to the finite event-plane resolution. Furthermore, even though we compute the hadron decays, they are done at the level of single-particle spectra, and thus all the direct correlations (non-flow) are absent in our calculations. We also note that typically in the experimental analysis the non-flow correlations are suppressed by choosing e.g.\ pseudorapidity gaps between the pairs of particles in Eq.~\eqref{eq:2plecumulant}.

For these reasons, for our purposes it is sufficient to define the cumulants directly through the flow-only limit. The $p_T$-integrated 2-particle cumulant flow coefficients are then defined as
\begin{equation}
v_n\{2\} \equiv \langle v_n^2\rangle_{ev}^{1/2},
\end{equation}
where the $v_n$ for a single event follows from Eq.~\eqref{eq:vn}, and the angular brackets denote the average over all the events in a given centrality class. Similarly, the event-averaged $p_T$-integrated 4-particle cumulant flow coefficients are defined as~\cite{Borghini:2001vi}
\begin{equation}
v_n\{4\} \equiv \left(2 \langle v_n^2\rangle_{ev}^{2} - \langle v_n^4\rangle_{ev}\right)^{1/4}.
\end{equation}

In addition to the $v_n\{2\}$ and $v_n\{4\}$, we also study the three-particle cumulant $v_4\{3\}$ measured by STAR \cite{Adams:2003zg}, defined as
\begin{equation}
v_4\{3\} \equiv \frac{\langle v_2^{2} v_4 \cos(4\left[\Psi_2 - \Psi_4\right])\rangle_{ev}}{\langle v_2^{2} \rangle_{ev}}.
\label{eq:v4_3ple}
\end{equation}

Originally, the higher-order cumulants were introduced to suppress the non-flow correlations~\cite{Borghini:2001vi}, but after the full realization of the importance of the event-by-event fluctuations~\cite{Alver:2010gr} it has become clear that different cumulants do not only have different sensitivity to non-flow correlations, but also measure different moments of the underlying probability distributions of the flow coefficients. 

\subsection{Event-plane correlations}

Different correlations between the flow coefficients and the event-plane angles give a rich variety of observables that can provide independent further constraints to the properties of the strongly interacting matter. In this paper, we consider also the correlations between the event-plane angles $\Psi_n$ of the different harmonics. In principle, one could define the correlations between the angles directly as $\langle \cos(k_1\Psi_1 + \cdots + n k_n\Psi_n)\rangle_{ev}$, with the $\Psi_n$ angles defined according to Eq.~\eqref{eq:EPangle}, but as was noted in Ref.~\cite{Luzum:2012da}, this leads to a similar ambiguity related to the event-plane resolution as for the event-plane $v_n\{{\rm EP}\}$ discussed above. For this reason it was suggested that it is better to define the event-plane correlations as 
\begin{multline}
 \langle \cos(k_1\Psi_1 + \cdots + n k_n\Psi_n)\rangle_{{\rm SP}} \equiv \\
\frac{\langle v_1^{|k_1|} \cdots v_n^{|k_n|} \cos(k_1\Psi_1 + \cdots + n k_n\Psi_n)\rangle_{ev}}{\sqrt{\langle v_1^{2|k_1|} \rangle_{ev} \cdots \langle v_n^{2|k_n|} \rangle_{ev}}},
 \label{eq:epcorrelation}
\end{multline}
where the $k_n$'s are integers with the property $\sum_n nk_n = 0$. This definition is actually equal to the low resolution limit of the (naive) definition above. These correlations were recently measured by the ATLAS Collaboration \cite{Aad:2014fla}, by using both definitions.

%%%%%%%%%%%%%%%%%%%%% SECTION %%%%%%%%%%%%%%%%%%%%%
\section{Results}
\subsection{Multiplicities, $p_T$--spectra and average $p_T$}

Once we have fixed the coefficients of the non-linear terms in Eq.~\eqref{eq:IShydro} from the kinetic theory calculations, and the width $\sigma=0.43$ fm of the gluonic $T_n$ from the HERA data,  we have essentially four free parameters $\{K_{\rm sat}, \beta, \mathrm{BJ/FS}, \eta/s(T)\}$ in our model. As shown in our previous studies \cite{Paatelainen:2012at,Paatelainen:2013eea}, the parameters $K_{\rm sat}$ and $\beta$ are strongly correlated and a continuum of equally well working pairs can be found, however so that the experimental data slightly favors larger values of $\beta$. For simplicity, to reduce the number of free parameters, we fix here $\beta=0.8$, and choose the BJ-case for the pre-thermal evolution discussed in Sec.~\ref{sec:iniforhydro}. We then tune the remaining parameter $K_{\rm sat}$ so that the charged hadron multiplicity $dN_{\rm ch}/d\eta_{\rm ps}$ matches the ALICE measurement in the most central Pb+Pb collisions, i.e., in the $0-5$ \% centrality class at the LHC. It should be emphasized that no further tuning is done for other centralities at the LHC, or for any of the RHIC results.

%%%%%%%%%%%%%%%%%%%%% FIGURE %%%%%%%%%%%%%%%%%%%%%
\begin{figure*}
\includegraphics[width=8.5cm]{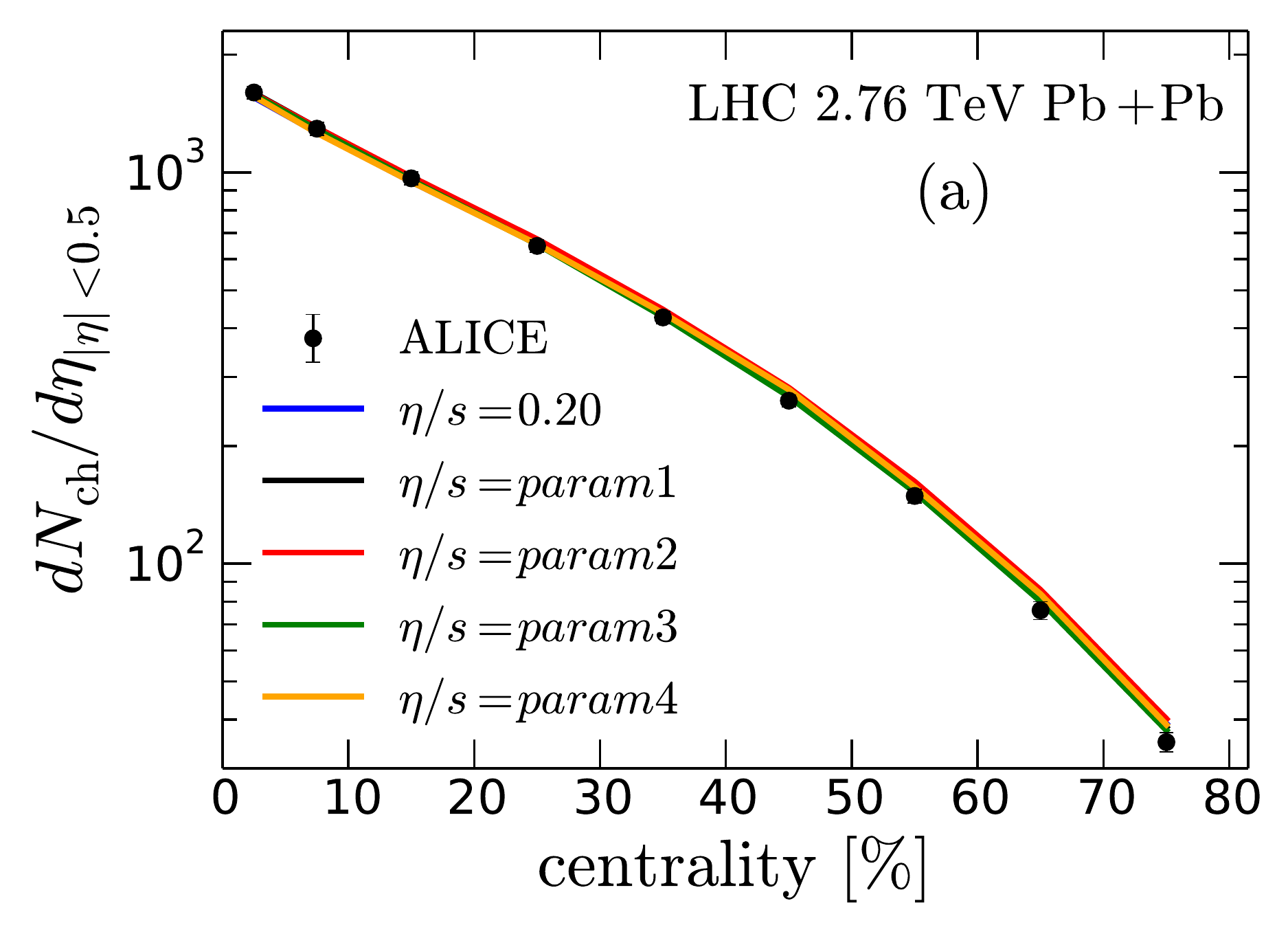}
\includegraphics[width=8.5cm]{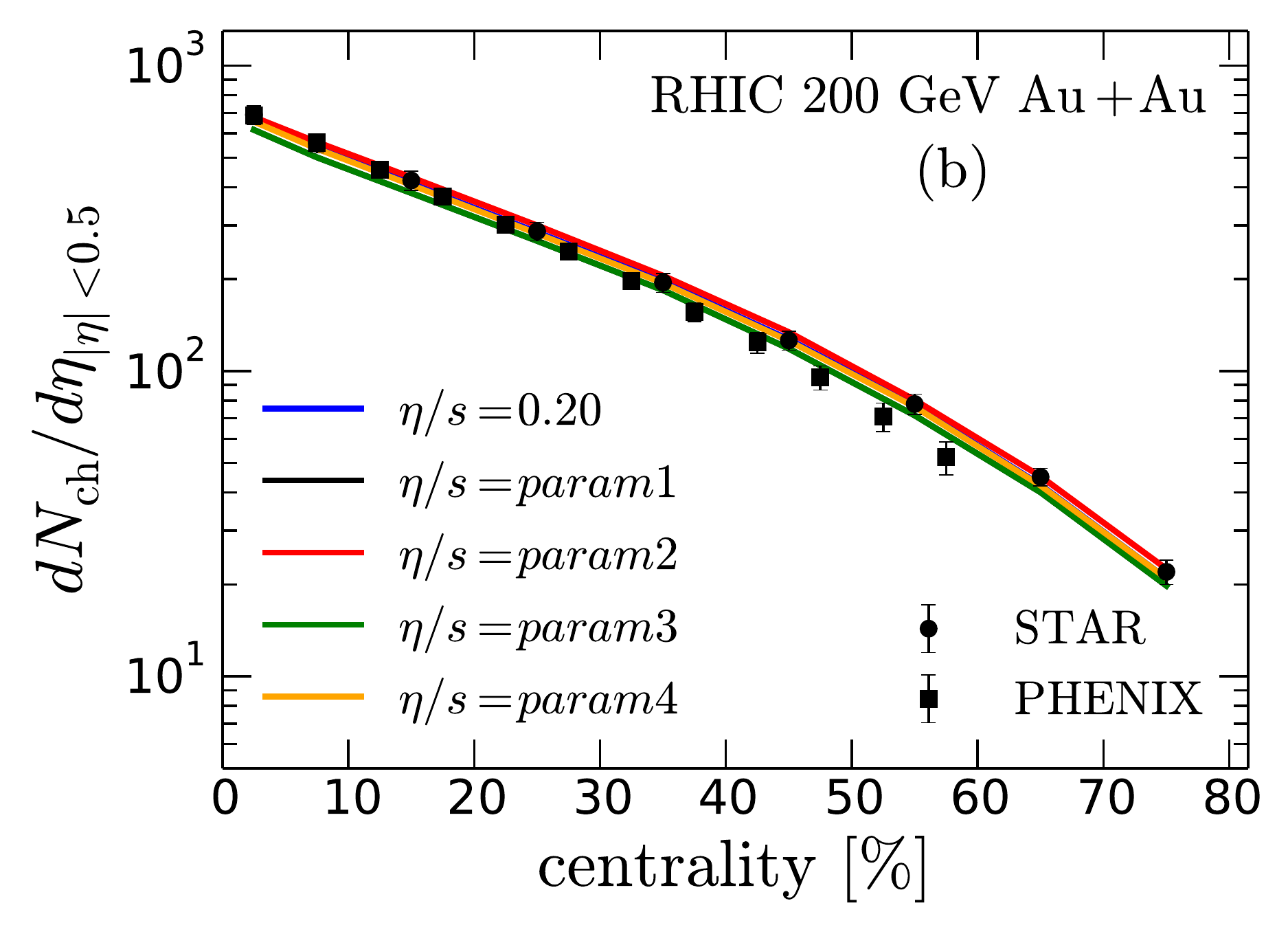}
\caption{(Color online) Centrality dependence of charged hadron multiplicities in $\sqrt{s_{NN}} = 2.76$  TeV Pb+Pb collisions at the LHC (panel (a)) and $200$ GeV Au+Au collisions at RHIC (panel (b)), computed for the five $\eta/s(T)$ parametrizations shown in Fig.~\ref{fig:etapers}. Experimental data are from ALICE \cite{Aamodt:2010cz}, STAR \cite{Abelev:2008ab} and PHENIX \cite{Adler:2004zn}. 
}
\label{fig:multiplicity_LHC}
\end{figure*}
%%%%%%%%%%%%%%%%%%%%% FIGURE %%%%%%%%%%%%%%%%%%%%%

As discussed in Sec.~\ref{sec:hydrosetup}, we consider the five different $\eta/s(T)$ parametrizations shown in Fig.~\ref{fig:etapers}. The viscous entropy production, different for each $\eta/s(T)$ case, needs to be compensated by (iteratively) adjusting $K_{\rm sat}$ for each parametrization. The obtained values of $K_{\rm sat}$ are shown for each $\eta/s$ parametrization in Table~\ref{tab:ksat}. The resulting centrality dependence of the charged particle multiplicity in Pb+Pb collisions at the LHC is shown in Fig.~\ref{fig:multiplicity_LHC}a and compared with the ALICE measurements~\cite{Aamodt:2010cz}. As can be seen from the figure, our calculation matches very well with the measured data, and in practice all the five $\eta/s$ parametrizations give an equally good agreement. 
\begin{table}[h]
\caption{The values of $K_{\rm sat}$ for different $\eta/s$ parametrizations.}
\begin{tabular*}{\columnwidth}{@{\extracolsep{\fill} } cccccc}
\hline
\hline
$\eta/s$	&  0.20 & param1 & param2 & param3  & param4  \\
$K_{\rm sat}$	&  0.63 &  0.50	 & 0.75   & 0.45    & 0.64   \\
\hline
\hline
\end{tabular*}
\label{tab:ksat}
\end{table}

Once the parameters are fixed at the LHC, the $\sqrt{s}$-, centrality- and also $A$-dependences follow from the calculation. The comparison of the corresponding calculation for Au+Au collisions at the top energy of RHIC is compared to the PHENIX~\cite{Adler:2004zn} and STAR~\cite{Abelev:2008ab} measurements in Fig.~\ref{fig:multiplicity_LHC}b. As can be seen from the figure, the agreement with the calculation and experimental data is again very good. We emphasize that here also the multiplicity in the most central collisions follows from the calculation, i.e., we do not change $K_{\rm sat}$ with $\sqrt{s}$ or $A$. Because $\eta/s(T)$ is also by definition independent of $\sqrt{s}$ and $A$, we are now in principle equipped to predict the multiplicities in any other collision systems, provided that the fluid dynamics and pQCD + saturation pictures are valid.

%%%%%%%%%%%%%%%%%%%%% FIGURE %%%%%%%%%%%%%%%%%%%%%
\begin{figure*}
\includegraphics[width=8.7cm]{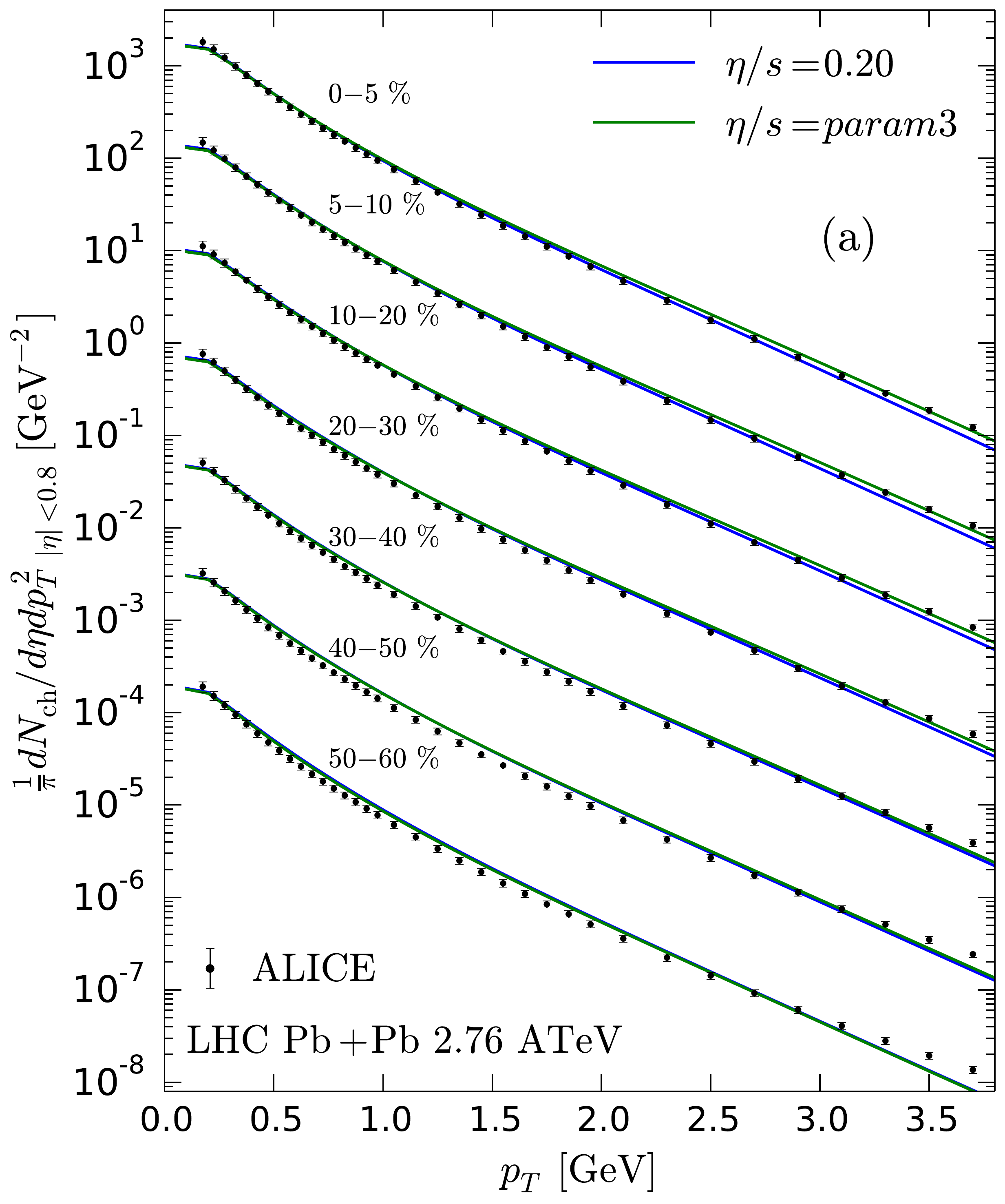}
\includegraphics[width=8.7cm]{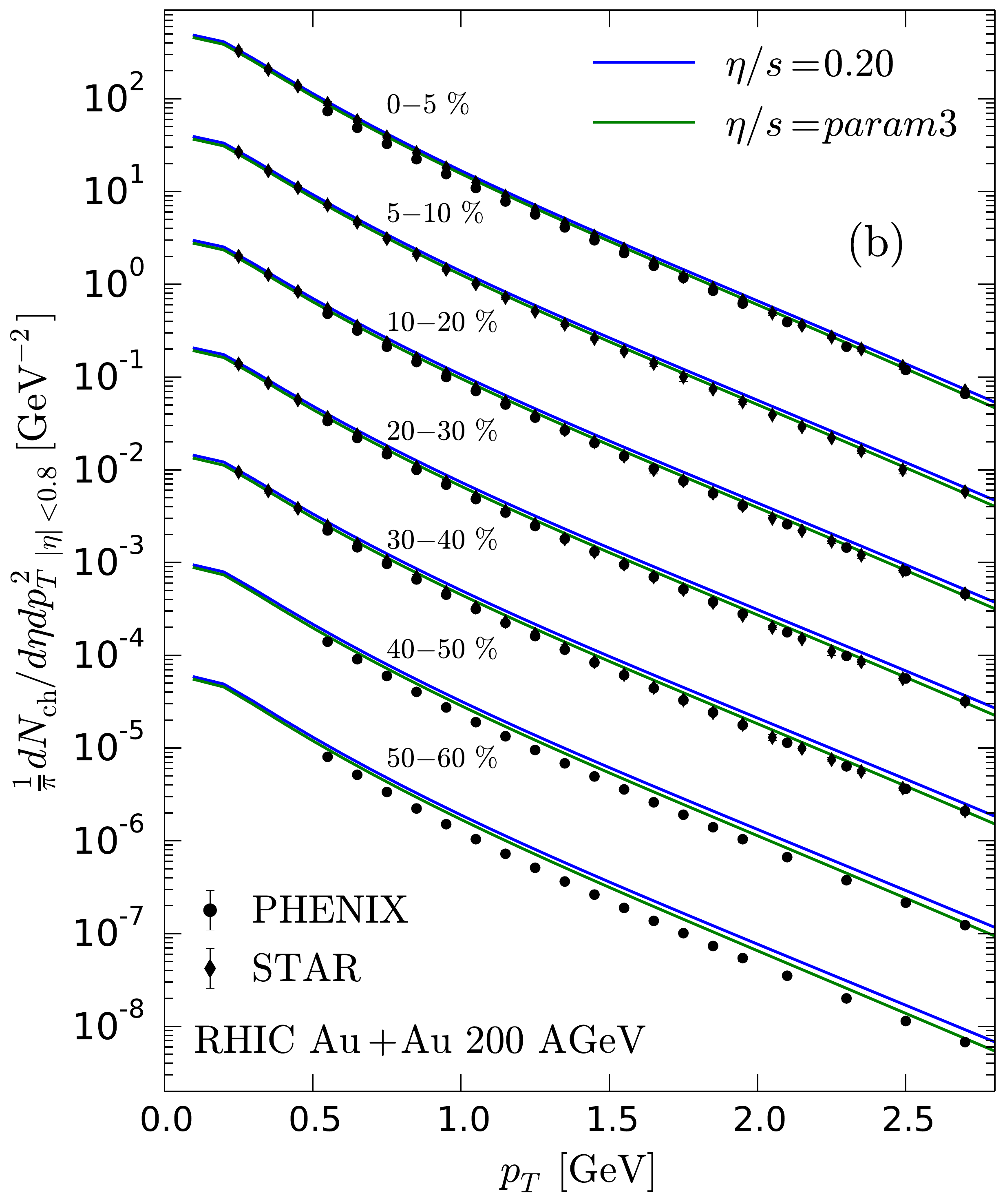}
\caption{(Color online) Transverse momentum spectra of charged hadrons in $\sqrt{s_{NN}} = 2.76$  TeV Pb+Pb collisions at the LHC (panel (a)) and $200$ GeV Au+Au collisions at RHIC (panel (b)), in the same centrality bins as in Fig.~\ref{fig:multiplicity_LHC},  
computed for the five $\eta/s(T)$ parametrizations shown in Fig.~\ref{fig:etapers}.  
Experimental data are from ALICE \cite{Abelev:2012hxa}, STAR \cite{Adams:2003kv} and PHENIX \cite{Adler:2003au}. For visibility, the curves and the data points have been shifted by increasing powers of 10.}
\label{fig:charged_spectra}
\end{figure*}
%%%%%%%%%%%%%%%%%%%%% FIGURE %%%%%%%%%%%%%%%%%%%%%
The comparison of the calculated $p_T$-spectra of charged hadrons with the ALICE measurement \cite{Abelev:2012hxa} in Pb+Pb collisions at the LHC is shown in Fig.~\ref{fig:charged_spectra}a, and the corresponding comparison with the STAR~\cite{Adams:2003kv} and PHENIX~\cite{Adler:2003au} data in Au+Au collisions at RHIC is shown in Fig.~\ref{fig:charged_spectra}b. As long as the multiplicities are well described, the $p_T$-spectra are quite insensitive to the $\eta/s$ parametrizations. In fact, the most important parameters that dictate the behavior of the $p_T$-spectra are the kinetic and chemical freeze-out temperatures $T_{\rm dec}$ and $T_{\rm chem}$. While the multiplicity ratios  of the identified hadrons, e.g. the pion-to-proton ratio, are best reproduced with $T_{\rm chem} \sim 150$ MeV, it tends to give too flat $p_T$-spectra, especially in the low-$p_T$ region, where fluid dynamics is expected to work best. This is the reason for our choice of a rather high $T_{\rm chem} = 175$ MeV. Although, the proton yields are somewhat overpredicted with this choice, we can, however, get a good description of the low-$p_T$ region of the charged hadron spectra, which we consider here more important than a detailed description of the hadronic chemistry. Inclusion of bulk viscosity could help to improve the overall agreement with the data, see e.g.\ Ref.~\cite{Ryu:2015vwa}.

%%%%%%%%%%%%%%%%%%%%% FIGURE %%%%%%%%%%%%%%%%%%%%%
\begin{figure*}
\includegraphics[width=8.5cm]{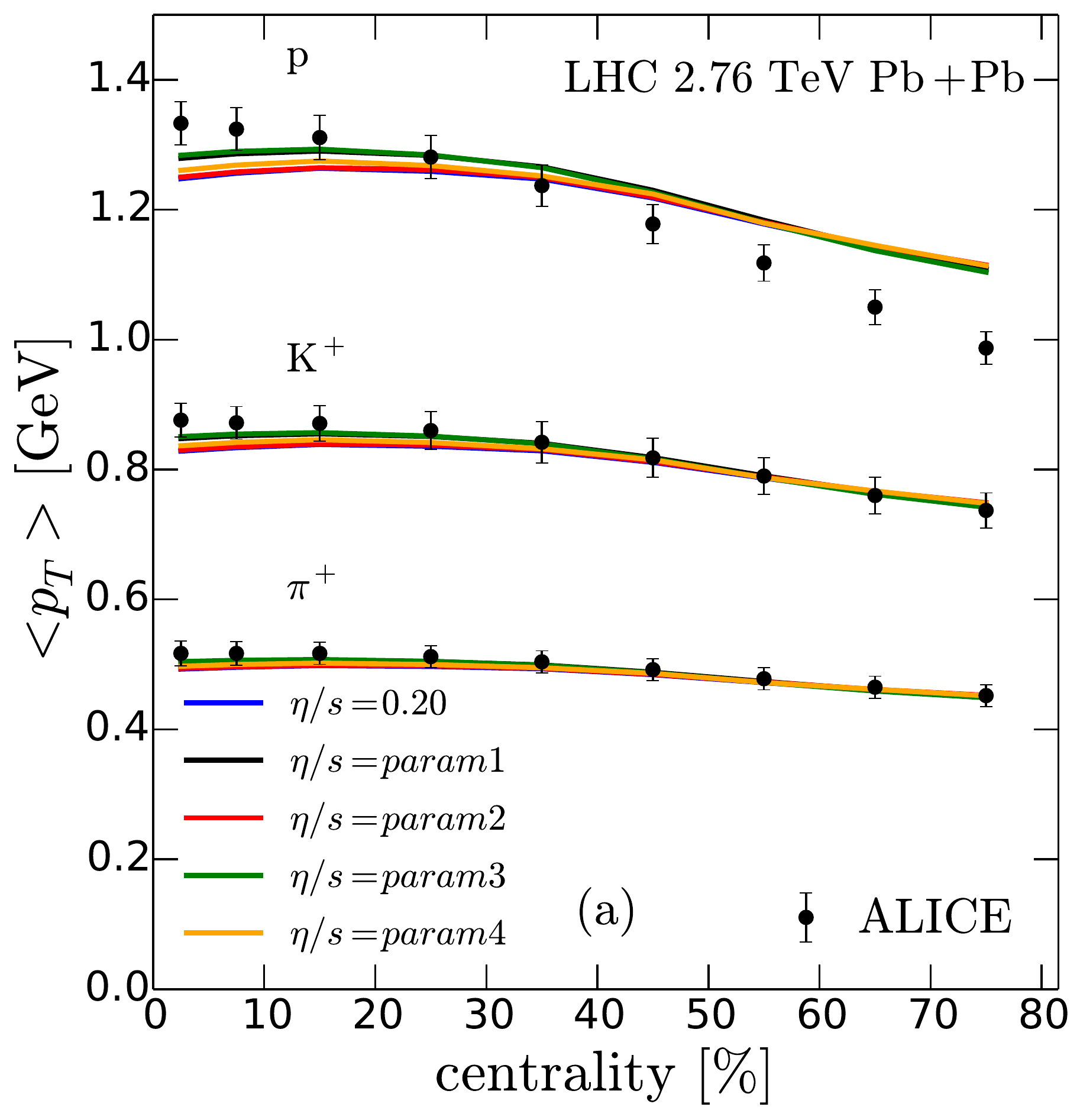}
\includegraphics[width=8.5cm]{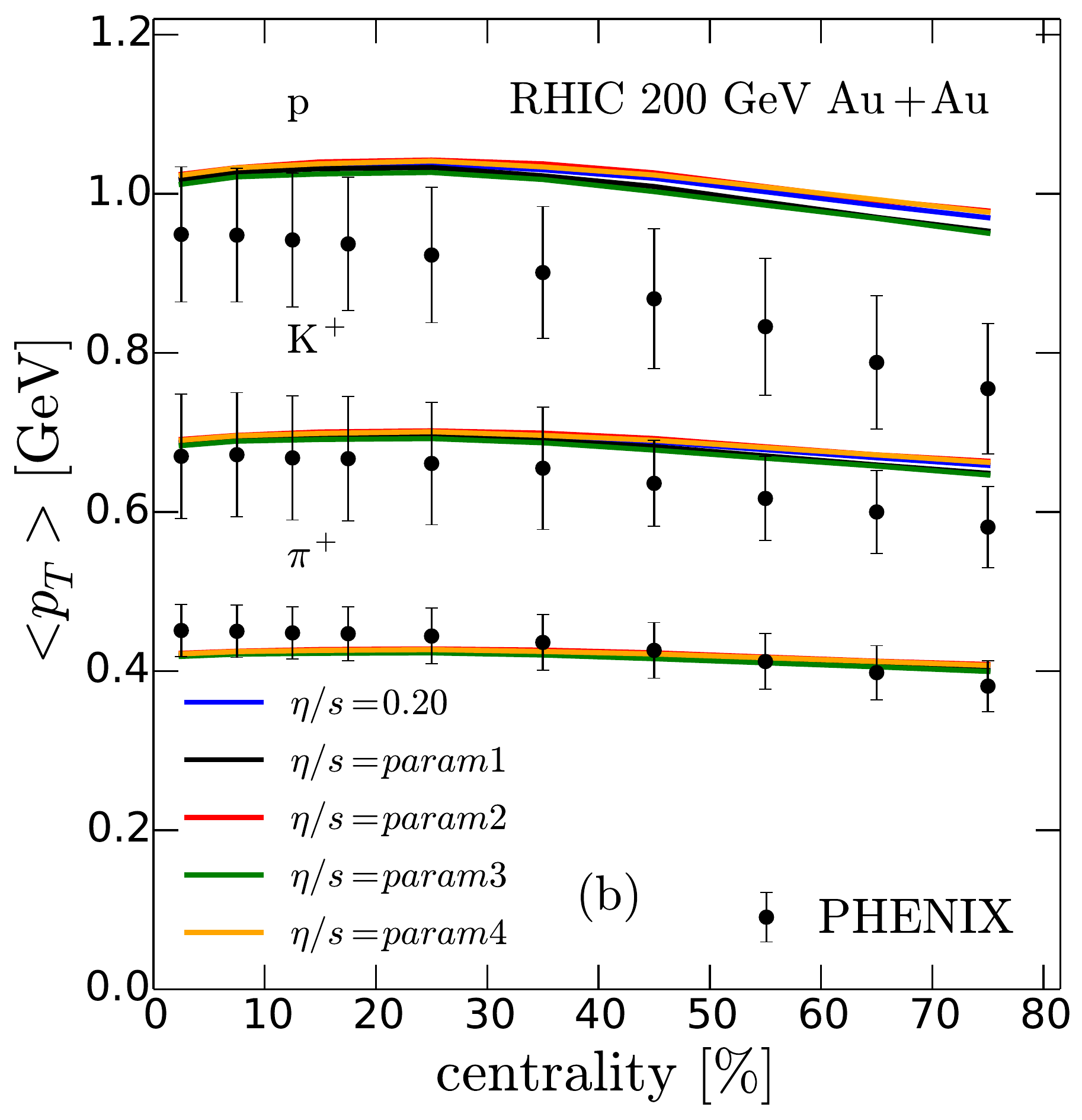}
\caption{(Color online) Centrality dependence of the average $p_T$ for pions, kaons and protons in $\sqrt{s_{NN}} = 2.76$  TeV Pb+Pb collisions at the LHC (panel (a)) and $200$  GeV Au+Au collisions at RHIC (panel (b)), computed for the five $\eta/s(T)$ parametrizations shown in Fig.~\ref{fig:etapers}.  Experimental data are from ALICE \cite{Abelev:2013vea} and PHENIX \cite{Adler:2003cb}.}
\label{fig:avept}
\end{figure*}
%%%%%%%%%%%%%%%%%%%%% FIGURE %%%%%%%%%%%%%%%%%%%%%
Figure \ref{fig:avept}a shows the average $p_T$ for pions, kaons and protons compared to the ALICE measurements~\cite{Abelev:2013vea}. The pions are at low-$p_T$ the most abundant particles, and the very good agreement of our results with the data reflects the fact that the low-$p_T$ region of the $p_T$-spectra is well enough described. The same conclusion holds for the average $p_T$ in Au+Au collisions at RHIC, shown in Fig.~\ref{fig:avept}b against the PHENIX~\cite{Adler:2003cb} data. While in both cases the average $p_T$ of pions is well reproduced, especially the centrality dependence of the proton $\langle p_T\rangle$ does not come out correctly. Whether this could be cured by a more detailed account of the chemical reactions in the hadron gas in the fluid-dynamical calculation, or whether a full microscopic treatment is needed, remains an open question. Overall, the agreement with the low-$p_T$ charged hadron spectra, and the very good agreement with the pion average $p_T$ gives us confidence that the $p_T$-integrated bulk observables for charged hadrons can be well described within our framework.

\subsection{Flow coefficients}

The viscosity does affect the multiplicities through the viscous entropy production, but this effect gets here compensated by the re-tuning of $K_{\rm sat}$ for each $\eta/s$ parametrization. Also, once the multiplicities are reproduced, the details of the $p_T$-spectra are quite insensitive to the values of $\eta/s$. Therefore, these quantities do not give a direct access to the determination of $\eta/s$ from the experimental data. The most direct constraint to the viscosity of the strongly interacting matter comes from the azimuthal structure of the hadron spectra.

%%%%%%%%%%%%%%%%%%%%% FIGURE %%%%%%%%%%%%%%%%%%%%%
\begin{figure*}
\includegraphics[width=8.7cm]{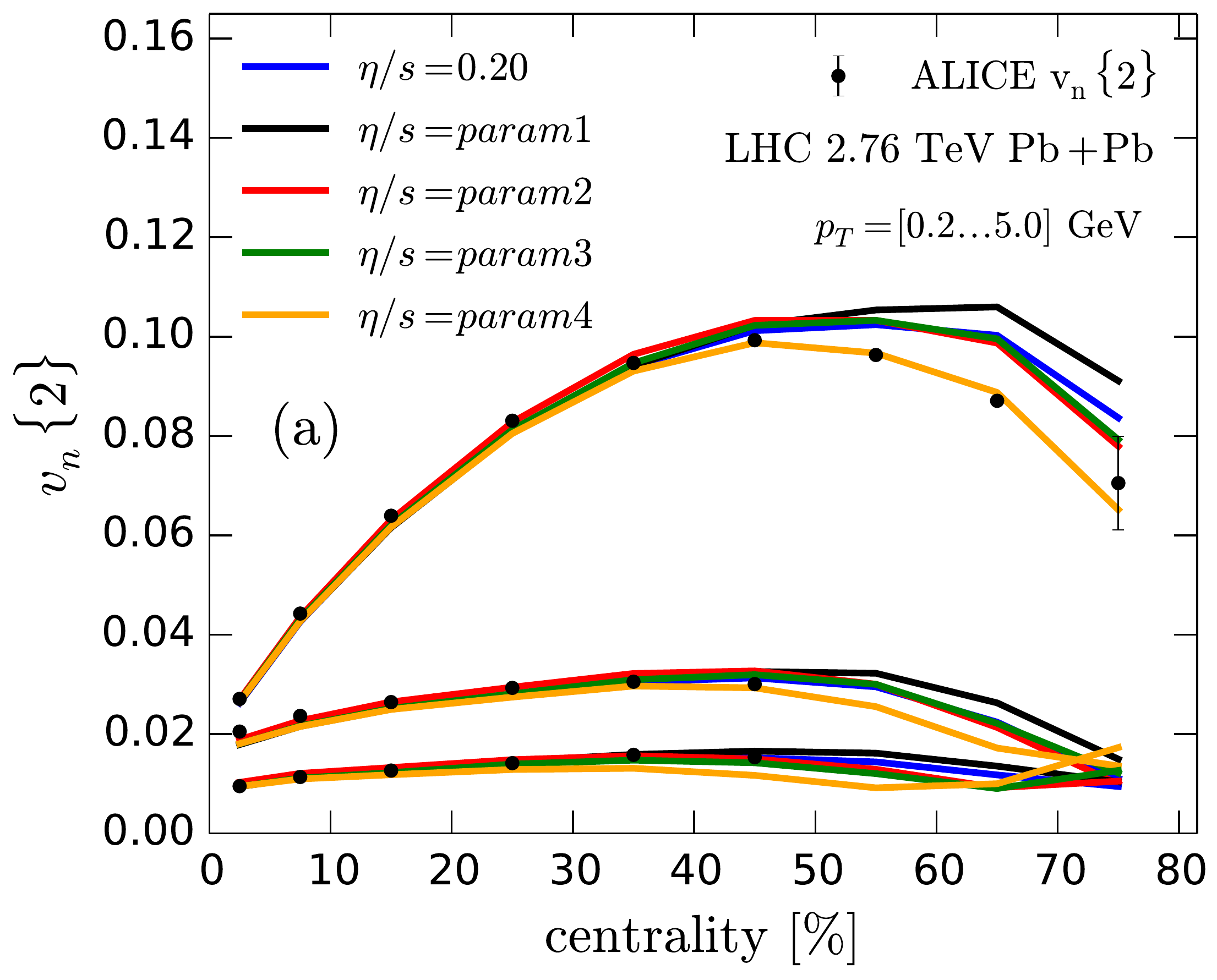}
\includegraphics[width=8.6cm]{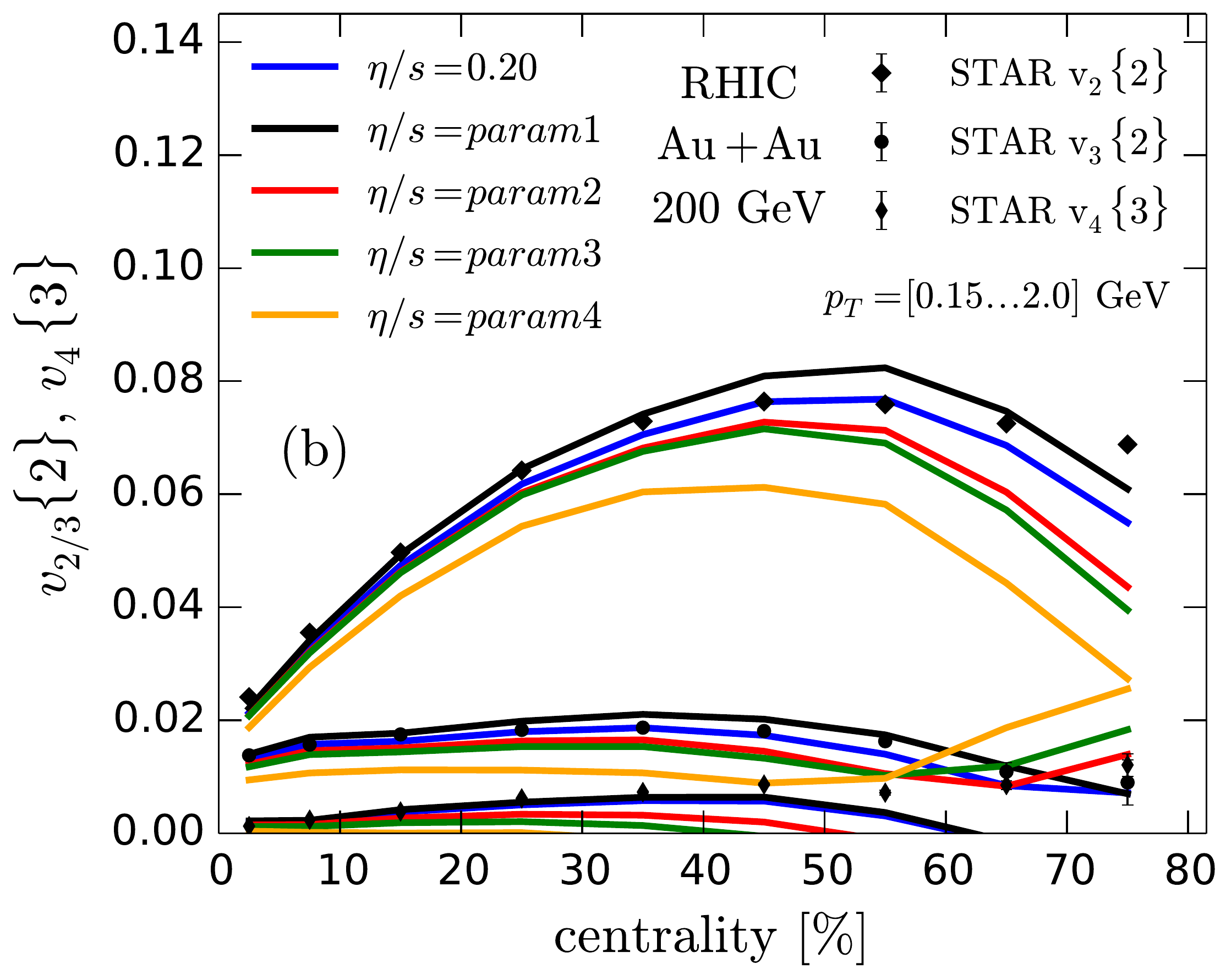}
\caption{(Color online) Centrality dependence of the flow coefficients $v_n\{2\}$  from the charged hadron 2-particle cumulants in $\sqrt{s_{NN}} = 2.76$  TeV Pb+Pb collisions at the LHC (panel (a)), and the coefficients $v_2\{2\}$, $v_3\{2\}$, and $v_4\{3\}$ from the charged hadron 2- and 3-particle cumulants in $200$  GeV Au+Au collisions at RHIC (panel (b)), computed for the five $\eta/s(T)$ parametrizations shown in Fig.~\ref{fig:etapers}.  Experimental data are from ALICE \cite{ALICE:2011ab} and STAR \cite{Adams:2004bi, Adamczyk:2013waa, Adams:2003zg}.
}
\label{fig:charged_vn}
\end{figure*}
%%%%%%%%%%%%%%%%%%%%% FIGURE %%%%%%%%%%%%%%%%%%%%%
The computed 2-particle cumulant $v_n\{2\}$ for charged hadrons at different centralities in Pb+Pb collisions at the LHC are shown in Fig.~\ref{fig:charged_vn}a against the ALICE data~\cite{ALICE:2011ab}. For the definitions, see  Sec.~\ref{sec:cumulants}. As the figure verifies, all the parametrizations of Fig.~\ref{fig:etapers} reproduce (by construction) the $v_n$'s at the LHC up to 40--50\% centralities very well. Thus, the LHC $v_n$ data alone do not allow to distinguish between the different $\eta/s$ temperature dependencies to such a precision. More notable differences appear only in the more peripheral collisions, where the uncertainties related to the fluid dynamics and its applicability, as well as to the initial state calculation, are large. 

One can also note that the higher harmonics measured at the LHC do not give directly additional constraints to the temperature dependence of the viscosity. The ratio of $v_3$ or $v_4$ to the elliptic flow coefficient $v_2$, however, depends strongly on the initial conditions, through the ratio of the initial eccentricities $\varepsilon_2/\varepsilon_n$. Therefore, the higher harmonics give an indirect constrain to the $\eta/s$, by restricting the possible initial states, see Ref.~\cite{Retinskaya:2013gca}. As seen in the figure, our approach with pQCD + saturation initial conditions describe the $v_n$'s very well.

So far, the $v_n$'s at the LHC give at most the upper limit for the minimum of $\eta/s$ (corresponding to the constant $\eta/s=0.20$), but even with these choices it varies between $\eta/s|_{\rm min} = 0.08$ and $0.20$, with a possibility that even smaller $\eta/s_{\rm min}$ could be tuned to fit the data. Furthermore, the location of the minimum is not constrained either. For the low- and high-temperature $\eta/s$ the uncertainties are even larger than for the minimum. It is then clear that further constraints are needed in order to pin down the temperature dependence of $\eta/s$.

A simultaneous analysis of other collision systems can provide further independent constraints for $\eta/s(T)$. As discussed in Refs.~\cite{Niemi:2011ix, Niemi:2012ry}, the viscous suppression of $v_n$'s depends differently on the temperature dependence of $\eta/s(T)$ at different collision energies. In $\sqrt{s_{NN}} = 200$ GeV Au+Au collisions at RHIC the $v_n$'s are practically independent of the high temperature, $T \gg T_c$, shear viscosity. At higher energies the high-temperature viscosity becomes gradually more important, while the influence of the hadronic viscosity decreases. 

In Fig.~\ref{fig:charged_vn}b we show the computed $v_2\{2\}$, $v_3\{2\}$, and $v_4\{3\}$ for charged hadrons in $\sqrt{s_{NN}} = 200$ GeV Au+Au collisions at RHIC compared to the STAR data~\cite{Adams:2004bi, Adamczyk:2013waa, Adams:2003zg}. As one can read from the figure, the same $\eta/s(T)$ parametrizations that gave an equally good fit to the $v_n$ data at the LHC are now clearly separated, demonstrating that the simultaneous RHIC and LHC analysis of $v_n$'s can be used at least to rule out some temperature dependencies. Here, especially, the $param4$ with a large hadronic viscosity fails to describe the data. Overall, the best agreement with the data is obtained with a constant $\eta/s=0.20$ and $\eta/s$ from $param1$ with the minimum at $T=150$ MeV. 

\subsection{Flow fluctuations}

A proper event-by-event description of heavy-ion collisions collisions should not only reproduce the event-averaged $v_n$'s but also their EbyE probability distributions $P(v_n)$. As we show here, and as earlier reported in Ref.~\cite{Niemi:2012aj}, it turns out that the probability distributions of the scaled $v_n$, defined as
\begin{equation}
 \delta v_n = \frac{v_n - \langle v_n\rangle_{\rm ev}}{\langle v_n\rangle_{\rm ev}},
 \label{eq:deltavn}
\end{equation}
are essentially independent of the details of the fluid dynamical evolution, but depend only on the corresponding eccentricity fluctuations of the initial state. Therefore, the current LHC data on $P(v_n)$ provide a direct constraint for the initial states \cite{Renk:2014jja} such as we compute here.

%%%%%%%%%%%%%%%%%%%%% FIGURE %%%%%%%%%%%%%%%%%%%%%
\begin{figure*}
\includegraphics[width=8.5cm]{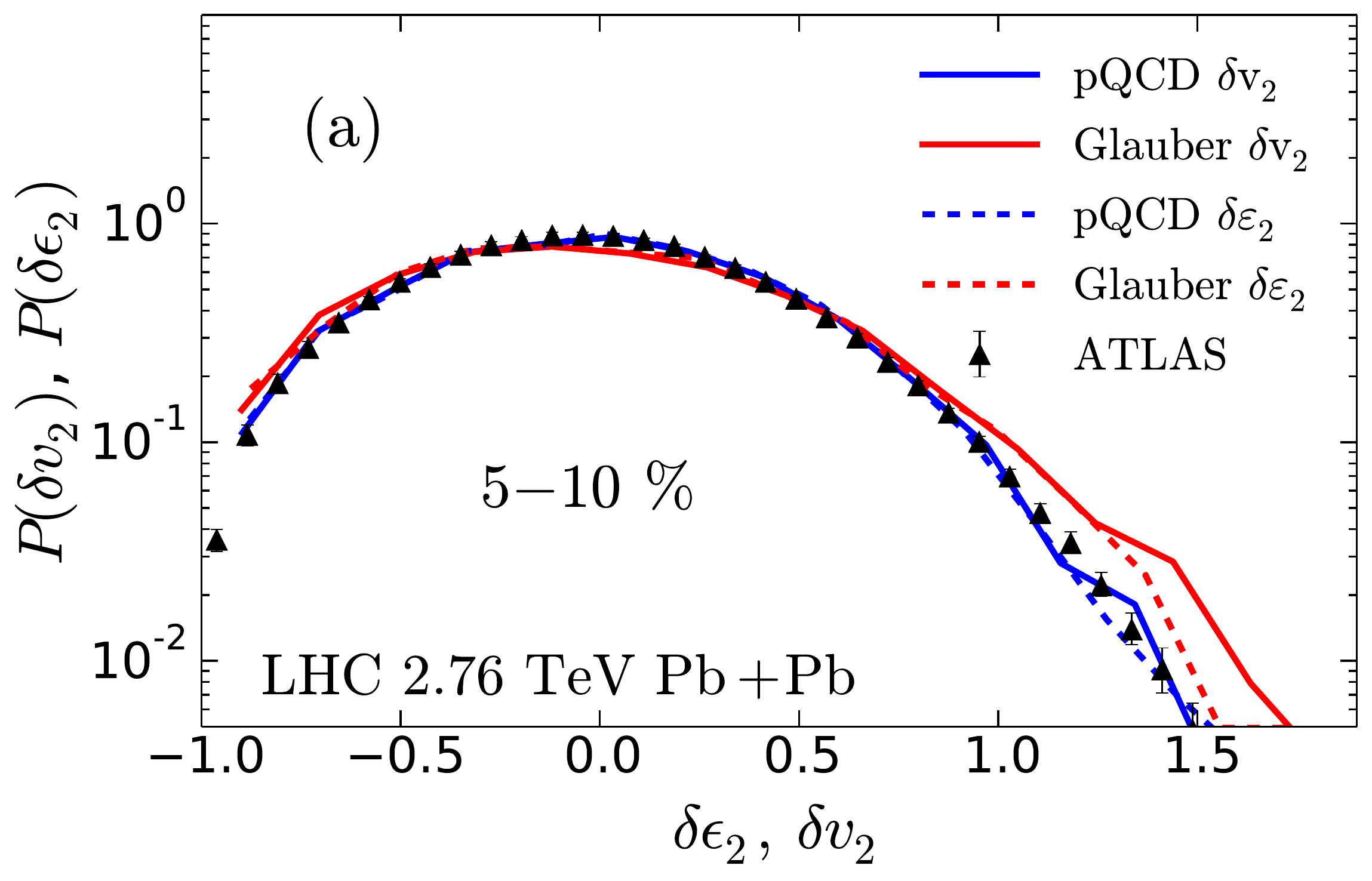}
\includegraphics[width=8.5cm]{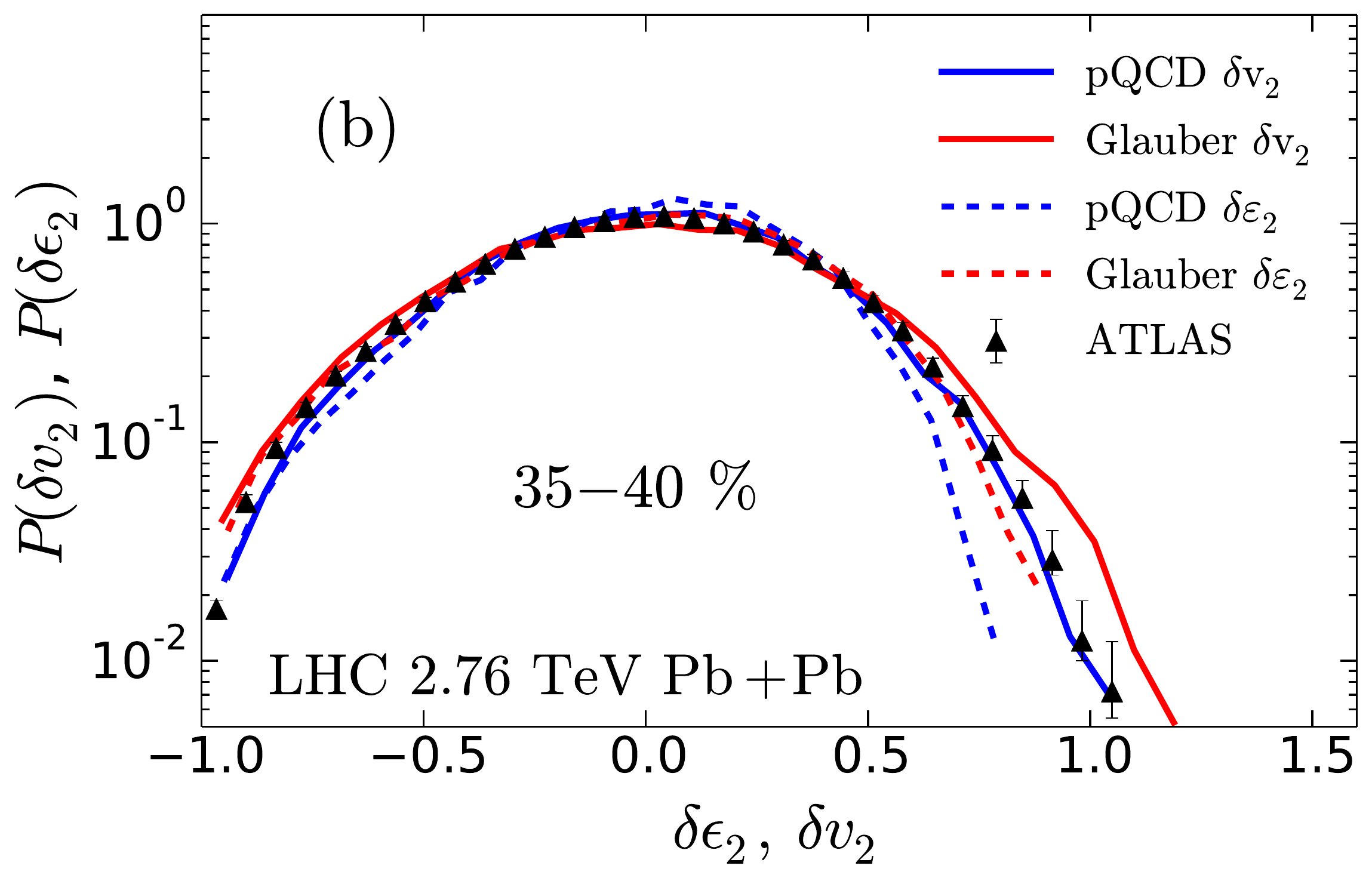}
\caption{(Color online) Panel (a): Fluctuation spectra of the final-state $v_{2}$ of charged hadrons (solid curves) and of the initial state $\varepsilon_2$ (dashed) in the $5-10$ \% centrality class in $\sqrt{s_{NN}} = 2.76$  TeV Pb+Pb collisions at the LHC, computed with the pQCD + saturation initial states and $\eta/s=0.20$, and with the Glauber-model initial states using $\eta/s=0.10$. The experimental data are from ATLAS \cite{Aad:2013xma}. Panel (b): The same but for the 35-40\% centrality class.}
\label{fig:v2_fluctuations_5_10}
\end{figure*}
%%%%%%%%%%%%%%%%%%%%% FIGURE %%%%%%%%%%%%%%%%%%%%%
In Figs.~\ref{fig:v2_fluctuations_5_10}a and \ref{fig:v2_fluctuations_5_10}b we show the computed $P(\delta v_2)$ fluctuation spectra compared to the ATLAS data~\cite{Aad:2013xma} in the $5-10$ \% and $35-40$ \% centrality classes, respectively. The pQCD+saturation initial state in this figure is computed with $\eta/s=0.20$. For comparison, we also show a calculation with the usual Glauber initial condition, where the energy density is proportional to a linear combination of a binary collision density $\rho_{\rm bin}$ and a wounded nucleon density $\rho_{\rm wn}$, i.e.\ $e \propto f \rho_{\rm bin} + (1-f)\rho_{\rm wn}$ with $f=0.15$ and $\eta/s=0.10$ to approximately match the measured centrality dependence of the multiplicity and $v_2$. The probability densities of the scaled eccentricities, $P(\delta \varepsilon_2)$ are also shown in the figures. 

As seen in the panel (a) of Fig.~\ref{fig:v2_fluctuations_5_10}, in the near central collisions the scaled $v_2$ distribution follows closely the distribution of the scaled $\varepsilon_2$ with both the EKRT and Glauber initial states. The pQCD-based initial conditions give a very good description of the ATLAS data, while the Glauber initial conditions result in a too wide distribution. In mid-peripheral collisions, shown in the panel (b) of Fig.~\ref{fig:v2_fluctuations_5_10}, the EKRT initial conditions give still a good description of the ATLAS data and the Glauber result is still too wide. However, as clearly seen in the figure, the scaled $v_n$ distributions do not anymore follow the eccentricity distribution, but the $v_2$ distributions are visibly wider than the $\varepsilon_2$ distributions, concretely demonstrating the necessity of fluid dynamics in describing the detailed response to the initial eccentricities, see also Ref.~\cite{Schenke:2014zha}. The fluctuation spectra of the higher harmonics $v_3$ and $v_4$ are also well reproduced with the pQCD+saturation initial conditions, but they do not show similar sensitivity to the initial conditions as the $v_2$ fluctuations.

%%%%%%%%%%%%%%%%%%%%% FIGURE %%%%%%%%%%%%%%%%%%%%%
\begin{figure}[t]
\includegraphics[width=8.5cm]{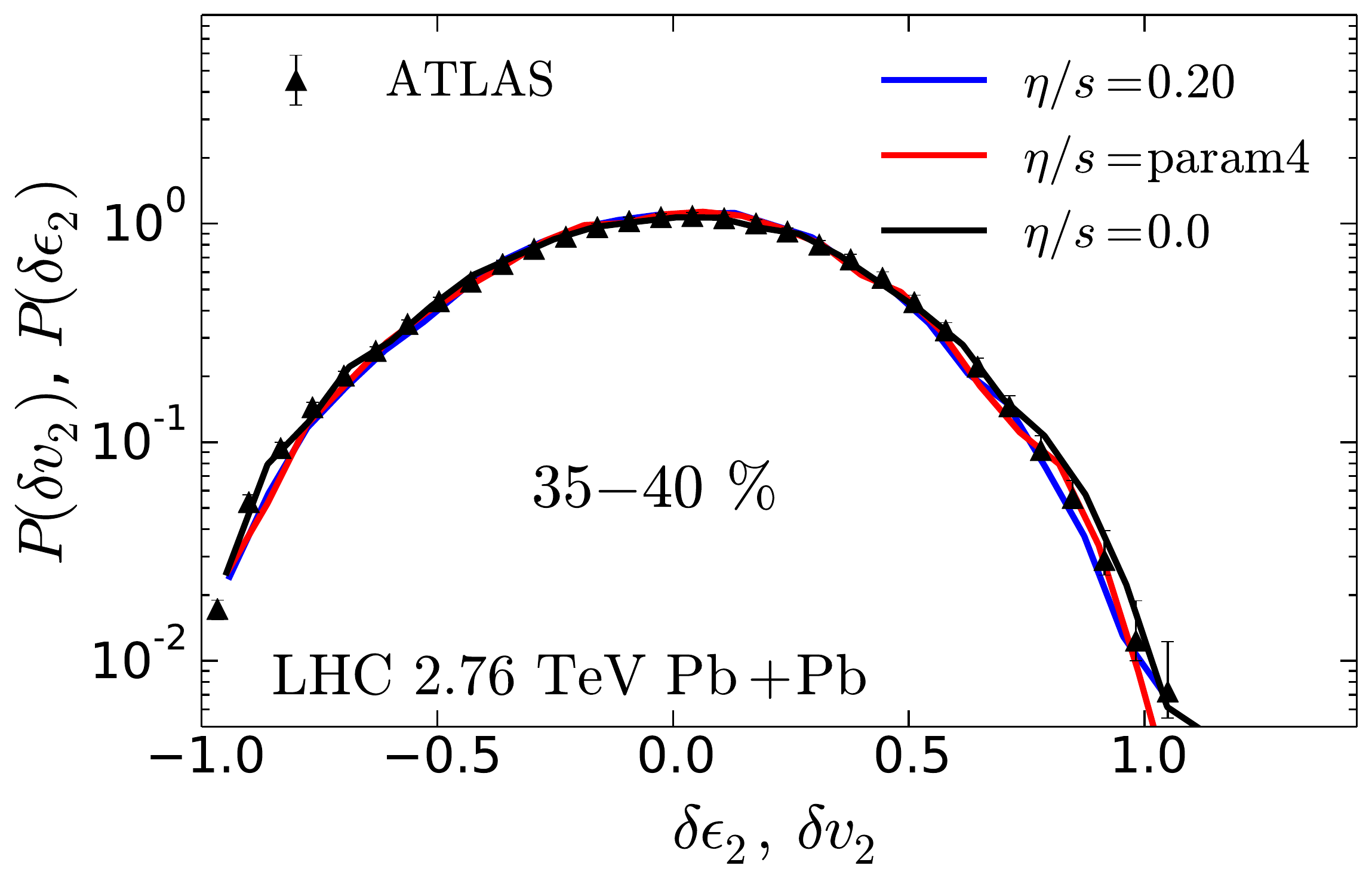}
\caption{(Color online) Fluctuation spectrum of $v_{2}$ of charged hadrons in the $35-40$ \% centrality class in $\sqrt{s_{NN}} = 2.76$  TeV Pb+Pb collisions at the LHC, computed with the pQCD + saturation initial states and with two different parametrizations of $\eta/s(T)$ and also using ideal fluid dynamics, $\eta/s=0$. The experimental data are from ATLAS \cite{Aad:2013xma}.}
\label{fig:v2_fluctuations_35_40_etas}
\end{figure}
%%%%%%%%%%%%%%%%%%%%% FIGURE %%%%%%%%%%%%%%%%%%%%%
Figure \ref{fig:v2_fluctuations_35_40_etas} shows the $P(\delta v_2)$ distribution of charged hadrons in the same $35-40$ \% centrality class with pQCD + saturation initial conditions as panel (b) of Fig.~\ref{fig:v2_fluctuations_5_10}, but with three different $\eta/s(T)$ parametrizations: $\eta/s=0.20$, $\eta/s=param4$, and $\eta/s=0$. As can be seen from the figure, the final $\delta v_2$ distribution is the same with all three $\eta/s$ parametrizations. This is true even in the perfect fluid limit $\eta/s=0$. This shows that even if the fluid dynamical evolution plays a crucial role in getting the final $v_2$ distributions correctly reproduced in the peripheral collisions, they are still a good probe of the initial conditions, because they do not depend on the \emph{details} of the fluid dynamical evolution. 

Then, a very interesting question is how directly the final-state $v_2$ distribution can reflect the initial state $\varepsilon_2$ distribution (and vice versa). If $v_2$ and $\varepsilon_2$ are, to a sufficient approximation, linearly correlated, $v_2 \propto \varepsilon_2$, then the scaled distributions $P(\delta v_2)$ and $P(\delta \varepsilon_2)$ are naturally identical. As seen from the panel (a) of Fig.~\ref{fig:v2_fluctuations_5_10}, this is the case in central collisions. However, as noticed from the panel (b), the distributions are not anymore the same in peripheral collisions, indicating that there must be  deviations from the linear relation. What complicates the initial state extraction from the $v_2$ fluctuation spectrum further is that the $\varepsilon_n \equiv \varepsilon_{n,n}$ are not actually sufficient to determine the full angular structure of the initial density profile, but in principle all of the $\varepsilon_{m,n}$ coefficients, defined in Eq.~\eqref{eq:eccentricities}, are needed. 

%%%%%%%%%%%%%%%%%%%%% FIGURE %%%%%%%%%%%%%%%%%%%%%
\begin{figure*}
\includegraphics[width=5.9cm]{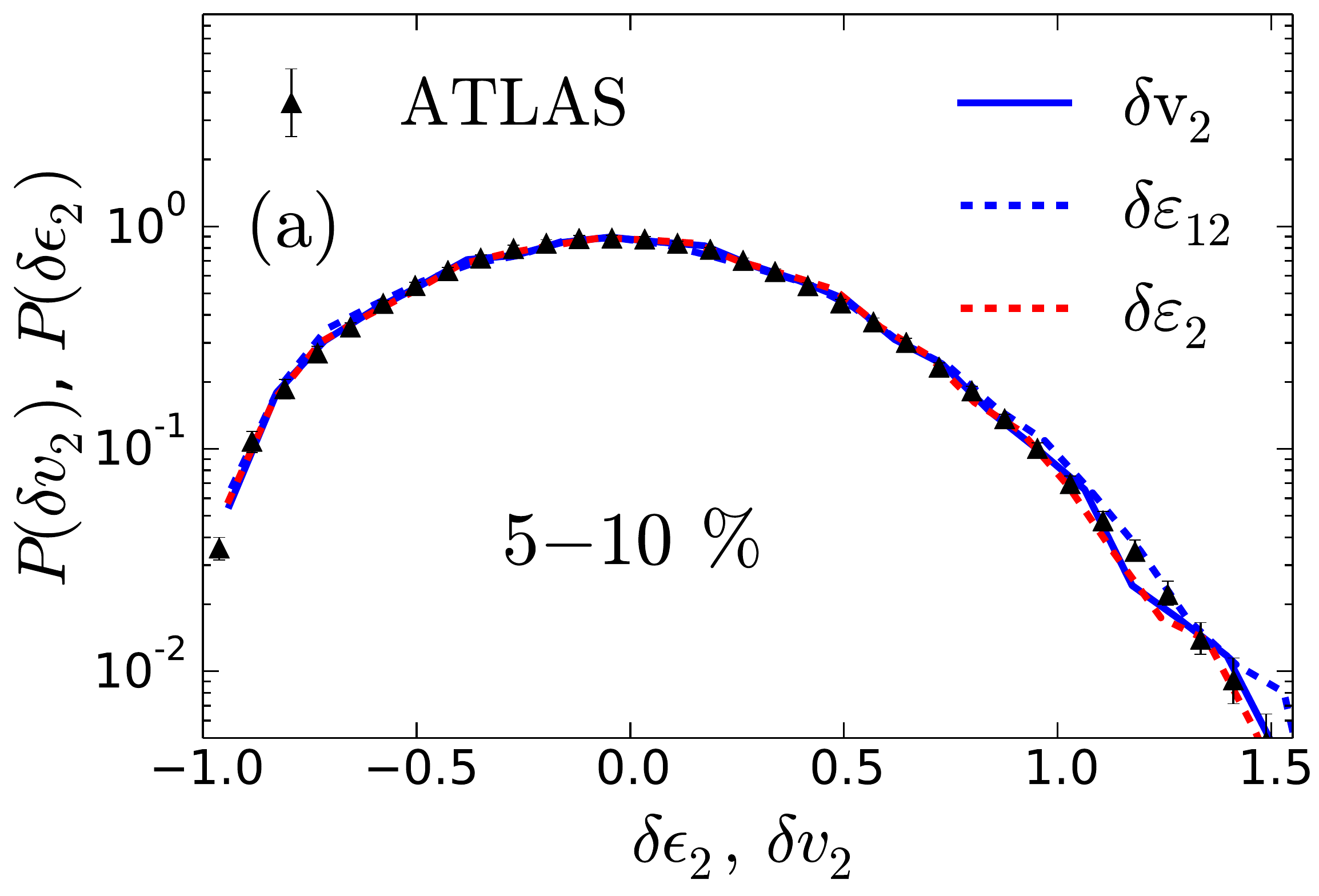}
\includegraphics[width=5.9cm]{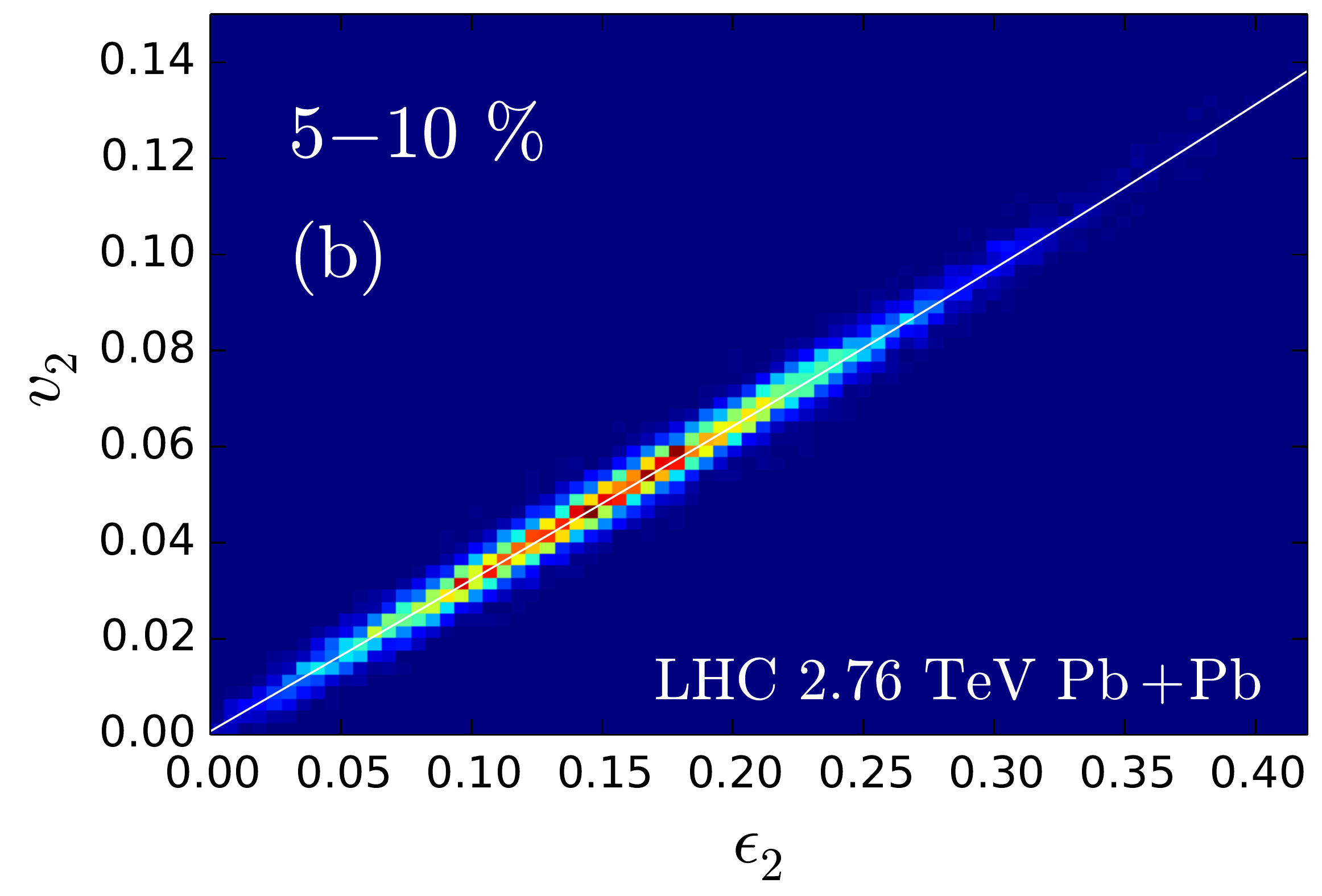}
\includegraphics[width=5.9cm]{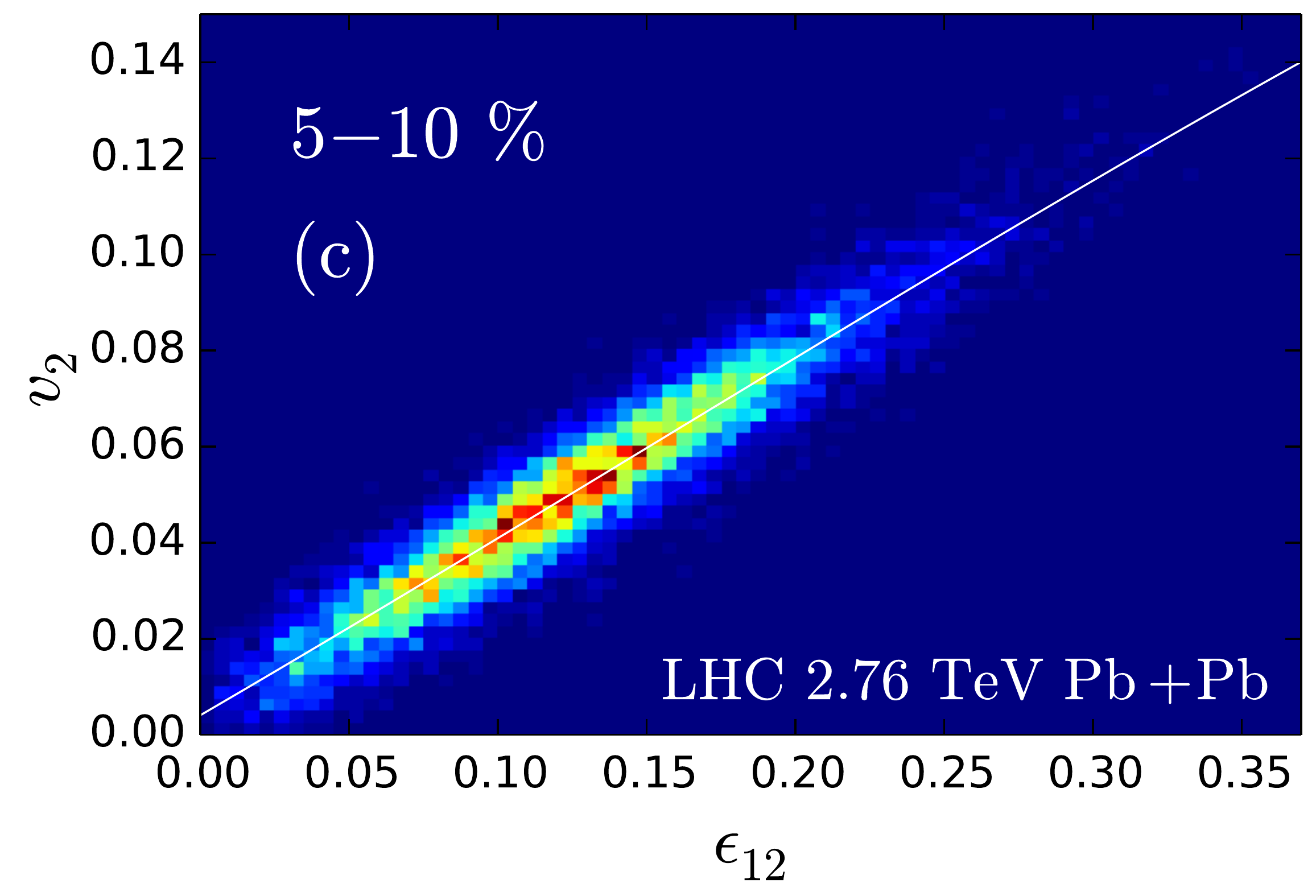}
%%%%%%%%%%%%%%%%%%%%% FIGURE %%%%%%%%%%%%%%%%%%%%%
%%%%%%%%%%%%%%%%%%%%% FIGURE %%%%%%%%%%%%%%%%%%%%%
\includegraphics[width=5.9cm]{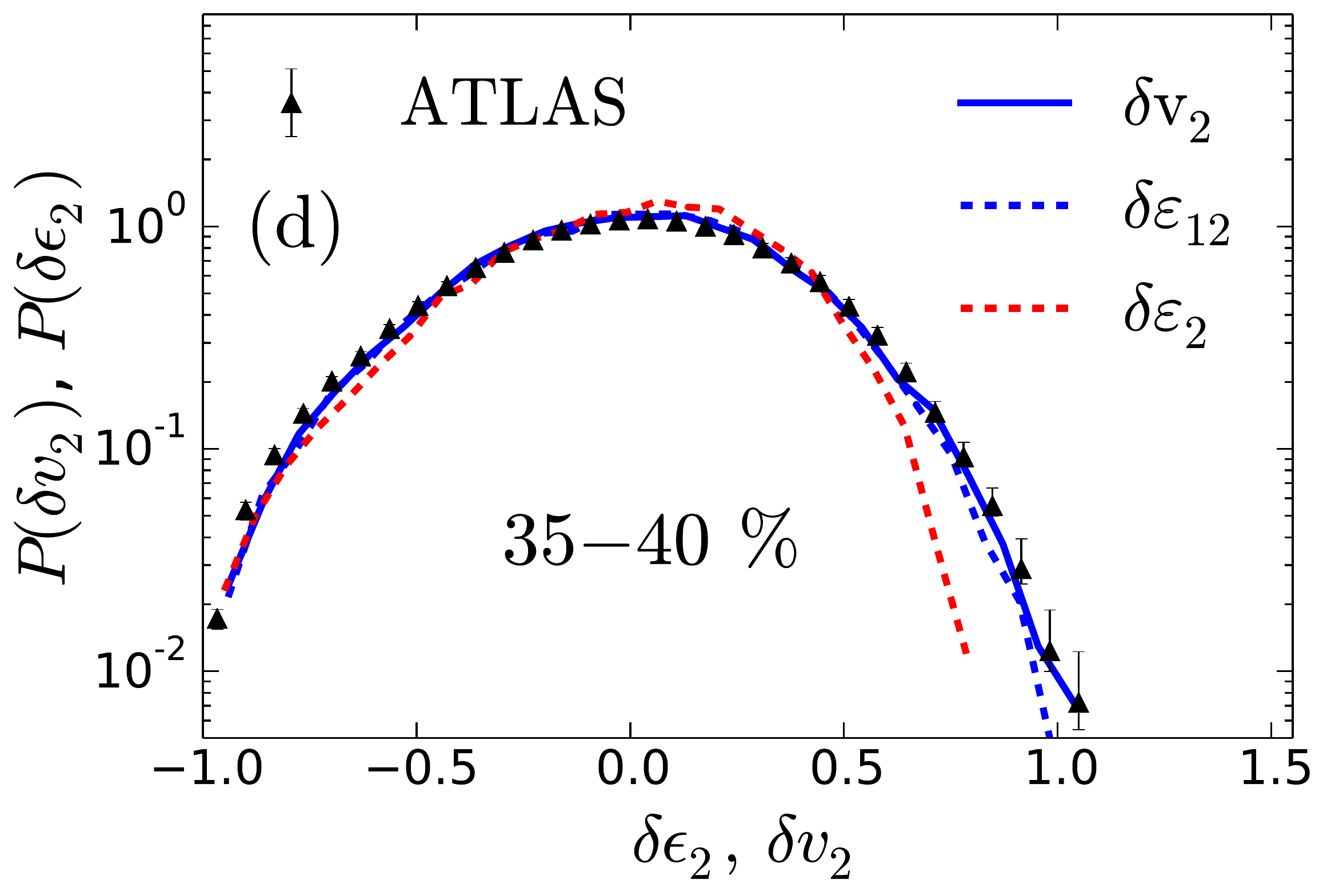}
\includegraphics[width=5.9cm]{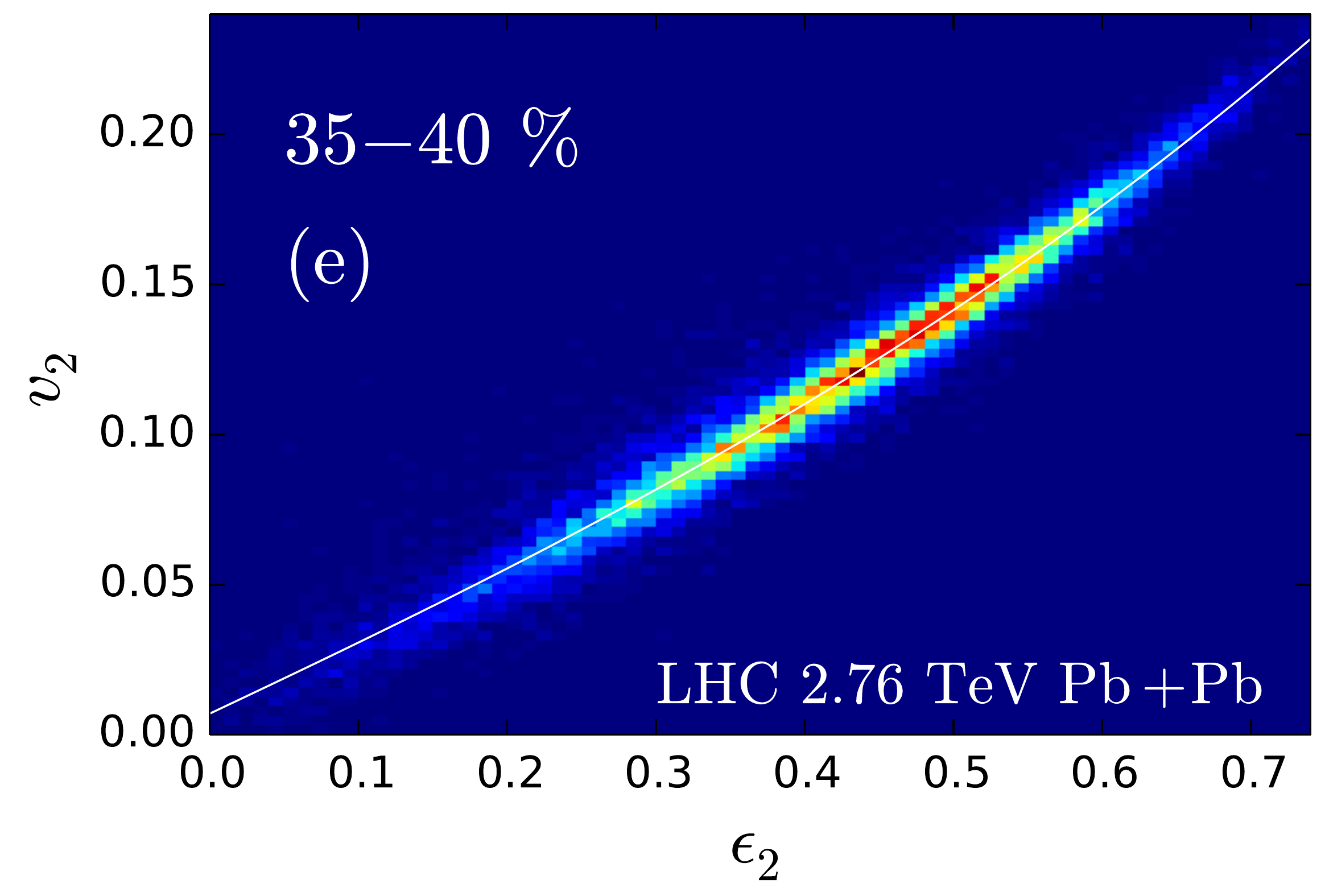}
\includegraphics[width=5.9cm]{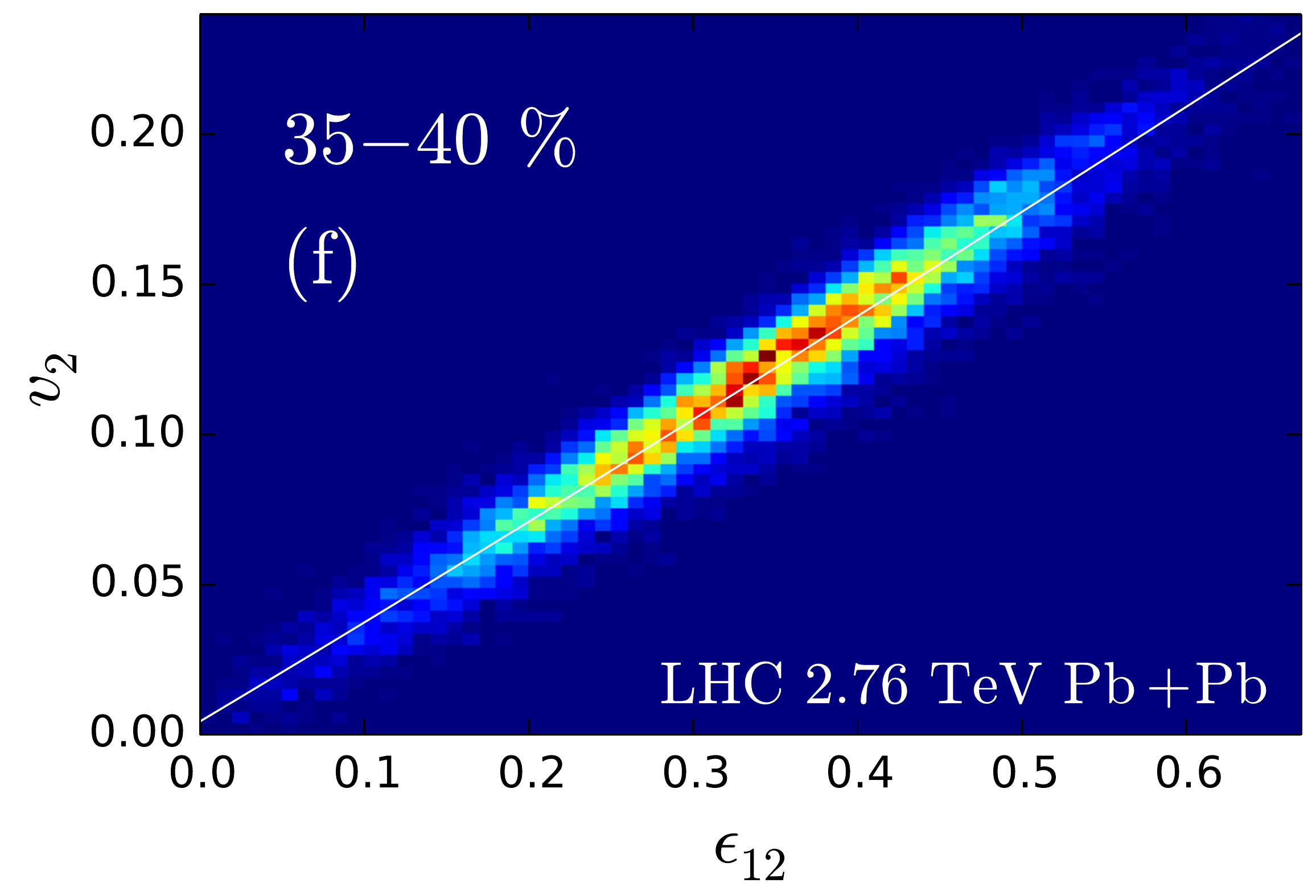}
%%%%%%%%%%%%%%%%%%%%% FIGURE %%%%%%%%%%%%%%%%%%%%%
%%%%%%%%%%%%%%%%%%%%% FIGURE %%%%%%%%%%%%%%%%%%%%%
\includegraphics[width=5.9cm]{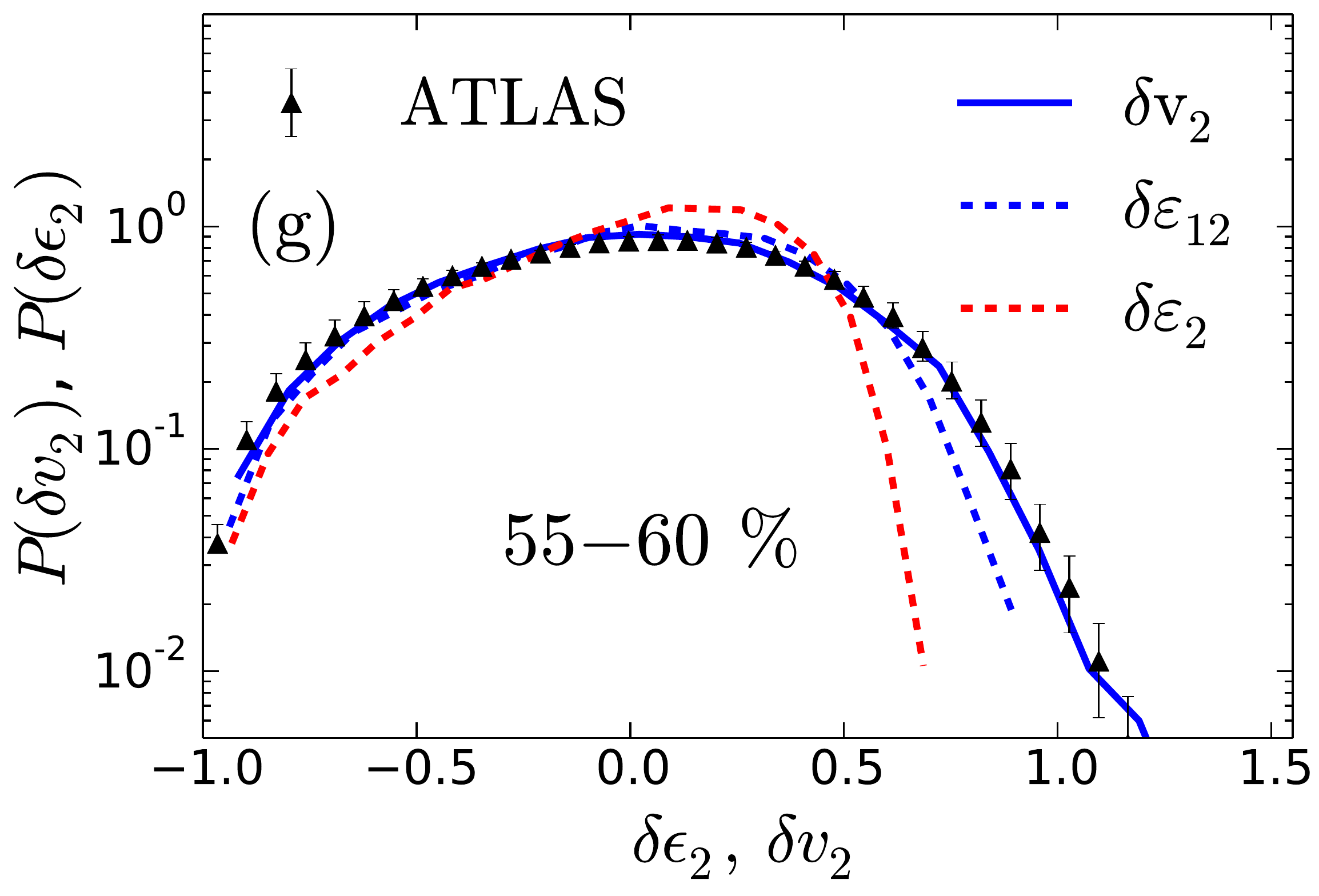}
\includegraphics[width=5.9cm]{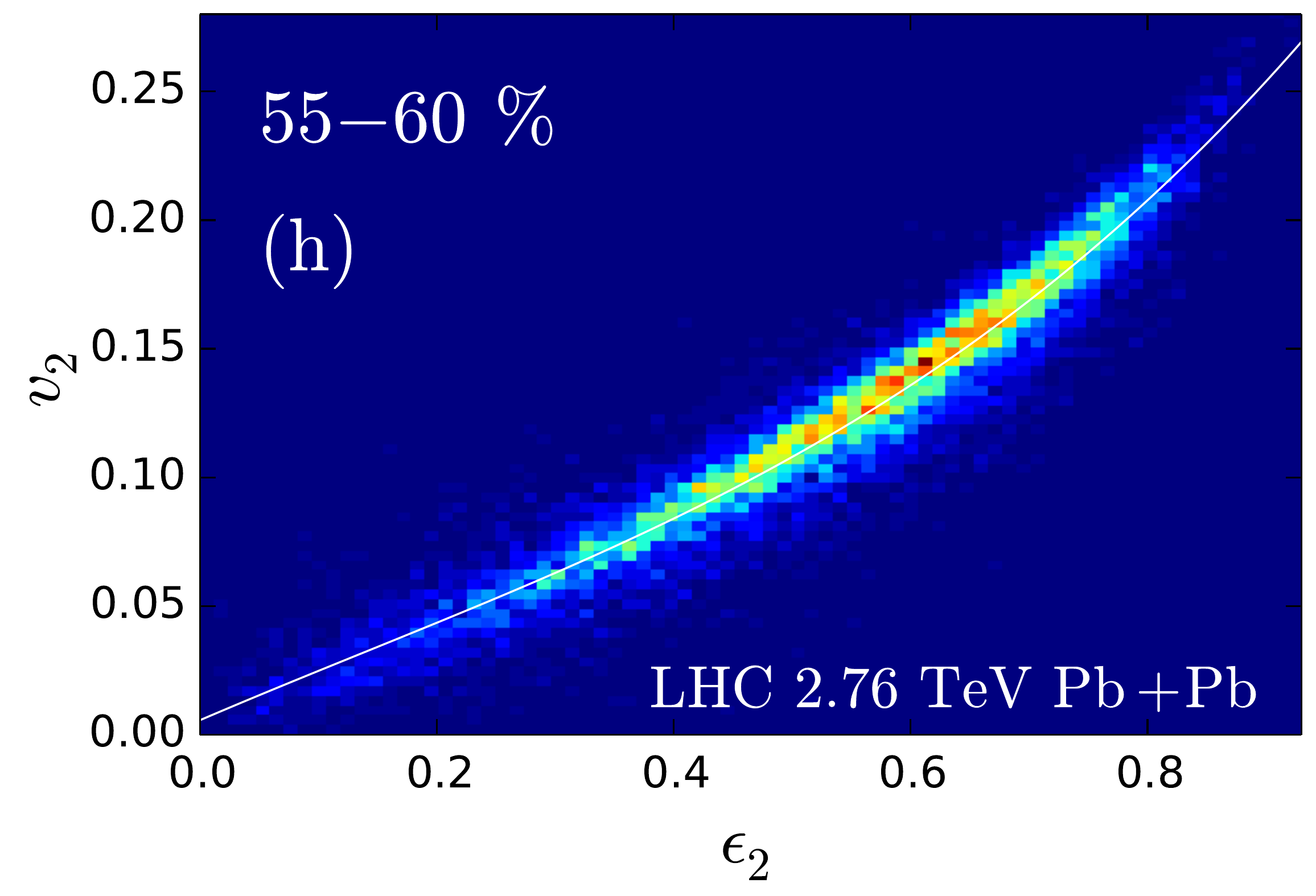}
\includegraphics[width=5.9cm]{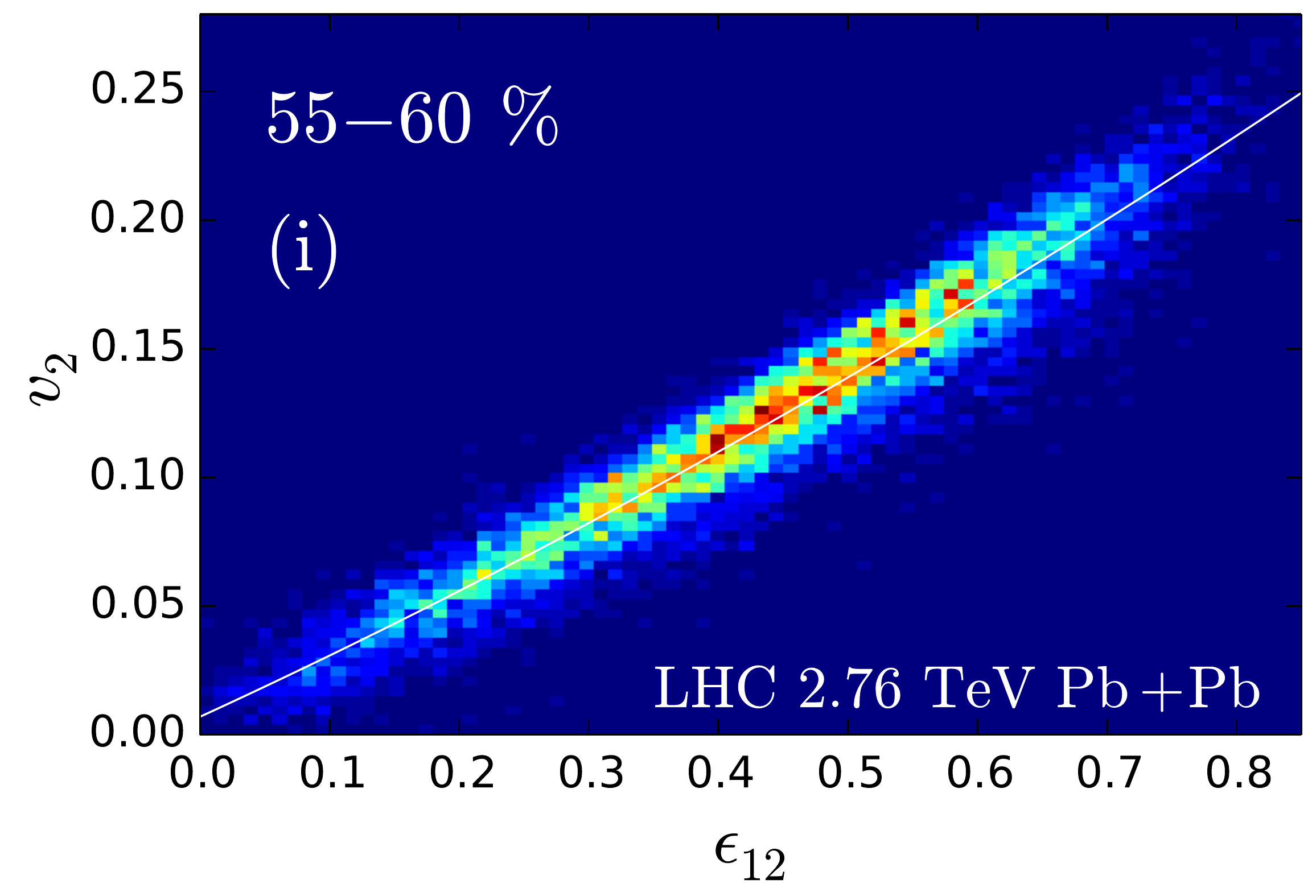}
\caption{(Color online) Left panels: Probability distributions of the charged hadron $\delta v_2$, and of the initial state $\delta \varepsilon_{2}$ and $\delta \varepsilon_{12}$, in the $5-10$ \% (top), $35-40$ \% (middle), and $55-60$ \% (bottom) centrality classes in $\sqrt{s_{NN}} = 2.76$  TeV Pb+Pb collisions at the LHC, computed with the pQCD + saturation initial states. The experimental data is from ATLAS \cite{Aad:2013xma}.
Middle panels: The correlation between $v_2$ and $\varepsilon_2$ as a two-dimensional histogram. 
Right panels: The correlation between $v_2$ and $\varepsilon_{12}$. 
The white lines in the middle and right panels are cubic polynomial fits, to guide the eye.
The statistics for these figures was 15k events for each centrality class}
\label{fig:ev_correlations}
\end{figure*}
%%%%%%%%%%%%%%%%%%%%% FIGURE %%%%%%%%%%%%%%%%%%%%%
In the left panels of Fig.~\ref{fig:ev_correlations} we show the probability distributions of $\delta v_2$, $\delta \varepsilon_2$, and $\delta \varepsilon_{1,2}$ in the $5-10$ \%, $35-40$ \%, and $55-60$ \% centrality classes, obtained with the pQCD + saturation initial conditions and $\eta/s=0.20$. The middle panels show the correlation between $v_2$ and $\varepsilon_2$, and the right panels show the correlation between $v_2$ and $\varepsilon_{1,2}$.  In the $5-10$ \% centrality class all three distributions are practically the same. The linear relation between $v_2$ and $\varepsilon_2$ holds very well, thus the corresponding fluctuation spectra fall on top of each other. As the top right panel indicates, the correlation between $v_2$ and $\varepsilon_{1,2}$ is visibly weaker, but the average $v_2$ computed at a fixed $\varepsilon_{1,2}$ still grows linearly with $\varepsilon_{1,2}$, so that again $P(\delta v_2)\approx P(\delta\varepsilon_{1,2})$. 

In the $35-40$ \% centrality class, the $(v_2, \varepsilon_2)$--correlation is still very strong, but there is already a clear deviation from a \emph{linear} correlation, and as a result the $v_2$ and $\varepsilon_2$ distributions are not anymore the same. However, the $(v_2, \varepsilon_{1,2})$--correlation is similar to the one in the near-central collisions, and the scaled $v_2$ distribution is practically the same as the scaled $\varepsilon_{1,2}$ distribution. In even more peripheral collisions, i.e., in the $55-60$ \% centrality class, the $(v_2, \varepsilon_2)$--correlations show even stronger deviations from a linear correlation, and there is a slight deviation from the linear $(v_2, \varepsilon_{1,2})$--correlation as well.

Overall, the $(v_2, \varepsilon_2)$--correlations are somewhat stronger than those of $(v_2, \varepsilon_{1,2})$ but exhibit a strong non-linear behavior in more peripheral collisions. On the other hand, the $(v_2, \varepsilon_{1,2})$--correlations stay more linear, and in central to mid-peripheral collisions the scaled $v_2$ distributions follow closely the scaled $\varepsilon_{1,2}$ distributions, but in more peripheral collisions also they start to deviate from each other. Based on the middle and r.h.s. panels we can also deduce why $\delta \varepsilon_2$ spectrum in peripheral collisions becomes narrower than that of $\delta \varepsilon_{1,2}$: for the averages $\langle \varepsilon_2\rangle
> \langle \varepsilon_{1,2}\rangle$ but the rare largest fluctuations are about the same magnitude, which for such largest absolute fluctuations means that 
$\varepsilon_2 - \langle \varepsilon_2\rangle < \varepsilon_{1,2} - \langle \varepsilon_{1,2}\rangle$, and for the scaled fluctuations even more strongly
$\delta\varepsilon_2 < \langle \varepsilon_{1,2}\rangle$.

At the moment, it is not clear whether one could find a more specific definition of the eccentricity that would always be able to predict the $v_2$ distributions, or if the non-linear correlations remain inevitably a necessary part of the analysis. However, we emphasize that in a full fluid-dynamical calculation as presented here, the different definitions of the initial state eccentricities do not play a role in obtaining the final-state observables: The agreement between the ATLAS data and our calculations is very good, systematically over a wide range of centralities.

Another way to get an access to the flow fluctuations are the flow cumulants. As discussed in Sec.~\ref{sec:cumulants}, if the non-flow contributions to the flow coefficients can be suppressed by the pseudorapidity gaps, the essential difference between $v_n\{2\}$ and $v_n\{4\}$ is that they measure the different moments of the probability distribution $P(v_n)$. In principle, the full set of cumulants provides the same information as the probability distributions themselves. Thus, if we describe the $v_2\{2\}$ measurements simultaneously with the full $v_n$ probability distributions, and the non-flow contributions are small, we should also agree with the $v_2\{4\}$ measurements. This turns out to be the case. 

\begin{figure*}
\includegraphics[width=8.5cm]{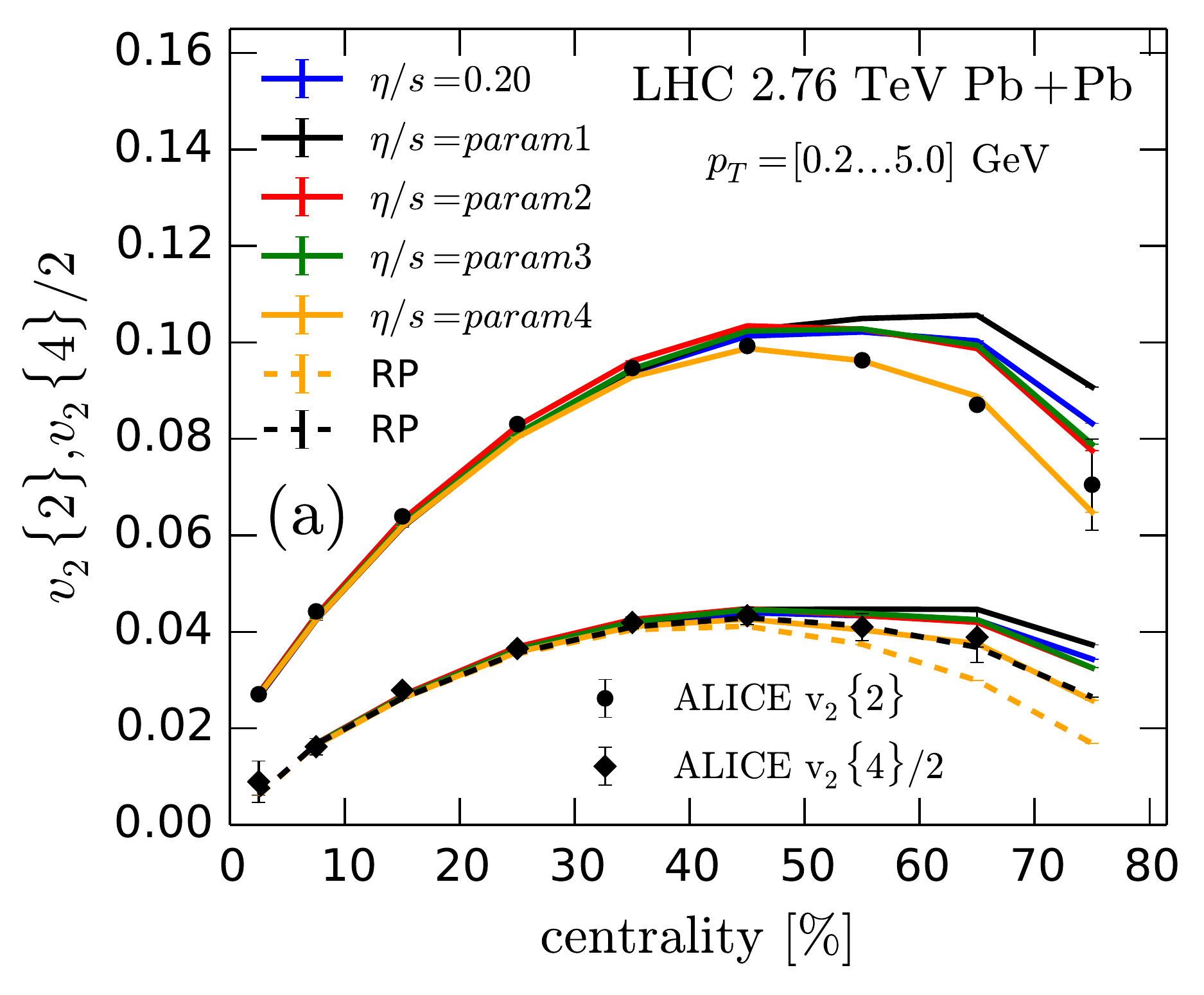}
\includegraphics[width=8.5cm]{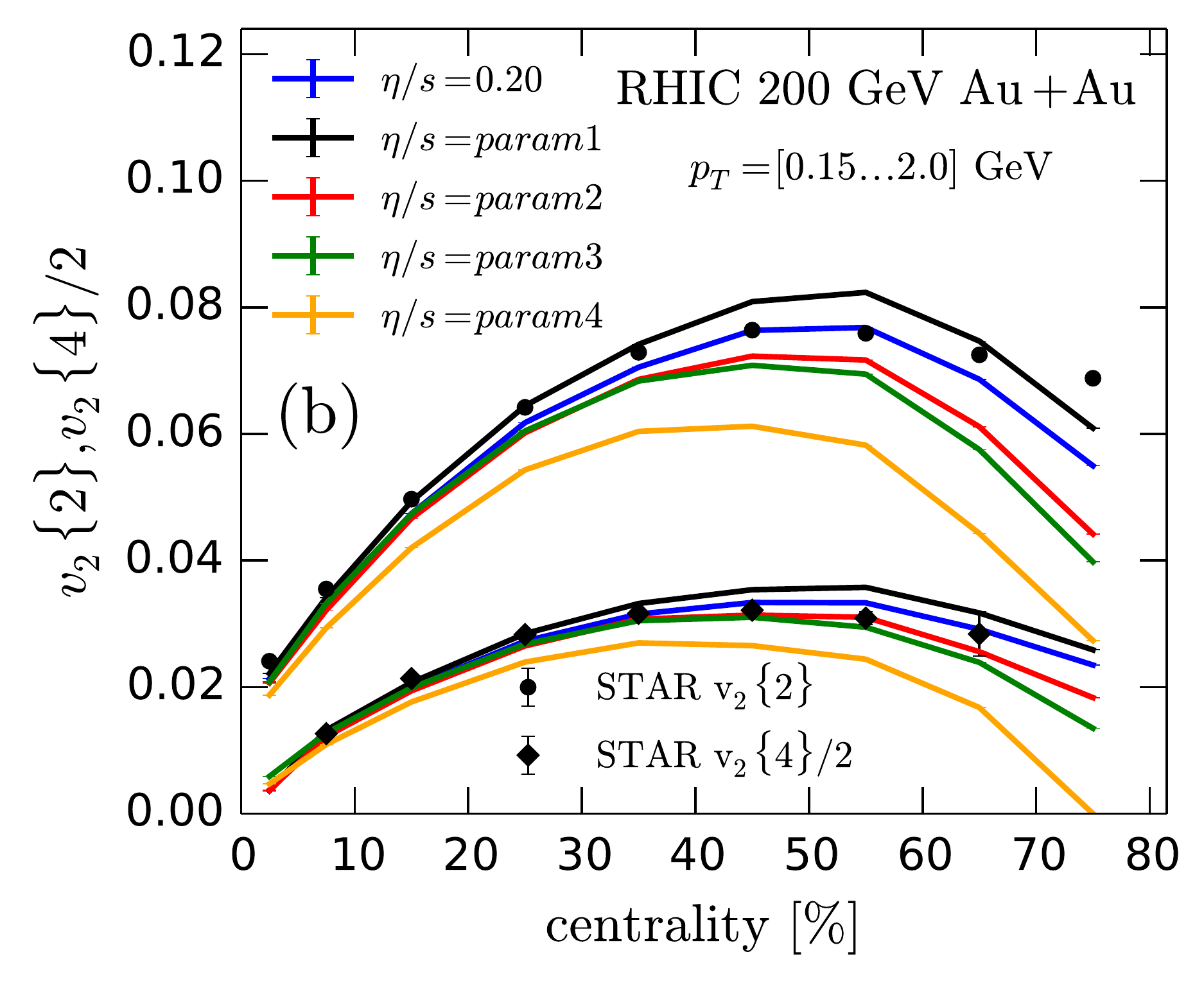}
\caption{(Color online) 2- and 4-particle cumulant flow-coefficients, $v_n\{2\}$ and $v_n\{4\}$, of charged hadrons in $\sqrt{s_{NN}} = 2.76$  TeV Pb+Pb collisions at the LHC (a), and in $200$  GeV Au+Au collisions at RHIC (b). The $v_n\{4\}$ results are divided by 2 for clarity. The dashed lines show the $v_n$ calculated with respect to the reaction plane (RP). The data are from ALICE\cite{ALICE:2011ab, Aamodt:2010pa} and STAR \cite{Adams:2004bi}, and the corresponding $p_T$ ranges are indicated.}
\label{fig:charged_vn_cumulants}
\end{figure*}

In Fig.~\ref{fig:charged_vn_cumulants}a we show the $v_2\{4\}$ of charged hadrons in Pb+Pb collisions at the LHC with different $\eta/s$ parametrizations, against the ALICE data~\cite{ALICE:2011ab, Aamodt:2010pa}. For comparison we also show the $v_2\{2\}$ results from Fig.~\ref{fig:charged_vn}. As one can see, the agreement with both measurements is equally good. Figure \ref{fig:charged_vn_cumulants}b shows the corresponding $v_2\{2\}$ and $v_2\{4\}$ in Au+Au collisions at RHIC compared to the STAR data~\cite{Adams:2004bi}. The measurements of the full probability distributions are not currently available at RHIC energies, but the fact that those $\eta/s$ parametrizations that give a good agreement with the $v_2\{2\}$ measurements also give an equally good agreement with $v_2\{4\}$ measurements already indicates that also at RHIC the main features of the probability distributions are correct in our approach with the pQCD + saturation initial conditions.

If the flow fluctuations are approximately Gaussian, then $v_2\{4\}$ is approximately equivalent to the $v_n\{\mathrm{RP}\}$ determined with respect to the reaction plane 
(the calculational plane whose $x$ axis is along the impact parameter) \cite{Voloshin:2007pc}. In Fig.~\ref{fig:charged_vn_cumulants}a we show also $v_n\{\mathrm{RP}\}$ with $\eta/s$ from $param1$ and $param4$. As one can see, $v_n\{\mathrm{RP}\}$ and $v_n\{4\}$ agree very well approximately up to the $40-50$ \% centralities. Looking back at the left panels of Fig.~\ref{fig:ev_correlations}, this result is expected, since towards peripheral collisions the fluctuation spectrum exhibits more clearly a non-Gaussian behavior.
%%%%%%%%%%%%%%%%%%%%% FIGURE %%%%%%%%%%%%%%%%%%%%%
%%%%%%%%%%%%%%%%%%%%% FIGURE %%%%%%%%%%%%%%%%%%%%%

%%%%%%%%%%%%%%%%%%%%% SECTION %%%%%%%%%%%%%%%%%%%%%
\subsection{Event-plane correlations}

%%%%%%%%%%%%%%%%%%%%% FIGURE %%%%%%%%%%%%%%%%%%%%%
\begin{figure*}
\includegraphics[width=17.5cm]{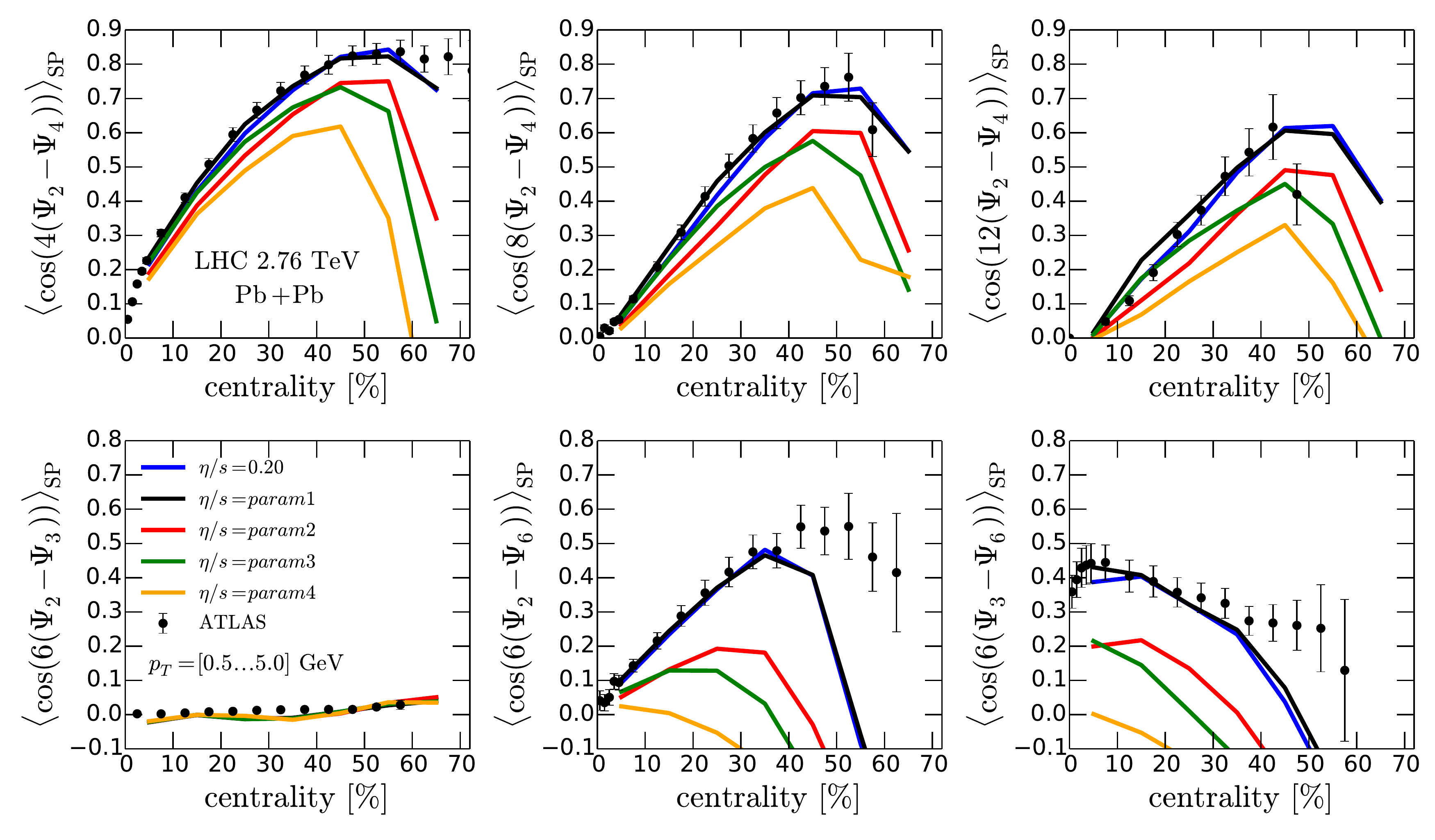}
\caption{(Color online) Correlations of two event-plane angles for charged particles in $\sqrt{s_{NN}} = 2.76$  TeV Pb+Pb collisions at the LHC, compared with the ATLAS data \cite{Aad:2014fla}.}
\label{fig:eventplane_correlation2}
\end{figure*}
%%%%%%%%%%%%%%%%%%%%% FIGURE %%%%%%%%%%%%%%%%%%%%%
Because fluid dynamics is a non-linear theory, there is no reason to expect that the linear relation e.g.\ between the eccentricities and flow coefficients, $v_n \propto \varepsilon_n$, holds in general or even that $v_n$ is created by a non-linear response to the $\varepsilon_n$ alone. In reality, the different $v_n$'s or $\Psi_n$'s do not evolve independently, but are correlated with each other, e.g. a large $v_2$ can create a large $v_4$ even if the initial $\varepsilon_4$ is zero. The evidence for this can be clearly seen in the measured event-plane correlations~\cite{Bilandzic:2012an, Aad:2014fla}, which show a strong correlation between various event-plane angles $\Psi_n$. 

%%%%%%%%%%%%%%%%%%%%% FIGURE %%%%%%%%%%%%%%%%%%%%%
\begin{figure*}
\includegraphics[width=17.5cm]{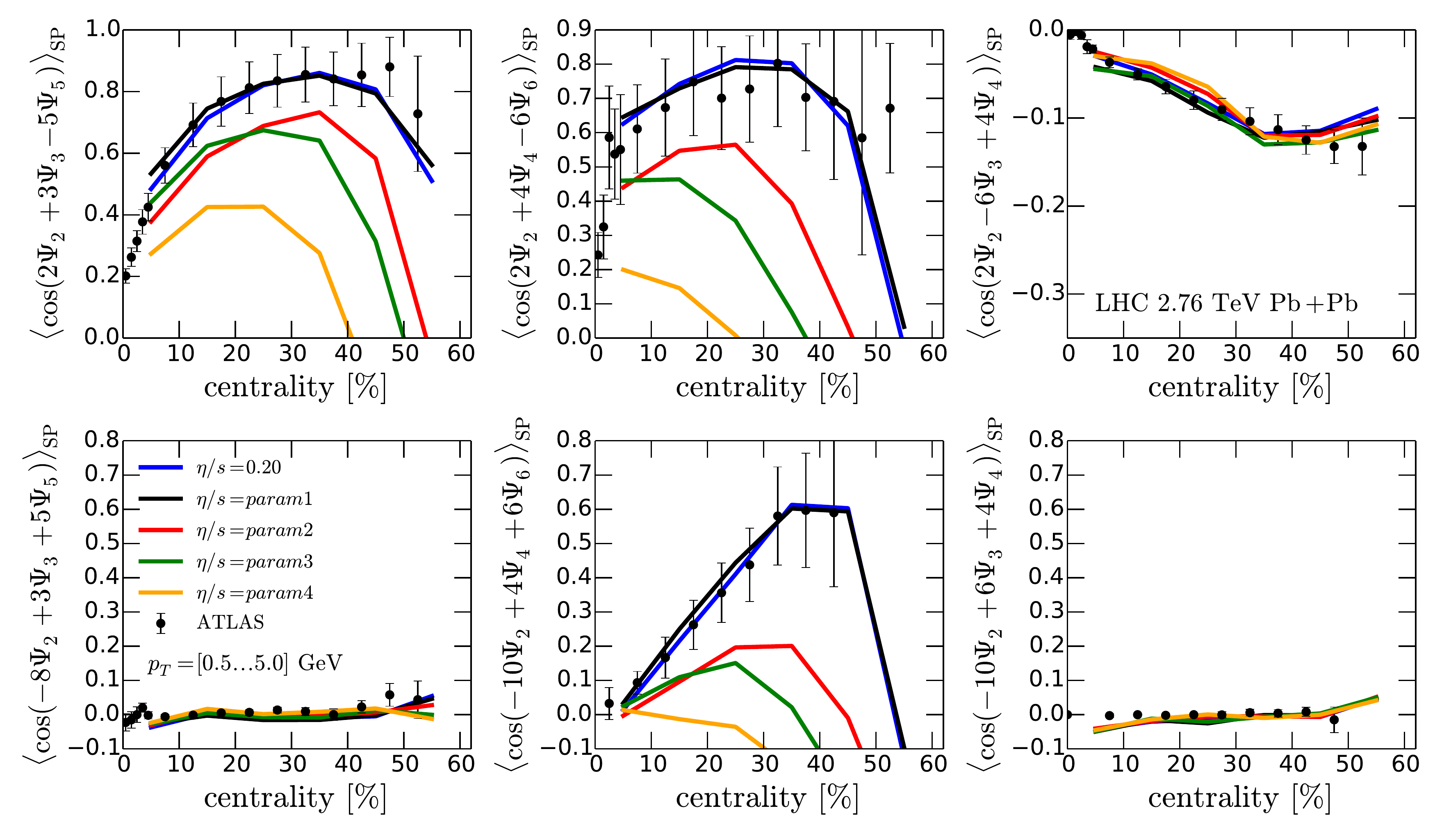}
\caption{(Color online) Correlations of three event-plane angles for charged particles in $\sqrt{s_{NN}} = 2.76$  TeV Pb+Pb collisions at the LHC, compared with the ATLAS data \cite{Aad:2014fla}.}
\label{fig:eventplane_correlation3}
\end{figure*}
%%%%%%%%%%%%%%%%%%%%% FIGURE %%%%%%%%%%%%%%%%%%%%%

Even though the correlation between the initial eccentricities creates correlations between $v_n$'s through a linear relation $v_n \propto \varepsilon_n$, even the signs of the measured correlations cannot be reproduced by this assumption. A generic behavior of the correlations can be explained by a linear response between the eccentricities defined through cumulants \cite{Teaney:2012ke} and $v_n$'s, but quantitatively the magnitude of the correlations indicates that a non-linear fluid dynamical evolution is essential to reproduce the measurements, see Ref.~\cite{Teaney:2013dta}. Furthermore, and most importantly for the present study, the event-plane correlations give independent constraints to the initial state and transport coefficients, even if the viscosity is tuned to reproduce the $v_2$ data~\cite{Qiu:2012uy}. 

In Fig.~\ref{fig:eventplane_correlation2} we show various event-plane correlations involving two different event-plane angles $\Psi_n$, defined by Eq.~\eqref{eq:epcorrelation}, in Pb+Pb collisions at the LHC, compared to the ATLAS measurements~\cite{Aad:2014fla}. As can be seen from the figure, the different $\eta/s$ parametrizations that give an equivalent agreement with the $v_n$ data at the LHC, can be clearly distinguished by the correlations. Only two cases, $\eta/s=0.20$ and $\eta/s=param1$, give a good agreement with the ATLAS data. Only in the peripheral collisions ($40-50$ \% centrality class and more peripheral) the correlations involving $\Psi_6$ are not reproduced. A further discussion on how viscosity affects the correlations is given in the next section. 

The ATLAS Collaboration has also measured correlations involving three different event-plane angles~\cite{Aad:2014fla}. As shown by Fig.~\ref{fig:eventplane_correlation3}, these are equivalently well reproduced in our framework by the same two parametrizations of $\eta/s$ as the two event-plane angle correlations above, but do not provide any further constraints to our setup so that $\eta/s=0.20$ and $\eta/s=param1$ parametrizations could be further separated by these measurements. It is to be emphasized that the same $\eta/s$ parametrizations that give the best fit to $v_n$ data at RHIC, also gives the best fit to the LHC event-angle correlators. 

In Au+Au collisions at RHIC the $v_4\{3\}$ measurement by the STAR Collaboration is actually similar to the event-plane angle measurement, as it involves also a correlation between the angles $\Psi_2$ and $\Psi_4$,  see the definition Eq.~\eqref{eq:v4_3ple}. This particular measurement, shown in Fig.~\ref{fig:charged_vn}b is also well described by the $\eta/s=0.20$ and $\eta/s=param1$ parametrizations.

Finally, we note that typically the required statistics (number of events) for the correlators is much higher than for the $v_n$ coefficients themselves, and it also depends strongly on the strength of the correlation. For example, for the correlation between $\Psi_2$ and $\Psi_3$, which is almost zero in Fig.~\ref{fig:eventplane_correlation2}d, the ATLAS Collaboration measures clearly a positive value, while our current statistics ($20k$ events for each $\eta/s$ parametrization) is not sufficient to accurately calculate such a small correlation but the statistical errors are larger than the signal itself.

\section{Discussion}

The dissipative suppression of the final azimuthal asymmetries of the spectra is a result of a combination of the dissipative effect into the fluid dynamical flow field, generated during the evolution, and the magnitude of the shear-stress tensor at the decoupling, i.e. the magnitude of the $\delta f$ corrections to the equilibrium distributions in Eq.~\eqref{eq:delta_f}. The relative contribution of these two effects depends on the $\eta/s$ parametrization and collision energy.

%%%%%%%%%%%%%%%%%%%%% FIGURE %%%%%%%%%%%%%%%%%%%%%
\begin{figure*}
\includegraphics[width=8.5cm]{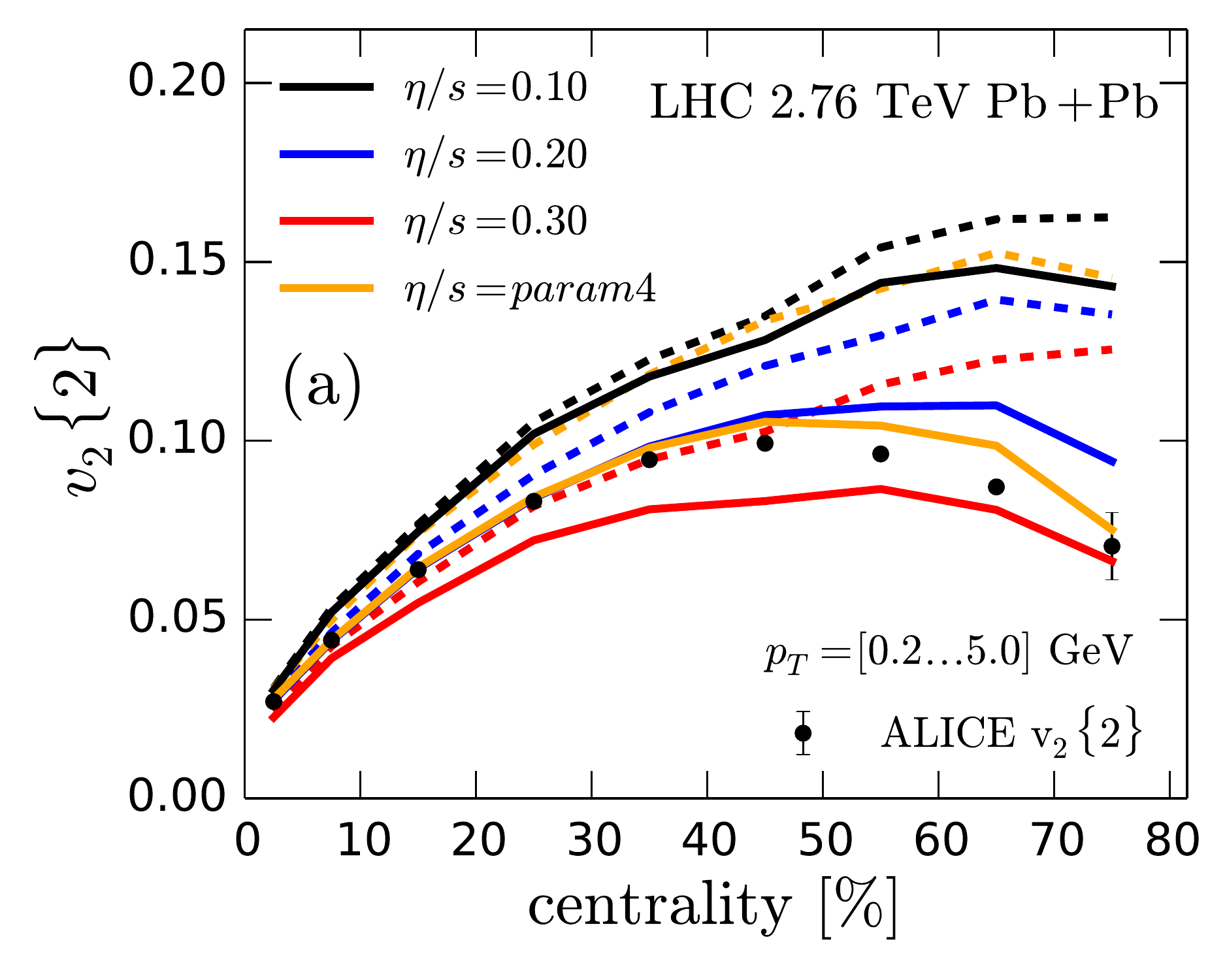}
\includegraphics[width=8.5cm]{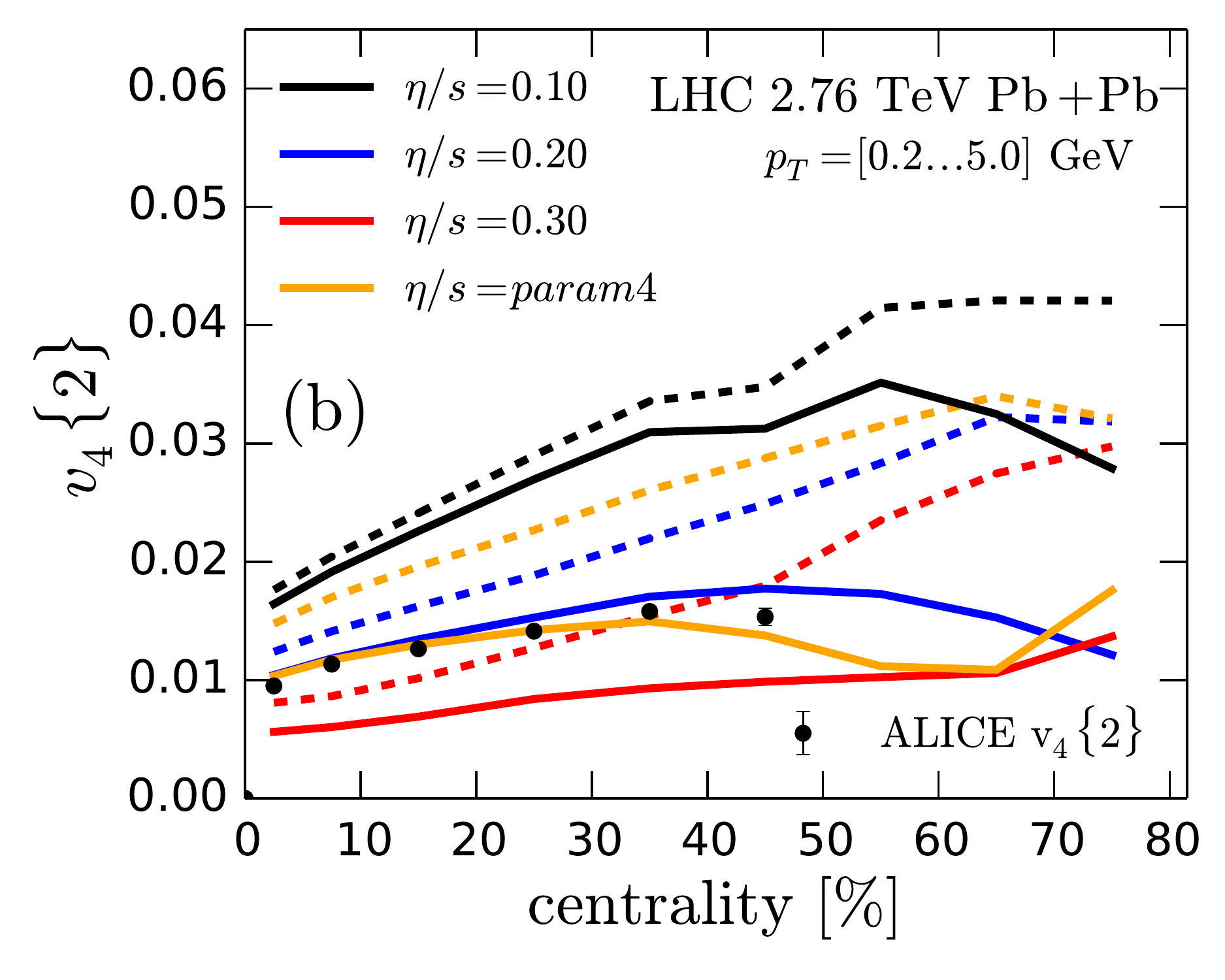}
\caption{(Color online) Left: Centrality dependence of the flow coefficients $v_2\{2\}$  from the charged hadron 2-particle cumulants in $\sqrt{s_{NN}}=2.76$ TeV Pb+Pb collisions at the LHC, for 4 different $\eta/s$ cases, with (solid curves) and without (dashed) the $\delta f$ corrections. Right: The same for $v_4$. The experimental data is from ALICE \cite{ALICE:2011ab} and the $p_T$ interval is indicated.}
\label{fig:charged_vn_deltaf}
\end{figure*}
%%%%%%%%%%%%%%%%%%%%% FIGURE %%%%%%%%%%%%%%%%%%%%%
In order to illustrate the effects of $\delta f$, we show in Fig.~\ref{fig:charged_vn_deltaf}a the centrality dependence of $v_2$ at the LHC, calculated with three different constant $\eta/s$ values ($\eta/s=0.1$, $0.2$ and $0.3$), and with $\eta/s$ from the parametrization $param4$ which has a large viscosity in the hadronic phase. The full results are shown with solid lines, and the results without the $\delta f$ contribution with dashed lines. Figure \ref{fig:charged_vn_deltaf}b shows the same, but for $v_4$. Note that for these checks of the  $(\eta,\delta f)$ systematics we do not include the decay contributions, so the solid lines for $\eta/s=0.2$ and $param4$ are not exactly the same as in Fig.~\ref{fig:charged_vn} but serve the purpose here. As one can see from the figures, the relative size of the $\delta f$ contribution increases with increasing $\eta/s$, and also from central to peripheral collisions. In addition it is larger for higher harmonics, i.e. relatively larger for $v_4$ than for $v_2$.

In this work, we have tested different temperature-dependent parametrizations of $\eta/s$ against the flow coefficient data from $A$+$A$ collisions at RHIC and the LHC. First we noticed that the $v_n$ measurements at the LHC alone do not give strong constraints on the temperature dependence of $\eta/s$ but all our different parametrizations give an equally good agreement with the LHC data. We emphasize that this is not trivially so, as the final azimuthal asymmetry is generated in different ways with the different $\eta/s(T)$ parametrizations. This can be seen by comparing $\eta/s=0.20$ and $\eta/s=param4$ curves in Fig.~\ref{fig:charged_vn_deltaf}. Both these parametrizations are tuned to reproduce the $v_n$ data at the LHC, but the $\delta f$ contribution is significantly larger with $\eta/s=param4$ due to the larger hadronic viscosity. Therefore, in order to reproduce the data, the viscosity effects during the evolution need to be weaker in this parametrization compared to the $\eta/s=0.20$ case. It turns out, however, that even if the relative contribution from $\delta f$ and from the evolution is different, the centrality dependence of $v_n$ is very closely the same with both parametrizations. One can see the differences only in very peripheral collisions where one has to be cautious about the applicability of fluid dynamics. Therefore, the current $v_n$ measurements at the LHC alone cannot reliably distinguish between the different temperature dependencies of $\eta/s$ at the level depicted in Fig.~\ref{fig:etapers}, or in other words, they cannot be used to distinguish the $\delta f$ contributions from the dissipative effects in the spacetime evolution of the flow field.

\begin{figure*}
\includegraphics[width=17.5cm]{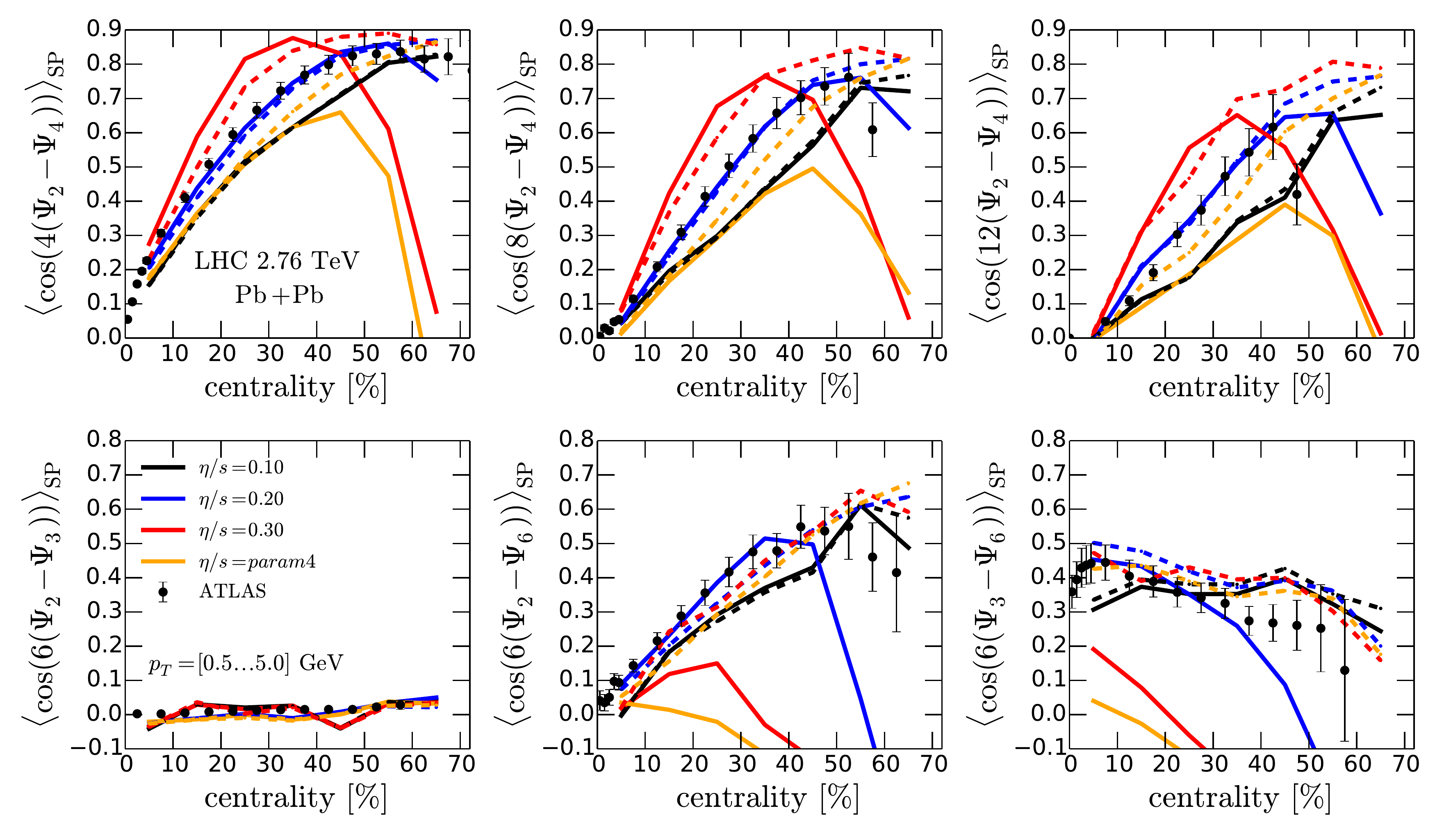}
\caption{(Color online) Correlations of two event-plane angles for charged particles in $\sqrt{s_{NN}}=2.76$ TeV Pb+Pb collisions at the LHC, computed with (solid) and without (dashed) the $\delta f$ corrections. The experimental data is from ATLAS \cite{Aad:2014fla}.}
\label{fig:eventplane_correlation_deltaf}
\end{figure*}

For $v_n$ both contributions, $\delta f$ and the dissipative effects in the evolution, work in the same direction, i.e., both suppress the flow coefficients. Interestingly, the same is not true for the event-plane correlations. While $\delta f$ still suppresses the correlations, increasing the viscosity during the evolution can enhance the correlations. This is can be seen in Fig.~\ref{fig:eventplane_correlation_deltaf}, where we show the event-plane correlations with the same $\eta/s$ parametrizations,  again with and without $\delta f$, as in the previous figure. In particular, one can see that the correlation between $\Psi_2$ and $\Psi_4$ gets clearly stronger when $\eta/s$ is increased from $0.1$ to $0.3$. The $\delta f$ contributions for these correlators remain small in the near-central and semi-peripheral collisions for all these $\eta/s$ parametrizations. Towards more peripheral collisions, however, the effect of $\delta f$ sets in, very quickly decorrelating the angles. 

Because of their different dependence on the viscosity, the event-plane correlations offer complementary information about the temperature dependence of $\eta/s$. Moreover, the weak dependence of $\langle \cos \left(N\left(\Psi_2 - \Psi_4\right)\right)\rangle$ on the $\delta f$ in central and mid-peripheral collisions gives confidence that these correlations actually probe the dissipation during the evolution. Furthermore, the relative $\delta f$ contribution to $v_n$ also changes with collision energy, e.g., the $\delta f$ contribution is generally larger at RHIC energy. Therefore, it is remarkable that the same $\eta/s$ parametrizations that give the best agreement with the LHC correlation data also give the best agreement with the $v_n$ data at RHIC. 

One should, however, keep in mind that large $\delta f$ is a result of large values of inverse Reynolds number $R_{\pi}^{-1}=\pi^{\mu\nu}/P_0$ at the decoupling, which means that the system is not close to local thermal equilibrium, and the larger the $R_{\pi}^{-1}$ the less reliable the fluid dynamical approximation becomes. Currently, it is not known to how large values of $R_{\pi}^{-1}$ we can go in the current fluid-dynamical picture, so that we can still reliably calculate the evolution. 

%%%%%%%%%%%%%%%%%%%%% FIGURE %%%%%%%%%%%%%%%%%%%%%
%%%%%%%%%%%%%%%%%%%%% FIGURE %%%%%%%%%%%%%%%%%%%%%

\begin{figure}
\includegraphics[width=8.5cm]{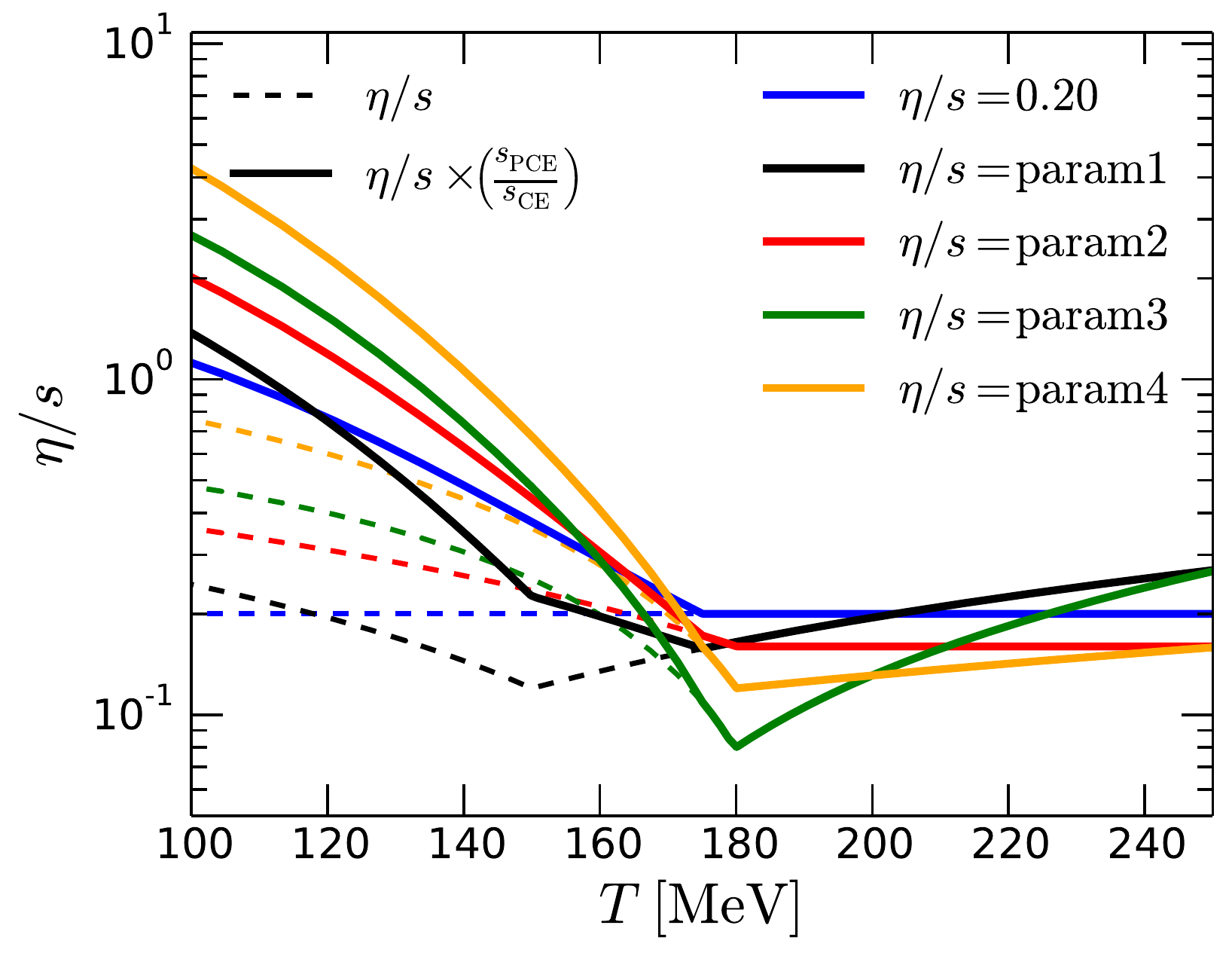}
\caption{(Color online) Parametrizations of the temperature dependence of the shear-viscosity to entropy ratio, scaled by the entropy density ratio of chemically frozen and chemical equilibrium system. The dashed curves show the original parametrizations of Fig.~\ref{fig:etapers}. }
\label{fig:etapers_scaled}
\end{figure}

Our results indicate that in order to keep the consistency with all the data shown here, the hadronic $\eta/s$ cannot be too large. At first this seems to be inconsistent with several microscopic calculations that show a strong increase of hadronic $\eta/s$ as temperature decreases, see e.g.\ Refs.~\cite{Prakash:1993bt,Csernai:2006zz,Gorenstein:2007mw,NoronhaHostler:2008ju,Wiranata:2013oaa}. However, it should be noted that in our case, below the chemical freeze-out temperature $T_{\rm chem}=175$ MeV, the entropy density in $\eta/s$ is not an entropy density of the system in full chemical equilibrium. Therefore, the comparison to microscopic calculations, see e.g.\ Refs.~\cite{Arnold:2003zc,Wesp:2011yy,Ozvenchuk:2012kh,Christiansen:2014ypa}, which typically assume a full chemical equilibrium at all temperatures, would require an estimate of how $\eta/s(T)=\eta/s_{\rm PCE}(T)$ is related to the full equilibrium $\eta/s_{\rm CE}$. In order to estimate the magnitude of this difference, we show in Fig.~\ref{fig:etapers_scaled} our $\eta/s$ parametrizations scaled with the ratio of entropy densities in chemically frozen system and system in chemical equilibrium, i.e. $\eta/s_{\rm CE} = \left(\eta/s_{\rm PCE}\right) \times \left( \frac{s_{\rm PCE}}{s_{\rm CE}} \right)$. At least in a simplified hadron gas $\eta$ itself depends only weakly on the chemical composition \cite{Wiranata:2014jda}, and the main difference between $\eta/s_{\rm PCE}$ and $\eta/s_{\rm CE}$ is due to the change in the entropy density. The original parametrizations are shown as dashed curves. As one can see, the entropy densities of the two systems at low temperatures are significantly different. For example, the constant $\eta/s=0.20$ scaled by the entropy ratio, is very close to the original non-scaled $\eta/s=param4$ parametrization, which is the one with the highest hadronic viscosity.

%%%%%%%%%%%%%%%%%%%%% FIGURE %%%%%%%%%%%%%%%%%%%%%
%%%%%%%%%%%%%%%%%%%%% FIGURE %%%%%%%%%%%%%%%%%%%%%

\section{Conclusions}

In this paper, we have developed an event-by-event framework of the NLO-improved pQCD + saturation + viscous fluid dynamics model \cite{Paatelainen:2013eea}. The main conclusions from the new EbyE EKRT framework are the following: 

1) We have now systematically tested the approach and successfully challenged it against a multitude of LHC and RHIC data. The centrality dependence of multiplicities, low-$p_T$ spectra, flow coefficients at the LHC and RHIC, and even the event-plane angle correlations at the LHC all come out in a beautiful agreement with experimental data. Especially the measured probability distributions of $\delta v_2$ at the LHC offer a stringent test for the computed pQCD + saturation initial states. We have also demonstrated the necessity of fluid dynamical evolution in describing the full centrality dependence of the measured $v_2$ fluctuation spectra. The multiobservable analysis which is performed simultaneously for the LHC and RHIC, together with our systematic fluid-dynamical cross-checks, suggests that the EbyE EKRT framework works remarkably well for collisions up to 40$\dots$50 \% centralities. 

2) At the same time, as the main goal of this paper, we obtain improved constraints to the QCD matter $\eta/s(T)$. We tested several parametrizations of $\eta/s(T)$, all tuned to reproduce the $v_n\{2\}$ in the mid-central collisions at the LHC. In practice, the centrality dependence of the $v_n$ coefficients at the LHC alone do not give strong constraints to the temperature dependence of $\eta/s$, but all our parametrizations shown in Fig.~\ref{fig:etapers} give an equally good agreement with the data. The differences show only in peripheral collisions, where the uncertainties of the framework also grow large. A simultaneous analysis of the flow coefficients at RHIC gives more stringent constraints, and of the parametrizations considered here the constant $\eta/s=0.20$ and $\eta/s=param1$, with a small hadronic viscosity and minimum $\eta/s$ at $T=150$ MeV, give an overall best agreement with the flow coefficients at the LHC and RHIC. Especially $\eta/s=param4$ with the largest hadronic viscosity gives too strong a suppression of the flow coefficients at RHIC.

3) The event-plane angle correlations which have been measured at the LHC, and which are here shown to probe especially the viscous effects in the space-time evolution of the QCD matter, provide most useful additional and also rather stringent constraints for $\eta/s(T)$. Remarkably, again the \textit{same} $\eta/s(T)$ which gives the best agreement with the RHIC $v_n$ data, reproduces also the LHC event-plane angle correlations best. To put a real statistical error bar onto $\eta/s(T)$ requires a full global analysis of the LHC and RHIC heavy-ion bulk data, see e.g.\ Refs.\ \cite{Novak:2013bqa,Bernhard:2015hxa,Pratt:2015zsa}. This is clearly beyond the scope of our study here but we consider the present paper as an important step towards such an analysis.

It is good to look back at the main uncertainties of the framework presented here. Our NLO calculation for the minijet $E_T$ is -- as an IR/CL-safe calculation and with the given PDFs, $p_0$, $\Delta y$ and $\beta$ -- rigorous. The saturation as we consider it here, is a conjecture but clearly it captures quite correctly the dominant features in the initial minijet production, from which we then compute the initial energy densities and formation times locally in the transverse plane. Our handling of the pre-thermal evolution from the local formation times to the starting time of the fluid dynamical simulation could in principle be improved by giving the initial minijet energy densities to the fluid dynamics as source terms at the locally varying formation times. Then, however, it is not clear whether the used fluid-dynamical picture is still valid as the density gradients and additional entropy generation at the earliest stages of evolution would become even larger than what they are in the present study with $\tau_0=0.2$~fm. Alternatively, one could develop an EbyE model also for the minijet production and feed the minijets obtained in each event into a parton cascade description such as BAMPS \cite{Uphoff:2014cba}, and extract the initial conditions for fluid-dynamics (including also the possible initial transverse flow now assumed to be zero) at a later time. On the fluid-dynamics side, the largest uncertainties are related to the treatment of the late hadronic evolution, e.g.\ chemical and kinetic decouplings. This might improve if one couples the fluid dynamics with a hadron cascade in the hadronic phase at high enough temperature. Then, however, one type of model uncertainties are replaced with uncertainties related to the matching conditions at the switching surface and uncertainties related to, e.g., the applicability of the cascade for very dense hadron systems, and also to the many unknown scattering cross sections one is forced to assume in such a simulation.

The evident next step in our NLO-improved pQCD + saturation EbyE framework is to consider also the dynamical fluctuations of initial gluon densities in the colliding nuclei, which are then reflected as additional fluctuations of the saturation scale and hence of the computed initial energy densities. The inclusion of these fluctuations will improve our description of ultra-central heavy-ion collisions, and also allow us to study the extremely interesting question of collectivity and flow p+Pb collisions at the LHC, see e.g.~Refs.\ \cite{Bozek:2013ska, Werner:2013ipa, Werner:2013tya, Nagle:2013lja,Kozlov:2014hya}. An interesting further question is the rapidity dependence of all the observables studied here. For this, one needs to develop a more complete EbyE framework by introducing a pQCD minijet event generator which is coupled with the determination of saturation in each event and which by construction also accounts for the different types of fluctuations.

\acknowledgments
We thank Pasi Huovinen, Tuomas Lappi, Matthew Luzum, Jean-Yves Ollitrault, Hannu Paukkunen, Dirk Rischke and Kimmo Tuominen for useful discussions. This work was financially supported by the the Helmholtz International Center for FAIR within the framework of the LOEWE program launched by the State of Hesse (HN), and the European Research Council grant HotLHC, No. ERC-2011-StG-279579 (RP). We acknowledge CSC -- IT Center for Science in Espoo, Finland, for the allocation of the computational resources.

\bibliographystyle{apsrev4-1}
\bibliography{ebyereferences}

\end{document}